\input harvmac
\input epsf
\noblackbox

\def\CN{{\cal N}}

\def\ket#1{|#1\rangle}


\def\unlockat{\catcode`\@=11}
\def\lockat{\catcode`\@=12}

\unlockat

\def\newsec#1{\global\advance\secno by1\message{(\the\secno. #1)}
\global\subsecno=0\global\subsubsecno=0\eqnres@t\noindent
{\bf\the\secno. #1}
\writetoca{{\secsym} {#1}}\par\nobreak\medskip\nobreak}
\global\newcount\subsecno \global\subsecno=0
\def\subsec#1{\global\advance\subsecno
by1\message{(\secsym\the\subsecno. #1)}
\ifnum\lastpenalty>9000\else\bigbreak\fi\global\subsubsecno=0
\noindent{\it\secsym\the\subsecno. #1}
\writetoca{\string\quad {\secsym\the\subsecno.} {#1}}
\par\nobreak\medskip\nobreak}
\global\newcount\subsubsecno \global\subsubsecno=0
\def\subsubsec#1{\global\advance\subsubsecno
\message{(\secsym\the\subsecno.\the\subsubsecno. #1)}
\ifnum\lastpenalty>9000\else\bigbreak\fi
\noindent\quad{\secsym\the\subsecno.\the\subsubsecno.}{#1}
\writetoca{\string\qquad{\secsym\the\subsecno.\the\subsubsecno.}{#1}}
\par\nobreak\medskip\nobreak}

\def\subsubseclab#1{\DefWarn#1\xdef
#1{\noexpand\hyperref{}{subsubsection}%
{\secsym\the\subsecno.\the\subsubsecno}%
{\secsym\the\subsecno.\the\subsubsecno}}%
\writedef{#1\leftbracket#1}\wrlabeL{#1=#1}}
\lockat

\def\IL{{\relax{\rm I\kern-.18em L}}}
\def\IH{{\relax{\rm I\kern-.18em H}}}
\def\IR{{\relax{\rm I\kern-.18em R}}}
\def\IE{{\relax{\rm I\kern-.18em E}}}
\def\IC{{\relax\hbox{$\inbar\kern-.3em{\rm C}$}}}
\def\IZ{{\relax\ifmmode\mathchoice
{\hbox{\cmss Z\kern-.4em Z}}{\hbox{\cmss Z\kern-.4em Z}}
{\lower.9pt\hbox{\cmsss Z\kern-.4em Z}}
{\lower1.2pt\hbox{\cmsss Z\kern-.4em Z}}\else{\cmss Z\kern-.4em
Z}\fi}}

\def\CN {{\cal N}}

\def\CF {{\cal F}}

\def\CO {{\cal O}}

\def\CH {{\cal H}}


\def\CN {{\cal N}}

\def\CO {{\cal O}}

\font\manual=manfnt \def\dbend{\lower3.5pt\hbox{\manual\char127}}

\def\IZ{{\relax\ifmmode\mathchoice
{\hbox{\cmss Z\kern-.4em Z}}{\hbox{\cmss Z\kern-.4em Z}}
{\lower.9pt\hbox{\cmsss Z\kern-.4em Z}}
{\lower1.2pt\hbox{\cmsss Z\kern-.4em Z}}\else{\cmss Z\kern-.4em
Z}\fi}}

\def\bbbone{{\mathchoice {\rm 1\mskip-4mu l} {\rm 1\mskip-4mu l}
          {\rm 1\mskip-4.5mu l} {\rm 1\mskip-5mu l}}}

\def\CN {{\cal N}}

\def\CO {{\cal O}}

\def\IOm{\relax\thinspace\inbar\kern1.95pt\inbar\kern-5.525pt\Omega}


\def\IZ{{\relax\ifmmode\mathchoice
{\hbox{\cmss Z\kern-.4em Z}}{\hbox{\cmss Z\kern-.4em Z}}
{\lower.9pt\hbox{\cmsss Z\kern-.4em Z}}
{\lower1.2pt\hbox{\cmsss Z\kern-.4em Z}}\else{\cmss Z\kern-.4em
Z}\fi}}
\def\IB{{\relax{\rm I\kern-.18em B}}}
\def\IC{{\relax\hbox{$\inbar\kern-.3em{\rm C}$}}}
\def\ID{{\relax{\rm I\kern-.18em D}}}
\def\IE{{\relax{\rm I\kern-.18em E}}}
\def\IF{{\relax{\rm I\kern-.18em F}}}
\def\IG{{\relax\hbox{$\inbar\kern-.3em{\rm G}$}}}
\def\IGa{{\relax\hbox{${\rm I}\kern-.18em\Gamma$}}}
\def\IH{{\relax{\rm I\kern-.18em H}}}
\def\II{{\relax{\rm I\kern-.18em I}}}
\def\IK{{\relax{\rm I\kern-.18em K}}}
\def\IP{{\relax{\rm I\kern-.18em P}}}

\def\inbar{\,\vrule height1.5ex width.4pt depth0pt}

\font\cmss=cmss10 \font\cmsss=cmss10 at 7pt
\def\IR{\relax{\rm I\kern-.18em R}}
\def\IT{\relax{\rm I\kern-.45em T}}

\def\Tr{{\rm Tr}}


\def\boxit#1{\vbox{\hrule\hbox{\vrule\kern8pt
\vbox{\hbox{\kern8pt}\hbox{\vbox{#1}}\hbox{\kern8pt}}
\kern8pt\vrule}\hrule}}
\def\mathboxit#1{\vbox{\hrule\hbox{\vrule\kern8pt\vbox{\kern8pt
\hbox{$\displaystyle #1$}\kern8pt}\kern8pt\vrule}\hrule}}


\def\inbar{\,\vrule height1.5ex width.4pt depth0pt}

\font\cmss=cmss10 \font\cmsss=cmss10 at 7pt
\def\IR{\relax{\rm I\kern-.18em R}}

\def\Tr{{\rm Tr}}


\def\Im{{\rm Im}}

\let\includefigures=\iftrue
\newfam\black
\includefigures

\input epsf
\def\plb#1 #2 {Phys. Lett. {\bf B#1} #2 }
\long\def\del#1\enddel{}
\long\def\new#1\endnew{{\bf #1}}
\let\<\langle \let\>\rangle

\def\figin{\epsfcheck\figin}\def\figins{\epsfcheck\figins}
\def\epsfcheck{\ifx\epsfbox\UnDeFiNeD
\message{(NO epsf.tex, FIGURES WILL BE IGNORED)}
\gdef\figin##1{\vskip2in}\gdef\figins##1{\hskip.5in} blank space instead
\else\message{(FIGURES WILL BE INCLUDED)}
\gdef\figin##1{##1}\gdef\figins##1{##1}\fi}
\def\DefWarn#1{}
\def\figinsert{\goodbreak\midinsert}
\def\ifig#1#2#3{\DefWarn#1\xdef#1{fig.~\the\figno}
\writedef{#1\leftbracket fig.\noexpand~\the\figno}
\figinsert\figin{\centerline{#3}}\medskip
\centerline{\vbox{\baselineskip12pt
\advance\hsize by -1truein\noindent
\footnotefont{\bf Fig.~\the\figno:} #2}}
\bigskip\endinsert\global\advance\figno by1}
\else
\def\ifig#1#2#3{\xdef#1{fig.~\the\figno}
\writedef{#1\leftbracket fig.\noexpand~\the\figno}
\figinsert\figin{\centerline{#3}}\medskip
\centerline{\vbox{\baselineskip12pt
\advance\hsize by -1truein\noindent
\footnotefont{\bf Fig.~\the\figno:} #2}}
\bigskip\endinsert
\global\advance\figno by1}
\fi

\input xy
\xyoption{all}
\font\cmss=cmss10 \font\cmsss=cmss10 at 7pt
\def\inbar{\,\vrule height1.5ex width.4pt depth0pt}
\def\IC{{\relax\hbox{$\inbar\kern-.3em{\rm C}$}}}
\def\IP{{\relax{\rm I\kern-.18em P}}}
\def\IF{{\relax{\rm I\kern-.18em F}}}
\def\IZ{\relax\ifmmode\mathchoice
{\hbox{\cmss Z\kern-.4em Z}}{\hbox{\cmss Z\kern-.4em Z}}
{\lower.9pt\hbox{\cmsss Z\kern-.4em Z}}
{\lower1.2pt\hbox{\cmsss Z\kern-.4em Z}}\else{\cmss Z\kern-.4em
Z}\fi}
\def\IR{{\relax{\rm I\kern-.18em R}}}
\def\IQ{\relax\hbox{\kern.25em$\inbar\kern-.3em{\rm Q}$}}

\def\pmb#1{\setbox0=\hbox{#1}%
 \kern-.025em\copy0\kern-\wd0
 \kern.05em\copy0\kern-\wd0
 \kern-.025em\raise.0433em\box0 }
\font\cmss=cmss10
\font\cmsss=cmss10 at 7pt
\def\rlx{\relax\leavevmode}
\def\Cop{\relax\,\hbox{$\inbar\kern-.3em{\rm C}$}}
\def\Rop{\relax{\rm I\kern-.18em R}}
\def\Nop{\relax{\rm I\kern-.18em N}}
\def\Pop{\relax{\rm I\kern-.18em P}}
\def\Zop{\rlx\leavevmode\ifmmode\mathchoice{\hbox{\cmss Z\kern-.4em Z}}
 {\hbox{\cmss Z\kern-.4em Z}}{\lower.9pt\hbox{\cmsss Z\kern-.36em Z}}
 {\lower1.2pt\hbox{\cmsss Z\kern-.36em Z}}\else{\cmss Z\kern-.4em
 Z}\fi}

\def\inbar{\,\vrule height1.5ex width.4pt depth0pt}
\def\IC{{\relax\hbox{$\inbar\kern-.3em{\rm C}$}}}
\def\IP{{\relax{\rm I\kern-.18em P}}}
\def\IF{{\relax{\rm I\kern-.18em F}}}
\def\IZ{\relax\ifmmode\mathchoice
{\hbox{\cmss Z\kern-.4em Z}}{\hbox{\cmss Z\kern-.4em Z}}
{\lower.9pt\hbox{\cmsss Z\kern-.4em Z}}
{\lower1.2pt\hbox{\cmsss Z\kern-.4em Z}}\else{\cmss Z\kern-.4em
Z}\fi}
\def\IR{{\relax{\rm I\kern-.18em R}}}
\def\IT{{\mathchoice {\setbox0=\hbox{$\displaystyle\rm
T$}\hbox{\hbox to0pt{\kern0.3\wd0\vrule height0.9\ht0\hss}\box0}}
{\setbox0=\hbox{$\textstyle\rm T$}\hbox{\hbox
to0pt{\kern0.3\wd0\vrule height0.9\ht0\hss}\box0}}
{\setbox0=\hbox{$\scriptstyle\rm T$}\hbox{\hbox
to0pt{\kern0.3\wd0\vrule height0.9\ht0\hss}\box0}}
{\setbox0=\hbox{$\scriptscriptstyle\rm T$}\hbox{\hbox
to0pt{\kern0.3\wd0\vrule height0.9\ht0\hss}\box0}}}}
\def\bbbti{{\mathchoice {\setbox0=\hbox{$\displaystyle\rm
T$}\hbox{\hbox to0pt{\kern0.3\wd0\vrule height0.9\ht0\hss}\box0}}
{\setbox0=\hbox{$\textstyle\rm T$}\hbox{\hbox
to0pt{\kern0.3\wd0\vrule height0.9\ht0\hss}\box0}}
{\setbox0=\hbox{$\scriptstyle\rm T$}\hbox{\hbox
to0pt{\kern0.3\wd0\vrule height0.9\ht0\hss}\box0}}
{\setbox0=\hbox{$\scriptscriptstyle\rm T$}\hbox{\hbox
to0pt{\kern0.3\wd0\vrule height0.9\ht0\hss}\box0}}}}
\def\K{{\cal{K}}}

\def\N{{\cal N}}


 \nref\Acharya{B. S. Acharya, ``A Moduli Fixing Mechanism in M-Theory,''
hep-th/0212294.}
\nref\ahar{O. Aharony, C. Sonnenschein, S. Yankielowicz and S. Theisen,
``Field Theory Questions for String Theory Answers'', Nucl. Phys. {\bf B493} (1997)
177, hep-th/9611222.
}
 \nref\alda{
G. Aldazabal, A. Font, L. E. Ibanez and G. Violero,
``$D=4$, ${\cal N}=1$, Type IIB Orientifolds,''
Nucl. Phys. {\bf B536} (1998) 29, hep-th/9804026.}
\nref\angelantonj{
C.~Angelantonj, I.~Antoniadis, G.~D'Appollonio, E.~Dudas and A.~Sagnotti,
``Type I Vacua with Brane Supersymmetry Breaking,''
Nucl.\ Phys.\  {\bf B572} (2000) 36, hep-th/9911081.}
\nref\angelreview{  C.~Angelantonj and A.~Sagnotti,
``Open Strings,''   Phys.\ Rept.\  {\bf 371} (2002) 1,
  [Erratum-ibid.\  {\bf 376} (2003) 339], hep-th/0204089.}
 \nref\AD{S. Ashok and M. R. Douglas, ``Counting Flux Vacua,'' JHEP
{\bf 0401} (2004) 060, hep-th/0307049.}
 \nref\aspbrane{P. S. Aspinwall,
``Enhanced Gauge Symmetries and $K3$ Surfaces,''
Phys. Lett. {\bf B357} (1995) 329, hep-th/9507012.}
 \nref\BB{V. Balasubramanian and P. Berglund, ``Stringy
Corrections to K\"ahler Potentials, Susy Breaking, and the
Cosmological Constant Problem,'' JHEP {\bf 0411} (2004) 085, hep-th/0408054.}
 \nref\BBtwo{V. Balasubramanian, P. Berglund, J. Conlon and
F. Quevedo, ``Systematics of Moduli Stabilisation in Calabi-Yau
Flux Compactifications,'' hep-th/0502058.}
 \nref\BeckerNN{ K.~Becker, M.~Becker, M.~Haack and J.~Louis,
``Supersymmetry Breaking and $\alpha'$-Corrections to Flux Induced
Potentials,'' JHEP {\bf 0206}, (2002) 060, hep-th/0204254.}
 \nref\BergmanRP{
  O.~Bergman, E.~G.~Gimon and S.~Sugimoto,
  ``Orientifolds, RR torsion, and K-theory,''
  JHEP {\bf 0105} (2001) 047, hep-th/0103183.}
 \nref\berkoozleigh{M. Berkooz and R. G. Leigh, ``A $D=4$ ${\cal N}=1$
Orbifold of Type I Strings'', Nucl. Phys. {\bf B483}
(1997) 187, hep-th/9605049.}
\nref\BerkoozIZ{  M.~Berkooz, R.~G.~Leigh, J.~Polchinski,
J.~H.~Schwarz, N.~Seiberg and E.~Witten,
``Anomalies, Dualities, and Topology of $D=6$ ${\cal N}=1$ Superstring Vacua,''
Nucl.\ Phys.\  {\bf B475} (1996) 115, hep-th/9605184.}
 \nref\berkovits{N. Berkovits,
``Quantum Consistency of the Superstring in $AdS_5\times S^5$ Background,'' hep-th/0411170.}
\nref\BKKMSV{M. Bershadsky, K. Intriligator, S. Kachru, D. R. Morrison, V. Sadov and C. Vafa,
``Geometric Singularities and Enhanced Gauge Symmetries'', Nucl. Phys. {\bf B481} (1996) 215,
hep-th/9605200.}
 \nref\bianchi{M. Bianchi, Ph.D. thesis, preprint ROM2F-92/13;
A. Sagnotti, ``Anomaly Cancellations and Open-String Theories'', hep-th/9302099.}
 \nref\BilloIP{ M.~Bill{\' o}, S.~Cacciatori, F.~Denef, P.~Fr{\' e}, A.~van
Proeyen and D.~Zanon, ``The 0-Brane Action in a General $D = 4$
Supergravity Background,'' Class.\ Quant.\ Grav.\  {\bf 16}, (1999) 2335, hep-th/9902100.}
  \nref\blumzaf{
J. D. Blum and A. Zaffaroni, ``An Orientifold from F-Theory'',
Phys. Lett. {\bf B387} (1996) 71, hep-th/9607019.}
 \nref\blumenone{R. Blumenhagen, M. Cvetic, F. Marchesano
and G. Shiu, ``Chiral D-Brane Models with Frozen Open String
Moduli,'' hep-th/0502095.}
 \nref\blumentwo{R. Blumenhagen, D. L\"ust and T. Taylor,
``Moduli Stabilization in
Chiral Type IIB Orientifold Models with Fluxes,'' Nucl. Phys.
{\bf B663} (2003) 319, hep-th/0303016.}
 \nref\B{C. Borcea, ``$K3$ Surfaces with Involution and Mirror Pairs of Calabi-Yau
Manifolds'', ``Mirror Symmetry II'', (1997) 717.}
 \nref\BP{R. Bousso and J. Polchinski, ``Quantization of Four-Form
Fluxes and Dynamical Neutralization of the Cosmological Constant,''
JHEP {\bf 0006} (2000) 006, hep-th/0004134.}
 \nref\BrunnerZM{
  I.~Brunner and K.~Hori,
  ``Orientifolds and Mirror Symmetry,''
  JHEP {\bf 0411} (2004) 005, hep-th/0303135.}
 \nref\BrunnerZD{
  I.~Brunner, K.~Hori, K.~Hosomichi and J.~Walcher,
  ``Orientifolds of Gepner Models,'' hep-th/0401137.}
\nref\ibanez{P. Camara, L. Ibanez and A. Uranga, ``Flux-Induced
Susy-Breaking Soft Terms on D7-D3 Brane Systems,'' Nucl. Phys.
{\bf B708} (2005) 268, hep-th/0408036\semi
P. Camara, L. Ibanez and A. Uranga, ``Flux Induced Susy Breaking
Soft Terms,'' Nucl. Phys. {\bf B689} (2004) 195, hep-th/0311241.}
\nref\CF{P. Candelas and A. Font, ``Duality Between the Webs of
Heterotic and Type II Vacua'', Nucl. Phys. {\bf B511} (1998) 295,
hep-th/9603170.}
 \nref\casu{J. Cascales and A. Uranga,
``Chiral $4D$ String Vacua with D-Branes
and NS-NS and R-R Fluxes,'' JHEP {\bf 0305} (2003) 011, hep-th/0303024.}
\nref\CheungAZ{Y.~K.~Cheung and Z.~Yin, ``Anomalies, Branes, and Currents,''
Nucl.\ Phys. {\bf B517} (1998) 69, hep-th/9710206.}
\nref\CrapsTW{
  B.~Craps and F.~Roose,
  ``(Non-)Anomalous D-Brane and O-Plane Couplings: The Normal Bundle,''
  Phys.\ Lett.\  {\bf B450} (1999) 358, hep-th/9812149.
}
 \nref\cvet{
M. Cvetic, T. Li and T. Liu, ``Standard-like Models as Type IIB
Flux Vacua,'' hep-th/0501041.}
 \nref\dabpark{A. Dabholkar and J. Park,
``A Note on Orientifolds and F-theory,'' Phys. Lett. {\bf B394} (1997) 302,
hep-th/9607041.}
 \nref\DasguptaCD{
  K.~Dasgupta, D.~P.~Jatkar and S.~Mukhi,
  ``Gravitational Couplings and $\IZ_2$ Orientifolds,''
  Nucl.\ Phys.\  {\bf B523} (1998) 465, hep-th/9707224.}
  \nref\DasguptaSS{
  K.~Dasgupta, G.~Rajesh and S.~Sethi,
  ``M Theory, Orientifolds and G-Flux,''
  JHEP {\bf 9908} (1999) 023, hep-th/9908088.
  }
  \nref\DelaOssaXK{
X.~de la Ossa, B.~Florea and H.~Skarke, ``D-Branes on Noncompact
Calabi-Yau Manifolds: K-Theory and Monodromy,'' Nucl.\ Phys.\  {\bf
B644} (2002) 170, hep-th/0104254.}
 \nref\ddone{F. Denef and M. R. Douglas, ``Distributions of Flux Vacua,''
JHEP {\bf 0405} (2004) 072, hep-th/0404116.}
 \nref\ddtwo{F. Denef and M. R. Douglas, ``Distributions of Nonsupersymmetric
Flux Vacua,'' hep-th/0411183.}
 \nref\DDF{F. Denef, M. R. Douglas and B. Florea, ``Building a Better
Racetrack,'' JHEP {\bf 0406} (2004) 034, hep-th/0404257.}
 \nref\DGKT{O. DeWolfe, A. Giryavets, S. Kachru and W. Taylor,
``Enumerating Flux Vacua with Enhanced Symmetries,'' hep-th/0411061.}
\nref\fractbrane{
D.-E. Diaconescu, M. R. Douglas and J. Gomis,
``Fractional Branes and Wrapped Branes,''
JHEP {\bf 9802} (1998) 013, hep-th/9712230.}
\nref\DFM{D.-E. Diaconescu, D. Freed and G. W. Moore, ``The M-Theory $3$-Form and $E_8$ Gauge Theory'',
hep-th/0312069.}
 \nref\dine{M. Dine, D. O'Neil and Z. Sun, ``Branches of the
Landscape,'' hep-th/0501214.}
 \nref\dougmoore{M. R. Douglas and G. W. Moore, ``D-Branes, Quivers and ALE Instantons'', hep-th/9603167.}
 \nref\Douglas{M. R. Douglas, ``The Statistics of String/M-Theory
Vacua,'' JHEP {\bf 0305} (2003) 046, hep-th/0303194.}
 \nref\fontib{
A. Font and L. Ibanez, ``SUSY-Breaking Soft Terms in a MSSM
Magnetized D7-Brane Model,'' hep-th/0412150.}
 \nref\fmorient{D.~Freed and G.~Moore, unpublished.}
 \nref\FW{D. Freed and E. Witten, ``Anomalies in String Theory with D-Branes'', hep-th/9907189.}
 \nref\Frey{A. Frey and J. Polchinski, ``${\cal N}=3$ Warped Compactifications,''
Phys. Rev. {\bf D65} (2002) 126009, hep-th/0201029.}
 \nref\gimonpol{E. Gimon and J. Polchiski, ``Consistency Conditions for Orientifolds and D-Manifolds'', Phys. Rev.
{\bf D54} (1996) 1667, hep-th/9601038.}
 \nref\GaZ{O. Ganor, ``A Note on Zeroes of Superpotentials in F-Theory'', Nucl. Phys. {\bf B499} (1997) 55. hep-th/9612077.}
 \nref\gkp{S. Giddings, S. Kachru and J. Polchinski, ``Hierarchies from
Fluxes in String Compactifications,'' Phys. Rev. {\bf D66} (2002) 106006,
hep-th/0105097.}
  \nref\GKTT{A. Giryavets, S. Kachru, P. Tripathy and S. Trivedi, ``Flux
Compactifications on Calabi-Yau Threefolds,'' JHEP {\bf 0404} (2004) 003,
hep-th/0312104.}
  \nref\gopamukh{R. Gopakumar and S. Mukhi, ``Orbifold and Orientifold Compactifications of F-Theory and M-theory to Six-
Dimensions and Four-Dimensions'', Nucl. Phys. {\bf B479} (1996) 260, hep-th/9607057.}
  \nref\GorlichQM{
L.~G\"orlich, S.~Kachru, P.~K.~Tripathy and S.~P.~Trivedi, ``Gaugino
Condensation and Nonperturbative Superpotentials in Flux
Compactifications,'' hep-th/0407130.}
  \nref\G{A. Grassi, ``Divisors on Elliptic Calabi-Yau 4-Folds and the Superpotential in F-Theory -- I'', math.AG/9704008.}
  \nref\GreenDD{ M.~B.~Green, J.~A.~Harvey and G.~W.~Moore, ``I-Brane
Inflow and Anomalous Couplings on D-Branes,'' Class.\ Quant.\ Grav.\
{\bf 14}, (1997) 47, hep-th/9605033.}
 \nref\Greene{B. R. Greene,
``D-Brane Topology Changing Transitions,''
Nucl. Phys. {\bf B525} (1998) 284, hep-th/9711124.}
  \nref\GH {P. Griffiths and J. Harris, ``Principles of Algebraic Geometry", J. Wiley \& Sons, 1978, New York.}
  \nref\GL{T. W. Grimm and J. Louis, ``The Effective Action of ${\cal N}=1$ Calabi-Yau Orientifolds'', hep-th/0403067.}
  \nref\GVW{S. Gukov, C. Vafa and E. Witten, ``CFT's from Calabi-Yau
Fourfolds,'' Nucl. Phys. {\bf B584} (2000) 69, hep-th/9906070.}
\nref\HananyFQ{
  A.~Hanany and B.~Kol,
  ``On Orientifolds, Discrete Torsion, Branes and M-Theory,''
  JHEP {\bf 0006} (2000) 013, arXiv:hep-th/0003025.}
 \nref\HosonoAV{
  S.~Hosono, A.~Klemm and S.~Theisen,
  ``Lectures on Mirror Symmetry,'' hep-th/9403096.}

 \nref\JL{H. Jockers and J. Louis, ``The Effective Action of
D7-Branes in ${\cal N}=1$ Calabi-Yau Orientifolds,'' Nucl. Phys.
{\bf B705} (2005) 167, hep-th/0409098\semi
H. Jockers and J. Louis, ``D-Terms and F-Terms from D7-Brane Fluxes,''
hep-th/0502059.}
 \nref\KPV{S. Kachru, J. Pearson and H. Verlinde, ``Brane/Flux Annihilation
and the String Dual of a Nonsupersymmetric Field Theory,'' JHEP {\bf 0206}
(2002) 021, hep-th/0112197.}
 \nref\KST{S. Kachru, M. Schulz and S. Trivedi, ``Moduli Stabilization
from Fluxes in a Simple IIB Orientifold,'' hep-th/0201028.}
 \nref\KKLT{S. Kachru, R. Kallosh, A. Linde and S. P. Trivedi, ``de Sitter Vacua in String Theory'', Phys. Rev. {\bf D68}
(2003) 046005, hep-th/0301240.}
\nref\Renata{R. Kallosh and D. Sorokin, ``Dirac Action on M5 and M2
Branes with Bulk Fluxes,'' hep-th/0501081.}

 \nref\Vafagcond{ S.~Katz and C.~Vafa, ``Geometric Engineering of ${\cal N} =
1$ Quantum Field Theories,'' Nucl.\ Phys. {\bf B497}, (1997) 196, hep-th/9611090.}
 \nref\KS{I. Klebanov and M. Strassler, ``Supergravity and a Confining
Gauge Theory: Duality Cascades and $\chi$SB Resolution of Naked
Singularities,'' JHEP {\bf 0008} (2000) 052, hep-th/0007191.}
\nref\KleinTF{  M.~Klein and R.~Rabadan,
``Orientifolds with Discrete Torsion,''
 JHEP {\bf 0007} (2000) 040, arXiv:hep-th/0002103.}
 \nref\KleinHF{  M.~Klein and R.~Rabadan,
``$D$ = 4, ${\cal N} = 1$ Orientifolds with Vector Structure,''
  Nucl.\ Phys.\  {\bf B596} (2001) 197, hep-th/0007087.}
\nref\KleinQW{  M.~Klein and R.~Rabadan,
``$\IZ_N\times\IZ_M$ Orientifolds with and without Discrete Torsion,''
JHEP {\bf 0010} (2000) 049, hep-th/0008173.}
\nref\K{K. Kodaira, ``On Compact Analytic Surfaces II'', Annals of Math. {\bf 77} (1963) 3.}
\nref\KM{J. Koll\'ar and S. Mori, ``Birational Geometry of Algebraic Varieties", Cambridge University Press,
 Cambridge, U.K., {\bf 134} (1998).}
 \nref\lrstwo{D.~L\"ust, S.~Reffert and S.~Stieberger, ``Flux-Induced Soft
    Supersymmetry Breaking in Chiral Type IIB Orientifolds with
    D3 / D7-Branes'', Nucl.~Phys. {\bf B706} (2005) 3,
    hep-th/0406092.}
 \nref\lrs{D.~L\"ust, S.~Reffert and S.~Stieberger, ``MSSM with Soft
Susy Breaking Terms from D7-Branes with Fluxes'', hep-th/0410074.}
 \nref\lmrs{D. L\"ust, P. Mayr, S. Reffert and S. Stieberger, ``F-Theory
    Flux, Destabilization of Orientifolds and Soft Terms on D7-Branes'',
    hep-th/0501139.
    }
 \nref\ES{A. Maloney, E. Silverstein and
A. Strominger, ``de Sitter Space in Noncritical String Theory,''
hep-th/0205316.}
 \nref\marshiu{
F. Marchesano and G. Shiu, ``Building MSSM Flux Vacua,'' JHEP {\bf 0411}
(2004) 041, hep-th/0409132.}
 \nref\Mayr{P. Mayr, ``On Supersymmetry Breaking in String Theory and its Realization
in Brane Worlds,'' Nucl. Phys. {\bf B593} (2001) 99, hep-th/0003198.}
 \nref\MinasianMM{
  R.~Minasian and G.~W.~Moore,
  ``K-Theory and Ramond-Ramond Charge,''
  JHEP {\bf 9711} (1997) 002, hep-th/9710230.
 }
 \nref\Moore{G.~W.~Moore, ``Les Houches Lectures on Strings and Arithmetic,''
hep-th/0401049.}
 \nref\MooreJV{ G.~W.~Moore,
``Anomalies, Gauss Laws, and Page Charges in M-Theory,'' hep-th/0409158.}
 \nref\N{V. V. Nikulin, ``Discrete Reflection Groups in Lobachevsky Spaces and Algebraic Surfaces'', Proc. ICM, Berkeley,
California (1986) 654.}
 \nref\parkrabur{
J. Park, R. Rabadan and A. M. Uranga,
``Orientifolding the Conifold,''
Nucl. Phys. {\bf B570} (2000) 38, hep-th/9907086.}
 \nref\Reid{ M. Reid,
``Canonical 3--Folds'' Journ\'{e}es de G\'{e}om\'{e}trie Alg\'{e}brique d'Angers (1980) 273.}
  \nref\poltensor{J. Polchinski, ``Tensors from $K3$ Orientifolds'', Phys. Rev. {\bf D55} (1997) 6423, hep-th/9606165.}
  \nref\polchinski{ J.~Polchinski, ``String Theory'' vol. I and II,
Cambridge University Press (1998).}
  \nref\RS{D. Robbins and S. Sethi, ``A Barren Landscape'', hep-th/0405011.}
  \nref\Saltman{A. Saltman and E. Silverstein, ``The Scaling of the
No-Scale Potential and de Sitter Model Building,'' JHEP {\bf 0411} (2004)
066, hep-th/0402135.}
  \nref\SenVD{ A.~Sen, ``F-theory and Orientifolds,'' Nucl.\ Phys.\
{\bf B475}, (1996) 562, hep-th/9605150. }
  \nref\Si{A. Sen, ``Orientifold Limit of F-Theory Vacua'',
Phys. Rev. {\bf D55} (1997) 7345, hep-th/9702165.}
 \nref\sensethi{A. Sen and S. Sethi,
``The Mirror Transform of Type I Vacua in Six Dimensions,''
Nucl.Phys. B499 (1997) 45-54; hep-th/9703157.}
 \nref\SethiZK{
  S.~Sethi,
  ``A Relation Between ${\cal N} = 8$ Gauge Theories in Three Dimensions,''
  JHEP {\bf 9811} (1998) 003, hep-th/9809162.}
  \nref\EvaCFT{E. Silverstein, ``AdS and dS Entropy from String
Junctions or the Function of Junction Conjunctions,''
hep-th/0308175.}
 \nref\StefanskiYX{
  B.~J.~Stefanski,
  ``Gravitational Couplings of D-Branes and O-Planes,''
  Nucl.\ Phys.\ B {\bf 548}(1999) 275, hep-th/9812088. }
  \nref\Susskind{L. Susskind, ``The Anthropic Landscape of String
Theory,'' hep-th/0302219.}
  \nref\TV{T. Taylor and C. Vafa, ``RR Flux on Calabi-Yau and Partial Supersymmetry
Breaking,'' Phys. Lett. {\bf B474} (2000) 130, hep-th/9912152.}
\nref\Sandip{P. Tripathy and S. Trivedi, ``D3 Brane Action and Fermion
Zero Modes in Presence of Background Flux,'' hep-th/0503072.}

 \nref\vafawitten{C.~Vafa and E.~Witten,
``On Orbifolds with Discrete Torsion,''
J.\ Geom.\ Phys.\  {\bf 15}, 189 (1995)
hep-th/9409188.}
 \nref\V{C. Voisin, ``Mirrors and Involutions on $K3$ Surfaces'', Journ\'{e}es de G\'{e}om\'{e}trie Alg\'{e}brique d'Orsay,
Ast\'erisque {\bf 218} (1993) 273.}
 \nref\W{E. Witten, ``Non-Perturbative Superpotentials in String Theory'', Nucl. Phys. {\bf B474} (1996) 343, hep-th/9604030.}
 \nref\WittenMD{ E.~Witten, ``On Flux Quantization in M-Theory and
the Effective Action,'' J.\ Geom.\ Phys.\  {\bf 22} (1997) 1,
hep-th/9609122.}
 \nref\WittenHC{ E.~Witten,
``Fivebrane Effective Action in M-Theory,'' J.\ Geom.\ Phys.\  {\bf
22} (1997) 103, hep-th/9610234.}
\nref\WittenXY{
  E.~Witten,
  ``Baryons and Branes in Anti de Sitter Space,''
  JHEP {\bf 9807} (1998) 006, hep-th/9805112.}
\nref\Wi{E. Witten, ``Duality Relations Among Topological Effects in String Theory'',
JHEP {\bf 0005} (2000) 031, hep-th/9912086.}
\Title{
\vbox{
\baselineskip12pt
\hbox{\rightline{hep-th/0503124}}
\hbox{\rightline{SLAC-PUB-11015}}
\hbox{\rightline{SU-ITP-05/07}}
}}
{\vbox{\vskip 2pt
\vbox{\centerline{Fixing All Moduli in a Simple F-Theory Compactification}}
}}
\vskip 4pt
\centerline{Frederik Denef\footnote{$^\natural$}{{{\tt denef@physics.rutgers.edu}}}$^1$,
Michael R. Douglas\footnote{$^\flat$}
{{{\tt mrd@physics.rutgers.edu}}}$^{1,\&}$,
Bogdan Florea\footnote{$^\sharp$}
{{{\tt florea@physics.rutgers.edu}}}$^1$, Antonella Grassi\footnote{$^\exists$}
{{{\tt grassi@math.upenn.edu}}}$^2$ and Shamit Kachru\footnote{$^\forall$}
{{{\tt skachru@stanford.edu}}}$^3$}
\bigskip
\centerline{{$^1$\it Department of Physics and Astronomy,
Rutgers University,}}
\centerline{\it Piscataway, NJ 08855-0849, USA}
\centerline{{$^\&$\it I.H.E.S., Le Bois-Marie, Bures-sur-Yvette, 91440 France}}
\centerline{{$^2$\it Department of Mathematics, University of Pennsylvania,}}
\centerline{{\it Philadelphia, PA 19104-6395, USA}}
\centerline{{$^3$\it Department of Physics and SLAC, Stanford University,}}
\centerline{{\it Stanford, California, CA 94305, USA}}
\bigskip
\smallskip
\noindent We discuss a simple example of an F-theory
compactification on a Calabi-Yau fourfold where background fluxes,
nonperturbative effects from Euclidean D3 instantons and gauge
dynamics on D7 branes allow us to fix all closed and open string
moduli. We explicitly check that the known higher order corrections
to the potential, which we neglect in our leading approximation,
only shift the results by a small amount. In our exploration of the
model, we encounter interesting new phenomena, including examples of
transitions where D7 branes absorb O3 planes, while changing
topology to preserve the net D3 charge.

 \vfill
 \vskip3mm
 \Date{March 2005}

\newsec{Introduction}

All well-studied quasi-realistic string compactifications come, in the
leading approximation, with large numbers of scalar moduli fields. It
has long been thought that this is an artifact of the leading
approximation, and that in many backgrounds non-perturbative effects
generate potentials which lift all of these fields.

More recently, it has been emphasized that such potentials can have a
very large number of minima, which combined with other choices made
in the constructions gives rise to an immense `landscape' of
string vacua
\refs{\BP,\ES,\Acharya,\KKLT,\Susskind,\Douglas}. Indeed, very
simple genericity arguments, based on knowledge of effective field
theory and of the contributions of various stringy ingredients to
the effective potential, suggest that this should be true.

Since typical constructions have dozens or hundreds of moduli fields,
and many different effects contributing to the scalar potential, the
explicit calculations required to verify this expectation are daunting
in almost all examples.  However it is important to pursue explicit
examples in as much detail as possible.  Such examples allow one both
to verify the hypothesis that there are many models with stabilized
moduli, and to gain further intuition about stringy potentials which
may be relevant in developing models of inflation or particle physics.

A proposal for stabilizing a wide class of type IIB or F-theory
compactifications was put forth in \KKLT, and fairly explicit
examples were provided in \DDF. In F-theory on an elliptic
Calabi-Yau fourfold, the moduli of interest are the complex
structure moduli and the K\"ahler moduli (which come from $h^{1,1}$
of the base of the elliptic fibration).  In the limit where one can
think of these constructions in terms of IIB Calabi-Yau
orientifolds, these moduli include the moduli of the Calabi-Yau
threefold, the axio-dilaton, and the positions of D7-branes in the
threefold geometry.

In the constructions described in \KKLT, one
approaches the problem of moduli stabilization in two steps. First,
one turns on background fluxes to stabilize the complex structure
moduli of the fourfold at some energy scale $E$ -- the fluxes can in
general stabilize both the complex structure and D7 brane moduli,
from the IIB perspective.  Then, one incorporates exponentially
small effects arising from Euclidean D3 instantons (corresponding to
M5 instantons wrapping vertical divisors of holomorphic Euler
characteristic\foot{This is often (but not quite accurately) called
the arithmetic genus in the physics literature in this context.}
$\chi_h=1$) or infrared gauge dynamics on D7 branes, to generate a
potential for the K\"ahler moduli at the scale $\tilde E$. At any
moderately large radius in string units, one finds $E >> \tilde E$,
and it is a good approximation to treat the complex moduli as fixed
when one evaluates the effects generating a potential for the
K\"ahler modes (so any complex structure-dependent determinants
multiplying the instanton action, should be evaluated at the
critical point of the flux potential). As described in \KKLT, as
long as the gravitino mass $e^{K/2}|W|$ resulting from the first
step is moderately small in string units, one has a small parameter
to use in the next step, which can result in radii stabilized within
the regime of convergence of the instanton expansion. In fact, as we
shall see explicitly, even ${\cal O}(1)$ values of $e^{K/2}|W|$ can
suffice in some concrete models.

In this paper, we provide an explicit example where this recipe is
carried out in complete detail.  We construct an orientifold of a
Calabi-Yau threefold and its F-theory dual fourfold, which contains
enough rigid $\chi_h=1$ divisors and pure non-abelian D7 brane gauge
theories, to generate a potential for all K\"ahler modes. The
Calabi-Yau threefold is the resolved orbifold $T^6/\IZ_2 \times
\IZ_2$, which has 51 K\"ahler moduli and 3 complex structure moduli.
It is not difficult to construct explicit flux vacua in the complex
structure sector of this model, and we provide some examples with
moderately small $e^K |W|^2$. We solve for the stabilized values of
all 51 K\"ahler moduli in the leading approximation, and show that
the subleading corrections which we neglected are indeed expected to
be quite small. Our example was chosen for its simplicity,
especially as regards its complex structure moduli space (since we
wished to provide explicit flux vacua),  and hence it admits far
fewer flux vacua than more generic fourfolds -- only ${\cal
O}(10^{13})$, in contrast to the more impressive numbers like ${\cal
O}(10^{300})$ that arise in more complicated examples. In suitable
cases with larger numbers of flux vacua, one expects similar
constructions to be even more controlled, but of course the
higher-dimensional complex structure moduli space becomes harder to
work with explicitly.

The organization of this paper is as follows. In \S2, we give a
first description of our main model. We find that the K\"ahler
moduli space has a rather intricate structure, with different phases
that meet in a singular orbifold geometry. We argue that each
blow-up mode of the orbifold which contributes a twisted K\"ahler
mode in the $4D$ theory, comes along with its own vertical divisor
of holomorphic Euler characteristic one, and that the gauge theory
sector is pure $SO(8)^{12}$ $\CN=1$ SYM without any matter. In \S3,
we describe the geometry of the F/M-theory dual fourfold in great
detail, and put the arguments of \S2 on firm mathematical footing.
In \S4, we discuss the relation of our model to some similar
toroidal orientifolds which have (in absence of flux) exactly
soluble worldsheet definitions at the orbifold point. In \S5, we
construct explicit flux vacua in this example, which stabilize the
complex structure moduli and provide the needed moderately small
$e^K|W|^2$. We choose to saturate the entire D3-tadpole by turning
on fluxes, so no wandering D3 branes are introduced in the geometry.
In \S6, we then include the leading non-perturbative effects, and
see that once one includes both the D3-instanton effects and the
gauge dynamics on the D7 branes in this model, all untwisted and
twisted K\"ahler moduli are stabilized.  We check the magnitude of
the known corrections to the leading result, from perturbative
\BeckerNN\ and worldsheet instanton corrections to the K\"ahler
potential as well as multi-instanton contributions to the
superpotential, and see that even with our only moderately small
$e^K |W|^2$ these corrections are expected to be at the one percent
level.  In the penultimate section, we briefly describe some other
explicit models with only a single K\"ahler mode which are good
candidates for stabilization. We conclude in \S8. In the appendix,
we describe our normalization conventions.

\newsec{Orientifold Geometry}

Our model is a type IIB orientifold of a
smooth Calabi-Yau threefold $Y$, which is a resolution of the orbifold
${\overline Y} = T^6/\IZ_2 \times \IZ_2$. The orbifold group acts as
 \eqn\toricdataA{\eqalign{ \matrix{ & z_1 & z_2 & z_3 \cr
        \alpha & + & - & - \cr
        \beta & - & + & - \cr
        \alpha\circ\beta & - & - & +. }\cr}}
There are $3 \times 16 = 48$ fixed lines under the action of a group
element, and $64$ fixed points where three fixed lines meet. Blowing
up the fixed lines resolves the geometry and introduces 48 twisted
K\"ahler moduli in addition to the 3 untwisted K\"ahler moduli descending
from $(T^2)^3$. There are just 3 complex
structure moduli, given by the modular parameters of the $T^2$
factors. Thus the resulting smooth Calabi-Yau threefold $Y$ has
$h^{1,1}=51$, $h^{2,1}=3$. This is one of the Borcea-Voisin models
\refs{\B,\V}.

We will see in the following that we can choose an orientifold
involution of $Y$ and D7-brane embeddings in such way that we get an
$SO(8)^{12}$ gauge group without matter, and a D3 tadpole $Q_3=-28$
(on the quotient space) induced by non-exotic O3-planes and 7-brane
curvature.

Before proceeding, let us briefly explain why we do not define our
model as a toroidal orientifold, {\it i.e.} by orientifolding the
orbifold ${\overline Y}$.  Despite the geometric singularities,
orientifolds of an orbifold can often be described directly by
perturbative string CFT's, and the case of $T^6 / \IZ_2 \times \IZ_2$
has been studied extensively.

At first sight, the toroidal orientifold appears promising:
the involution $\Omega: z_i \to -z_i$, which (modulo the $\IZ_2
\times \IZ_2$ equivalences) has $3 \times 4 = 12$ O7 fixed planes
and 64 fixed points at the triple intersections of the O7 planes,
each of which corresponds to an O3 plane, roughly matching the
smooth geometry we are about to describe.  This comparison will
be carried much further in \S4, where it is shown that the
$SO(8)^{12}$ gauge group and many other features of our model can
be realized by a toroidal orientifold.

However, in general, not every large radius orientifold model has an
orbifold limit with a well-behaved CFT description. When trying to
go continuously from large radii to the orbifold point,
nonperturbative massless states could arise from D-branes wrapping
vanishing cycles, rendering perturbation theory singular. In the
parent $\CN=2$ theories, such singularities are of complex
codimension one in moduli space, so one can go around them and
continuously connect various phases. However, after orientifolding,
these singularities become of {\it real} codimension 1, preventing
different phases from being connected smoothly. Therefore,
orientifolds constructed geometrically in the large radius regime do
not necessarily all have orbifold CFT counterparts, and in
particular may have different discrete properties.

Indeed, we will see that this is the case for our model in \S4.
For this reason, and because we will stabilize all radii at finite
distance away from the orbifold point anyway, we will {\it first} blow
up the orbifold into a smooth, large Calabi-Yau threefold $Y$, using
standard techniques in algebraic geometry.  We then consider the
string theory whose world-sheet definition is the sigma model with
this Calabi-Yau target space, and {\it then} define an orientifold of
this model.  The result is similar to a toroidal orientifold but
realizes discrete choices not possible in the CFT orientifold framework.
Still, readers familiar with that framework may find it
useful to read \S4 to get an overview of the construction before
proceeding.

In the following, we will first study a local description valid near
the orbifold fixed points. We give a completely explicit description
of the resolution, the orientifold involution, and the brane
embeddings. We also review how intersection numbers, important for
example to derive the K\"ahler potential, are computed in this
setup. We then move on to the compact Calabi-Yau and discuss its
lift to F-theory on an elliptically fibered Calabi-Yau fourfold
(which we define as M-theory in the limit of vanishing elliptic
fiber area). Finally, we calculate the D3 tadpole for our model, and
note an interesting geometrical transition where a 7-brane stack
``eats'' an O3-plane while changing its topology to preserve the net
D3-brane charge.

\subsec{Local model}

To understand the resolved geometry and the orientifold involution,
it is useful to consider first a local model of the singularities,
given by ${\overline X}=\IC^3/{\IZ_2\times \IZ_2}$. The resolution of this
orbifold can be described explicitly as a toric variety, following
the general construction outlined in \DelaOssaXK.

\ifig\CZZsing{Top view of fan of $\IC^3/\IZ_2\times \IZ_2$
(unresolved).} {\epsfxsize4cm\epsfbox{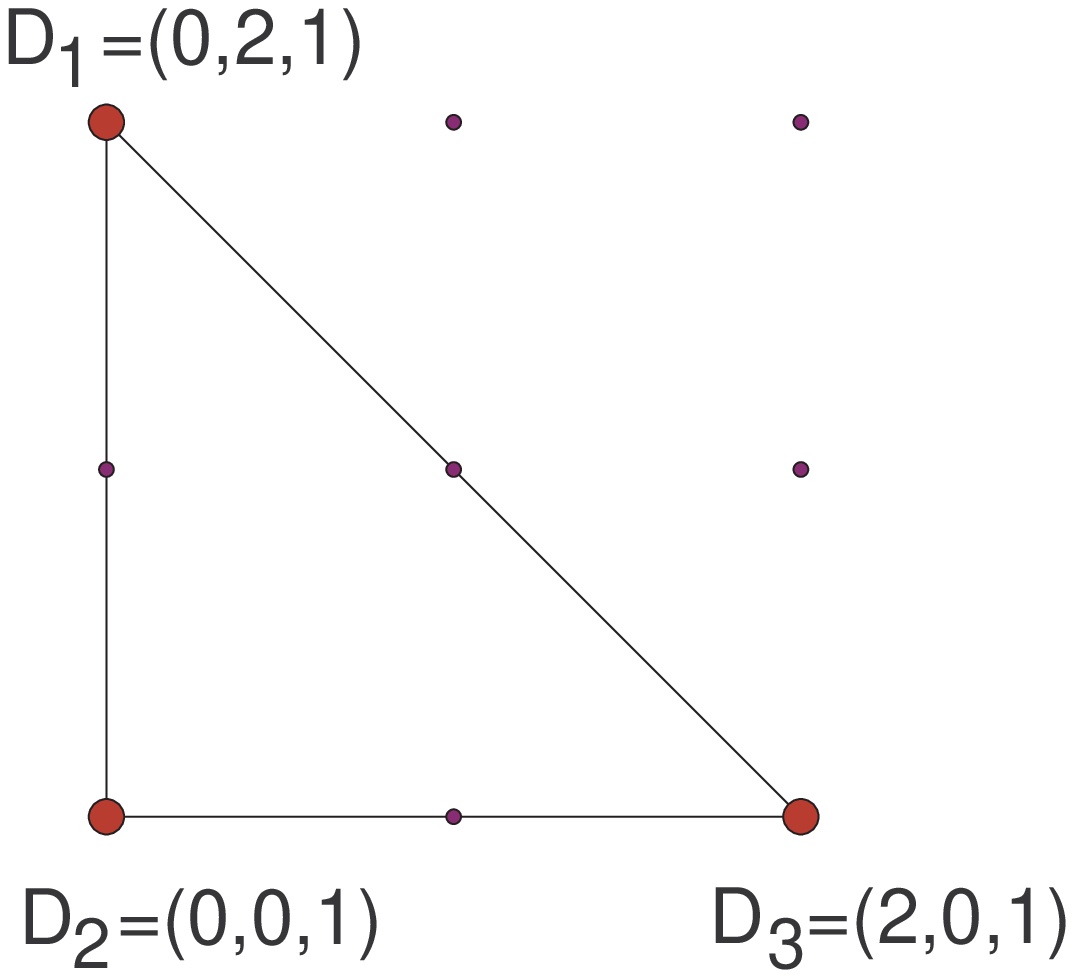}}

The data underlying any three dimensional toric variety is given by
a lattice $N=\IZ^3$ and the choice of a fan $\Delta$, which is
a collection of cones generated by lattice points in $N$, satisfying
the condition that every face of a cone is also a cone, and that the
intersection of two cones is a face of each.

The singular orbifold $\IC^3/\IZ_2 \times \IZ_2$ is described by the
simple fan given in \CZZsing, consisting of a single 3-dimensional
cone generated by the lattice vectors $D_1=(0,2,1)$, $D_2=(0,0,1)$
and $D_3=(2,0,1)$. As usual for Calabi-Yau varieties, the third
component of each vector equals 1, so we can restrict our attention
to the other two coordinates, as we did in the figure. To each
vertex $D_i$ a complex variable $z_i$ is assigned, and to each
dimension $r=1,\ldots,3$ a monomial $U_r \equiv \prod_i
z_i^{(D_i)_r}$. In this case, $U_1=z_3^2$, $U_2=z_1^2$, $U_3=z_1 z_2
z_3$. The toric variety $\bar{X}$ is then simply given by all
$(z_1,z_2,z_3)$ not in a certain set $F$, modulo complex rescalings
that leave the $U_r$ invariant. The excluded set $F$ is given by the
values of $z_i$ which have simultaneous zeros of coordinates not
belonging to the same cone. Since there is only one three
dimensional cone here, $F$ is empty. The only rescalings that
leave the $U_r$ invariant are given precisely by \toricdataA. Thus,
${\overline X}$ is indeed $\IC^3/\IZ_2 \times \IZ_2$.

\ifig\fancthztwztw{Fan for $(a)$ asymmetric and $(b)$ symmetric
resolution of $\IC^3/\IZ_2\times \IZ_2$.}
{\epsfxsize3.7in\epsfbox{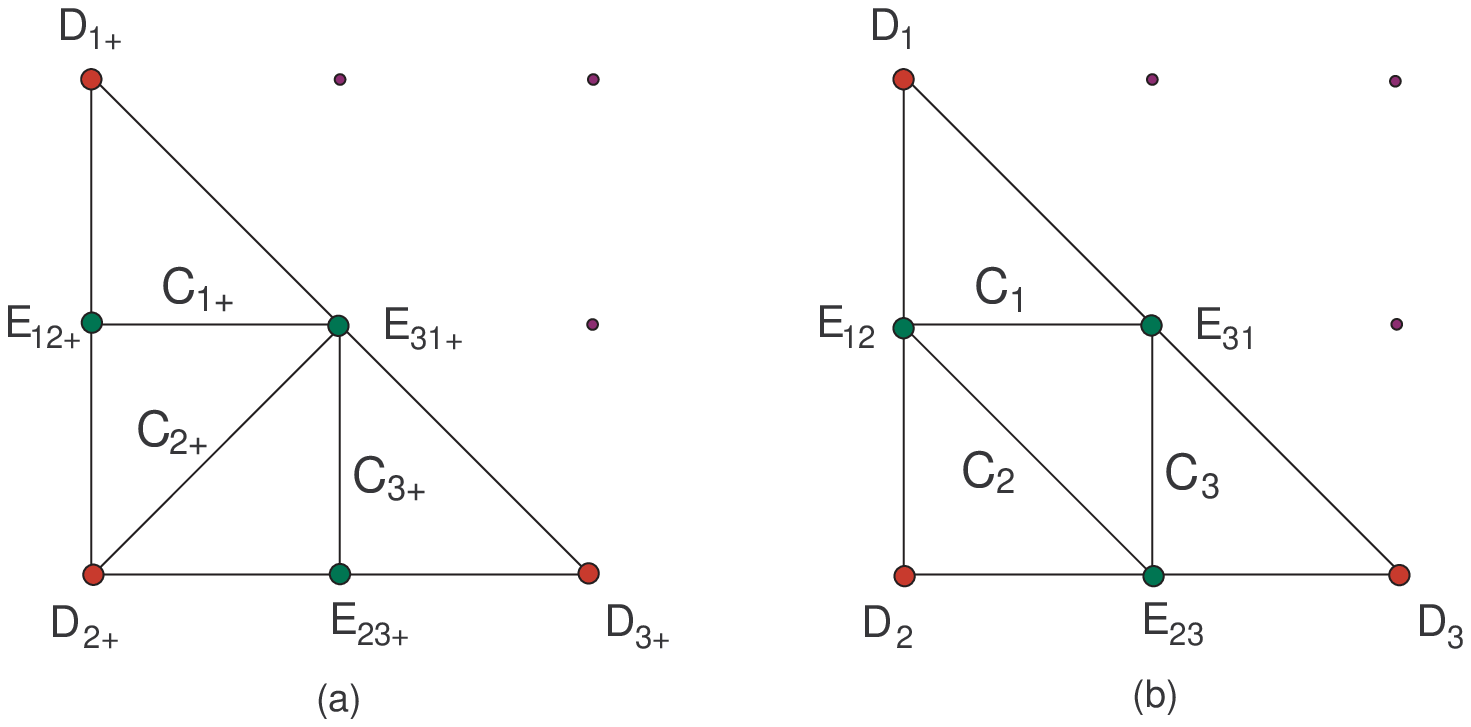}}

\ifig\divscthztwztw{Dual graph for $(a)$ asymmetric and $(b)$ symmetric
resolution.} {\epsfxsize4.5in\epsfbox{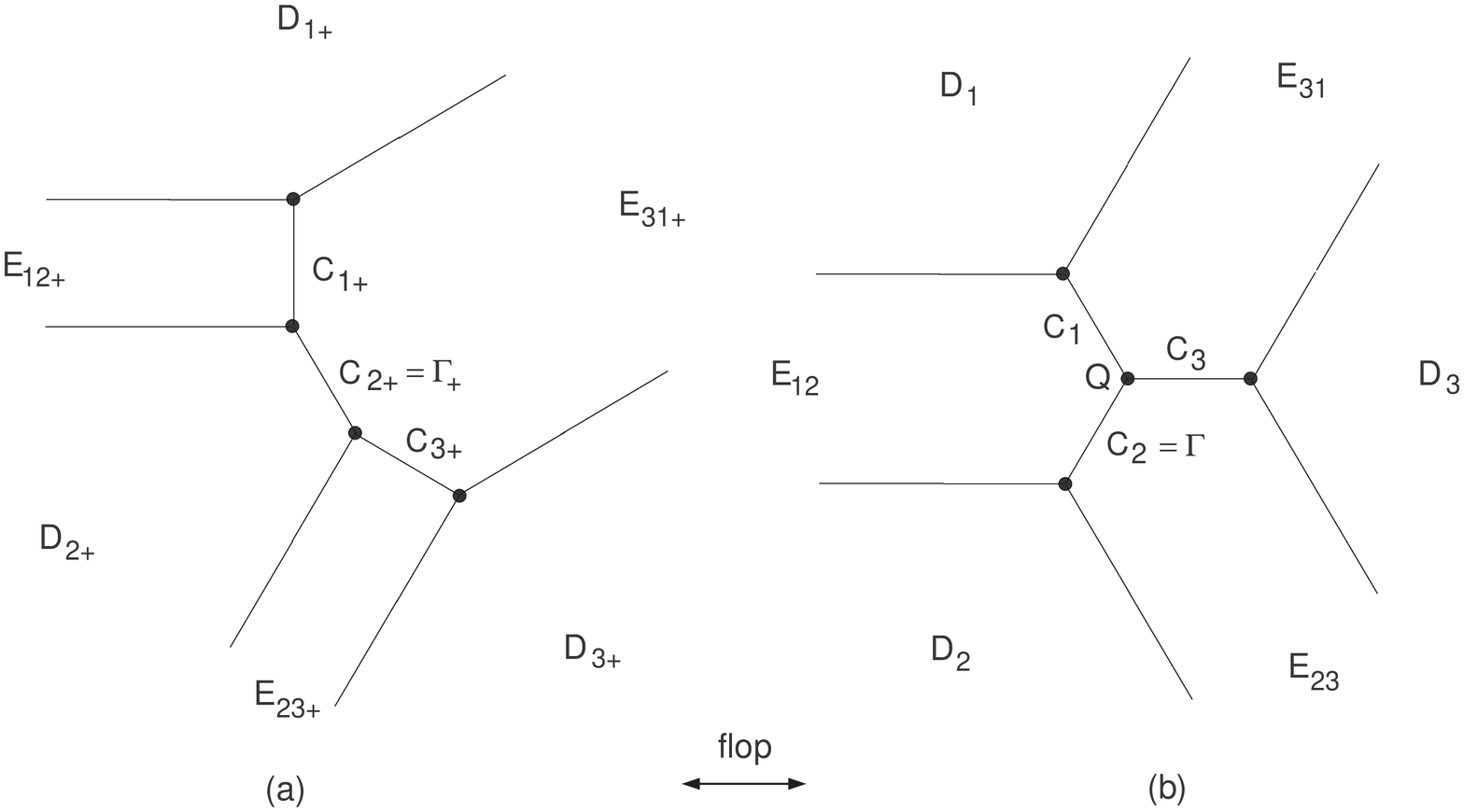}}

The fact that ${\overline X}$ is singular can be traced back to the
fact that the top dimensional cone generators do not span the full
lattice $N$, since $\det(D_{i,r})=4$. To resolve the variety, one
has to refine the fan such that all top dimensional cones have
determinant 1. There are two distinct ways of doing this in the case
at hand, one symmetric and one asymmetric, as shown in
\fancthztwztw\ $(b)$ resp.\ $(a)$. The dual graphs are shown in
\divscthztwztw. As we will review below, the vertices in
\fancthztwztw\ can be thought of as divisors, the lines as curves at
the intersections of two divisors, and the faces as points at the
intersections of three divisors. In the dual graphs, the role of
faces and vertices is interchanged.

 \vskip4mm
 \noindent {\it Symmetric resolution}
 \vskip2mm

As shown in \fancthztwztw\ $(b)$, there are now 6 vertices and 4
cones $(D_1,E_{12},E_{31})$, $(D_2,E_{12},E_{2,3})$,
$(D_3,E_{23},E_{3,1})$, $(E_{12},E_{23},E_{31})$ all of determinant
1. The vertices are given by the matrix
 \eqn\verticessymmres{
 \matrix{D_1 & D_2 & D_3 & E_{23} & E_{31} & E_{12} \cr
         0   & 0   & 2   & 1      & 1      & 0      \cr
         2   & 0   & 0   & 0      & 1      & 1      \cr
         1   & 1   & 1   & 1      & 1      & 1 }.
 }
We associate complex variables $z_i$ to the $D_i$ and $y_i$ to the
$E_{jk}$. The powers in the monomials $U_r$ are simply given by the
rows of this matrix, i.e.\ $U_1=z_3^2 y_1 y_2$, $U_2=z_1^2 y_2 y_3$,
$U_3=z_1 z_2 z_3 y_1 y_2 y_3$. The rescalings leaving the $U_r$
invariant are
 \eqn\rescalings{
  (z_1,z_2,z_3,y_1,y_2,y_3) \to (\lambda_1 z_1,\lambda_2 z_2,\lambda_3 z_3,
  {\lambda_1 \over \lambda_2 \lambda_3} y_1,
  {\lambda_2 \over \lambda_3 \lambda_1} y_2,
  {\lambda_3 \over \lambda_1 \lambda_2} y_3 ),
 }
with $\lambda_i \in \IC^*$. The set of excluded points $F$ is again
given by simultaneous zeros of coordinates not in the same cone, for
example $(z_1,y_1)=(0,0)$ and $(z_2,y_1,z_3)=(0,0,0)$ are excluded,
but $(z_2,y_1)=(0,0)$ is not. The toric variety is thus given by
$X=(\IC^6 \backslash F)/{\IC^*}^3$, with the $\IC^*$ actions given
in \rescalings.

To each vertex corresponds a toric divisor, by setting the
associated coordinate equal to zero. Curves are obtained by
intersecting divisors, i.e.\ setting two coordinates to zero. To
avoid being on the excluded locus $F$, the corresponding vertices
must be part of the same cone, in other words they have to be joined
by a line in \fancthztwztw. Compact curves correspond to internal
lines. In the case at hand, there are three such curves, which we
denote by $C_i$, where $C_1 = E_{31} \cdot E_{12}$ and cyclic
permutations thereof. Finally, triple intersection points of
divisors are obtained by setting the 3 coordinates associated to a
single cone to zero. Thus the triple intersection number is 1 for 3
distinct divisors belonging to the same cone, and 0 otherwise.

The divisors $D_i$ are the original divisors we had in the
unresolved variety, the $E_{ij}$ are the exceptional divisors
produced by the resolution, and the $C_i$ are the exceptional
curves. The latter have topology $\IP^1$. This can be seen as
follows. $C_1$ for example is given by $y_2=y_3=0$. To avoid the
excluded set $F$, we must take $z_2 \neq 0$, $z_3 \neq 0$ and
$(z_1,y_1) \neq (0,0)$. This allows choosing a gauge with
$z_2=z_3=1$, so
 \eqn\exccurvetop{
  C_1 = \{ (z_1,1,1,y_1,0,0) | (z_1,y_1) \in \IC^2 \backslash (0,0) \} \,
  / \,
  (z_1,y_1) \sim \lambda (z_1,y_1),
 }
which is of course $\IP^1$.

 \vskip4mm
 \noindent {\it Orientifold action and D-brane embedding}
 \vskip2mm

Let us now look at the orientifold involution. There are several
choices. We choose $\Omega:(z,y) \to (-z,y)$. The fixed points are
then given by the $(z,y)$ for which
 \eqn\orfixed{
  (-z,y) = (\lambda_1 z_1,\lambda_2 z_2,\lambda_3 z_3,
  {\lambda_1 \over \lambda_2 \lambda_3} y_1,
  {\lambda_2 \over \lambda_3 \lambda_1} y_2,
  {\lambda_3 \over \lambda_1 \lambda_2} y_3 )
 }
for some $\lambda_i$. The following possibilities arise:

(1) If all $z_i \neq 0$, we need $\lambda_i=-1$ and therefore
$y=-y=0$. Because $(y_1,y_2,y_3)$ belong to the same cone, this is
an allowed point. Thus, we get an isolated fixed point $Q$ that can
be represented by $(1,1,1,0,0,0)$. This corresponds to an O3-plane.

(2) If say $z_1=0$, then to avoid the excluded set $F$, we must take
$z_2 \neq 0$, $z_3 \neq 0$, $y_1 \neq 0$. Then \orfixed\ implies
$\lambda_2=\lambda_3=-1$ and $\lambda_1=1$, and imposes no further
constraints. Therefore the entire divisor $D_1: z_1=0$ is fixed.
This gives us an O7-plane. Similarly, there will be O7 planes on
$D_2$ and $D_3$. The topology of these divisors is easily
determined. Since $z_2$, $z_3$ and $y_1$ are all nonzero, we can fix
the scaling gauge by setting these variables equal to 1. The divisor
is then parametrized by the remaining variables $y_2$ and $y_3$
without further identifications, so it is a copy of $\IC^2$.

The action of $\Omega$ on the exceptional $\IP^1$s is also
straightforward to determine. After a gauge transformation
$\lambda_i=-1$, the orientifold action can be written as $(z,y) \to
(z,-y)$, which acts on \exccurvetop\ as
 \eqn\Poneaction{
  [z_1,y_1] \to [z_1,-y_1].
  }
That is, the $\IP^1$s are mapped to themselves in an orientation
preserving way, with fixed points at the poles, where the $\IP^1$
intersects the O3 or O7 planes. Note that one cannot wrap a closed
string once around a pole of the quotient $\IP^1/\IZ_2$, since the
endpoints of a string can only be identified by an orientifold
$\IZ_2$ if the orientation of the string is reversed. Therefore the
minimal closed string instanton wraps $\IP^1/\IZ_2$ twice (or the
original $\IP^1$ once). The instanton phase should furthermore be
invariant under the orientifold action $B \to -B$, which implies
 $$e^{2 \pi i \int_{\IP^1} B} = \pm 1.$$

We still have to specify how we embed D-branes in this geometry. We
will put D7-branes on top of the O7-planes such that D7 tadpole is
canceled locally. We choose the O7-planes to be non-exotic, so each
induces $-8$ units of D7-brane charge in the Calabi-Yau $X$ (or $-4$
in the quotient $X/\IZ_2$), and we need a stack of 8 coincident
D7-branes on each $D_i \subset X$ to cancel this. This gives rise to
an $SO(8)$ gauge group on each divisor $D_i$. Note that since the
$D_i$ are disjoint in the resolved manifold, there are no massless
bifundamentals from strings stretching between the D7-branes. To
decide if there is massless adjoint matter, we need to know the
topology of the $D_i$ in the compact geometry. We will get to this
further on.

 \vskip4mm
 \noindent {\it Intersection numbers}
 \vskip2mm

To construct the K\"ahler potential on moduli space, we will need
the triple intersection numbers of the divisors, including triple
intersections involving identical divisors. These numbers also
determine self-intersections of curves inside divisors, which
characterize the local geometry. In this subsection we will review
how to obtain these numbers. The reader who is only interested in
the results can safely skip this part however.

We can derive the intersection numbers in the local model for
compact intersections. Linear combinations of divisors whose
associated line bundle is trivial on the noncompact variety $X$ will
give zero compact intersections with any combination of other
divisors. Denoting the divisors collectively as $R_a$,
$a=1,\ldots,6$, and the corresponding coordinates by $x_a$, we have
that $f=\prod_a x_a^{n_a}$ is a section of the line bundle $R=\sum
n_a R_a$. The invariant monomials $U_r$ are functions, so the
corresponding $R$ is trivial. For the purpose of computing compact
intersections, this implies three linear relations between our
divisors\foot{Despite the abuse of notation, these relations between
divisors should not be confused with the relations between the
corresponding vertices \verticessymmres!}: $2 D_1 + E_{31} +
E_{12}=0$, and so on. We should emphasize that this relation does
not mean that this linear combination of divisors will also be
trivial in the compact geometry. Rather, it means that this linear
combination does not intersect the compact curves, and hence can be
moved away from the origin --- in the compact geometry, such a
divisor corresponds to a ``sliding divisor'' such as $R_1: \{ z_1 =
c \} \cup \{ z_1 = -c \}$. These divisors descend directly from the
unresolved $T^6/\IZ_2 \times \IZ_2$, and are in this sense
independent of the blowup.

At any rate, these relations together with the triple intersection
numbers of distinct divisors obtained directly from the fan are
sufficient to determine all compact triple intersection numbers.
This gives for example $E_{12}^2 \cdot E_{23}=-(2D_1+E_{31})\cdot
E_{12}\cdot E_{23} = -1$. From this, we also obtain the intersection
numbers of the divisors with the compact curves $C_i$ defined above.
These curves form a basis of the Mori cone i.e.\ the cone of
effective holomorphic curves in $X$. We get for their intersections:
 \eqn\moriconesymmres{
 \matrix{                       & D_1 & D_2 & D_3 & E_{23} & E_{31} & E_{12} \cr
 E_{31} \cdot E_{12} \equiv C_1 & 1   & 0   & 0   & 1      & -1     & -1     \cr
 E_{12} \cdot E_{23} \equiv C_2 & 0   & 1   & 0   & -1     & 1      & -1     \cr
 E_{23} \cdot E_{31} \equiv C_3 & 0   & 0   & 1   & -1     & -1     & 1 }
 }
Note that the entries are precisely the charges of the rescalings
\rescalings. Indeed, the Mori cone intersection numbers always form
a basis of the rescaling charges. This is an elementary algebraic
consequence of the various definitions we made.

The triple intersection numbers also give the self-intersection
numbers of the curves inside the exceptional divisors, for example
$C_1^2|_{E_{12}} = E_{12} \cdot E_{31}^2 = -1$.

Finally, apart from one subtlety, it is straightforward to deduce
the intersection numbers of the orientifold quotient $X/\IZ_2$. The
subtlety is the following. Denote the projection from $X$ to
$X/\IZ_2$ by $\pi$. Naively one might think one should take the
toric divisors $\widetilde{R}_a$ of the quotient, considered as 2-forms,
to be related to those of the double cover by $R_a = \pi^*
\widetilde{R}_a$. This is correct for the divisors $E_{ij}$, but not for
the $D_i$, for which we should take $D_i = \pi^* \widetilde{D}_i/2$.
This can be seen as follows. Because $D_i$ is fixed by the $\IZ_2$,
its volume in $X$ must equal the volume of $\widetilde{D}_i$ in the
quotient. But the volume of $\widetilde{D}_i$ is given by
 $$
  \int_{X/\IZ_2} \widetilde{D}_i \wedge {J \wedge J \over 2}=
  {1 \over 2} \int_{X} \pi^* \widetilde{D}_i \wedge {J \wedge J \over 2}
 $$
which is {\it half} the volume of $\pi^* \widetilde{D}_i$. So we must
take $D_i = \pi^* \widetilde{D}_i/2$ to correct for this. For the
divisors $E_{ij}$ on the other hand, whose volume does indeed get
halved, there is no such correction factor of 2. Thus we get for
example
 \eqn\intersectex{\eqalign{
  \widetilde{D}_1 \cdot \widetilde{E}_{31} \cdot \widetilde{E}_{12}
  &= {1 \over 2} \int_X (2 D_1) \wedge E_{12} \wedge E_{31} = D_1
  \cdot E_{12} \cdot E_{31} = 1 \cr
  \widetilde{E}_{12} \cdot \widetilde{E}_{23} \cdot \widetilde{E}_{31} &= {1
  \over 2} \int_X E_{12} \wedge E_{23} \wedge E_{31} = {1 \over 2} E_{12}
  \cdot E_{23} \cdot E_{31} = {1 \over 2}.
  }}
The half integral triple intersection product is possible because
the intersection point coincides with the $\IZ_2$ fixed point
singularity $Q$ (the O3). For the intersections of the Mori cone
generators, we thus get
 \eqn\moriconesymmresquotient{
 \matrix{  & \widetilde{D}_1 & \widetilde{D}_2 & \widetilde{D}_3 & \widetilde{E}_{23} & \widetilde{E}_{31} & \widetilde{E}_{12} \cr
 \widetilde{E}_{31} \cdot \widetilde{E}_{12} \equiv \widetilde{C}_1 & 1   & 0   & 0   & 1/2      & -1/2     & -1/2     \cr
 \widetilde{E}_{12} \cdot \widetilde{E}_{23} \equiv \widetilde{C}_2 & 0   & 1   & 0   & -1/2     & 1/2      & -1/2     \cr
 \widetilde{E}_{23} \cdot \widetilde{E}_{31} \equiv \widetilde{C}_3 & 0   & 0   & 1   & -1/2     & -1/2     & 1/2 }
 }

 \vskip4mm
 \noindent {\it Asymmetric resolution}
 \vskip2mm

The asymmetric resolution in \fancthztwztw\ $(a)$ can be treated in
a completely
analogous manner. The vertices of the fan remain the same, so the scalings
\rescalings\ remain the same too. The cones themselves do change, so
the excluded region $F$ will be different, as well as the
intersection products. The generators of the Mori cone and their
intersections are now given by\foot{To avoid cluttering, we drop the
index `$+$' here. In section 3, where the relation between the two
resolutions will be studied in more detail, the `+' index will be
reinstated.}
 \eqn\moriconeasymmres{
 \matrix{                       & D_1 & D_2 & D_3 & E_{23} & E_{31} & E_{12} \cr
 E_{31}\cdot E_{12} \equiv  C_1 & 1   & 1   & 0   & 0      & 0      & -2     \cr
 D_2 \cdot E_{31}   \equiv  C_2 & 0   & -1  & 0   & 1      & -1     & 1      \cr
 E_{23} \cdot E_{31}\equiv  C_3 & 0   & 1   & 1   & -2     & 0      & 0 }
 }
From this, we again get the self-intersections of the curves in the
divisors: $C_1^2|_{E_{12}}=E_{31}^2 \cdot E_{12}=0$,
$C_1^2|_{E_{31}}=-2$, $C_2^2|_{E_{31}}=-1$, $C_2^2|_{D_2}=-1$.
At the level of the
intersections, the curves $C_i$ are related to those of the
symmetric resolution $C_i'$ by $C_1'=C_1+C_2$, $C_2'=-C_2$,
$C_3'=C_3+C_2$. These relations are characteristic of a flop;
indeed, the symmetric and asymmetric resolutions are related by
flopping the curve $C_2$.

The orientifold action is again $\Omega:(z,y) \to (-z,y)$. As in the
symmetric resolution, the divisors $D_i$ support O7-planes. Now
however there is no isolated fixed point: $y=0$ lies in the excluded
set $F$. All $\IP^1$'s are acted on by $\Omega$ as in \Poneaction,
except $C_2$, which is pointwise fixed, since it is embedded in an
O7-plane. The triple intersections of the quotient are obtained by
the rules given earlier (i.e.\ add an overall factor of $1/2$ and
$D_i \to 2 D_i$). This gives for the Mori cone
\eqn\moriconeasymmresquotient{
 \matrix{& \widetilde{D}_1 & \widetilde{D}_2 & \widetilde{D}_3 & \widetilde{E}_{23} & \widetilde{E}_{31} & \widetilde{E}_{12} \cr
 \widetilde{E}_{31}\cdot \widetilde{E}_{12} \equiv  \widetilde{C}_1 & 1   & 1   & 0   & 0      & 0      & -1     \cr
 \widetilde{D}_2 \cdot \widetilde{E}_{31}   \equiv  \widetilde{C}_2 & 0   & -2  & 0   & 1      & -1     & 1      \cr
 \widetilde{E}_{23} \cdot \widetilde{E}_{31}\equiv  \widetilde{C}_3 & 0   & 1   & 1   & -1     & 0      & 0 }
 }

\subsec{Compact model}

To get the compact model $Y$, one simply glues the 64 local models
together, with transition functions determined by the transition
functions between the $z$-coordinates in the original $T^6$. This
gives $3 \times 16 = 48$ exceptional divisors $E_{i\alpha,j\beta}$
and $3 \times 4 = 12$ O7-planes on divisors $D_{i\alpha}$. Here
$i,j=1,\ldots,3$ (with $i<j$)
and $\alpha,\beta=1,\ldots,4$. On each O7-plane, we
furthermore put an $SO(8)$ stack of D7-branes. This locally cancels
the D7-tadpole, so the axio-dilaton is constant on $Y$. In the
symmetric resolution, there are 64 O3-planes. In the asymmetric
resolution, these are absent.\foot{By asymmetric resolution in the
compact model, we mean the resolution obtained by blowing up each
local patch in the same asymmetric way. In principle there could be
mixed symmetric/asymmetric resolutions, but we will not consider
these.}

The global topology of the various divisors is easily deduced. Let
us consider for example the divisors $D_{i\alpha}$ in the symmetric
resolution. The topology of the resolved manifold with the
exceptional divisors removed is the same as the topology of
$T^6/\IZ_2 \times \IZ_2$ with its singularities removed. In this
space, the divisors $D_{i\alpha}$ have topology $T^2/\IZ_2 \times
T^2/\IZ_2$ with the singularities removed, that is $\IP^1 \times
\IP^1$ with four points removed in each $\IP^1$ factor. In each
local patch, this looks like $\IC \times \IC$ with the origin in
each $\IC$ factor removed. From the explicit construction of the
local model given above, it is clear that in the resolved space, the
origin of each $\IC$ factor is simply put back as a point (as
opposed to being replaced by some exceptional curve). Therefore, in
the resolved compact model, the divisors $D_{i\alpha}$ are simply
$\IP^1 \times \IP^1$.

For the topology of the exceptional divisors $E_{i\alpha,j\beta}$ we
get similarly $\IP^1 \times \IP^1$ blown up in 4 points
(corresponding to the four intersections of a fixed line with the
fixed planes in $T^6/\IZ_2 \times \IZ_2$). In the asymmetric
resolution on the other hand the $D_1$ and $D_3$ divisors still have
topology $\IP^1 \times \IP^1$, but the $D_2$ divisors are now $\IP^1
\times \IP^1$ blown up in 16 points. The $E_{12}$ and $E_{23}$
divisors are $\IP^1 \times \IP^1$, and $E_{31}$ is $\IP^1 \times
\IP^1$ blown up in 8 points.

All these divisors evidently have $h^{1,0}=h^{2,0}=0$, since $\IP^1
\times \IP^1$ has this property, and blowing up only changes
$h^{1,1}$. This has important consequences:

(1) There is no massless adjoint matter in the $SO(8)^{12}$ gauge
theory. Since moreover the $D_{i \alpha}$ do not intersect, there is
no massless bifundamental matter either. So the gauge theory is pure
$\CN=1$ super Yang-Mills, and in particular will give rise to
gaugino condensation and the generation of a nonperturbative
superpotential for the K\"ahler moduli governing the size of the
$D_{i\alpha}$.

(2) D3-instantons wrapping the exceptional divisors will have the
minimal number of fermionic zero modes, and therefore contribute to
the superpotential. To make this more precise, we need to consider
the dual M-theory on a smooth Calabi-Yau fourfold, where the
D3-instantons lift to M5-instantons. In this context it has been
shown that if the M5 wraps a divisor satisfying
$h^{1,0}=h^{2,0}=h^{3,0}=0$ (which in particular implies that its
holomorphic Euler characteristic $\chi_h \equiv \sum_i (-1)^i \,
h^{0,i}$ equals 1), there is a contribution to the superpotential
\W. In section 3, we will prove in detail that this is indeed the
case for the lifts of the D3-instantons wrapped on the exceptional
divisors. We also give a short argument below.

There are in fact other consistency conditions that need to be
fulfilled. We will discuss these in section 6.

\subsec{M/F-theory description of the model}

Type IIB string theory on the $Y/\IZ_2$ orientifold is dual to
M-theory on an elliptically fibered Calabi-Yau fourfold with base
$B=Y/\IZ_2$, in the limit of vanishing fiber area. The dual fourfold
is easily constructed in this case \SenVD: it is simply $Z = (Y
\times T^2)/\IZ_2$, where the $\IZ_2$ acts as our orientifold
involution $\Omega$ on $Y$, and as $z \to -z$ on $T^2$. This gives a
singular fourfold, with elliptic fibers degenerating to a $D_4$
singularity on top of the divisors $D_{i\alpha}$, and, in the
symmetric resolution, a degenerate fiber with four terminal $\IZ_2$
singularities on top of each fixed point in $B$. It can be
considered as a partial resolution of $T^8/\IZ_2^3$. Again this is
an example of a Borcea-Voisin model \refs{\B,\V}.

To rigorously address the question whether the lifts of the
D3-instantons have the required properties mentioned in the previous
subsection, one needs to resolve this fourfold in a way that
preserves the elliptic fibration. This is somewhat tricky, and will
be the subject of section 3. The basic idea is simple however. On
the double cover $T^2 \times Y$ of $Z$, the M5-brane lift of a D3
instanton wrapped on a divisor $E$ is $\widetilde{W}=T^2 \times E$.
As argued in section 2.2, $h^{i,0}(E)=0$ for $i>0$ and $E$ any of
the divisors of interest discussed there. So the only harmonic
$(i,0)$-form on $\widetilde{W}$ is the $(1,0)$-form $dz$ living on
$T^2$. Considering now the quotient $W=\widetilde{W}/\IZ_2$ in $Z$,
we see that $dz$ is odd and thus gets projected out. Moreover,
blowing up the quotient singularities of $W$ will only change
$h^{1,1}$. Hence, also after resolving the fourfold, $h^{i,0}(W)=0$.

Another important point in the arguments we will give for the
nonvanishing of the instanton conributions is the fact that the
M5-branes under considerations have trivial third cohomology. This
can be argued similarly. On $\widetilde{W}=T^2 \times E$, the third
cohomology is given by the product of $H^1(T^2)$ and $H^2(E)$. But
quotienting by $\IZ_2$ projects out every such class because the
elements of $H^2(E)$ are even and those of $H^1(T^2)$ are odd.
Furthermore, blowing up will not add any new 3-cycles. So $H^3(W)$
is trivial also after resolving the fourfold. A more precise
discussion will be given in section 3.

\subsec{D3 tadpole and O3-curvature transition}

We now compute the D3 tadpole $Q_3$ measured in the quotient
$Y/\IZ_2$ (as usual, in the double cover $Y$, $Q_3$ is twice this).
In the symmetric resolution, we have O3-planes. Choosing these to be
non-exotic, their contribution to the D3-brane charge is
 \eqn\Othreecontrib{
  Q_3({\rm O}3^-) = - {1 \over 4 } \times 64 = -16.
 }
The 7-branes also contribute to the D3 tadpole, through the
``anomalous'' couplings of RR-fields to worldvolume curvature
\refs{\GreenDD,\CheungAZ,\StefanskiYX,\CrapsTW,\BrunnerZM,\BrunnerZD}.
In a Calabi-Yau threefold, a single D-brane wrapped around a divisor
$D$ thus contributes a D3-charge $-\chi(D)/24$, and an O7-plane
$-4\chi(D)/24$. The total contribution from an O7 + $SO(8)$ D7 stack
wrapped on $D$ is therefore $-12 \chi(D)/24$ in $Y$, and half of
that in the quotient. Hence the total 7-brane contribution is
 \eqn\Dsevencontrib{
  Q_3(7) = -{1 \over 4} \sum_{i,\alpha} \chi(D_{i\alpha}).
 }
In the symmetric resolution, $D_{i\alpha}$ has topology $\IP^1
\times \IP^1$, so $\chi(D_{i\alpha})=4$ and
 $$
  Q_3(7,{\rm symm}) = - 12.
 $$
There can also be contributions from the (half-integral quantized)
$B$-field to various tadpoles, as well as from gauge instantons on
the D7-branes, but we will take $(B-F)|_{D_i}=0$ here, in which case
there are no tadpole contributions of this kind.
Combining O3 and 7-brane contributions in this case gives
 $$
  Q_3({\rm tot})= -28.
 $$
In the asymmetric resolution $\chi(D_1)=\chi(D_3)=4$ and
$\chi(D_2)=4+16=20$, so
 $$
  Q_3({\rm tot}) = Q_3(7,{\rm asymm}) = - 4 - 4 - 20 = -28.
 $$
This agreement of tadpoles in symmetric and asymmetric resolutions
can be understood locally: when flopping one local patch from
\divscthztwztw\ $(a)$ to $(b)$, one O$3$ disappears from the
corresponding orientifold, so $Q_3$ increases by $1/4$ in
\Othreecontrib, but at the same time the Euler characteristic of the
7-branes wrapped around $D_2$ changes: a point gets blown up, which
increases $h^{1,1}$ and therefore $\chi$ by 1, so $Q_3$ decreases by
$1/4$ in \Dsevencontrib, and the total charge $Q_3$ is conserved. If
this transformation can be realized physically, this is a rather
interesting phenomenon, in which a 7-brane stack ``eats'' an O3 and
blows up to conserve the net D3-charge.

This value of the tadpole fits nicely with the fourfold description.
In this picture, the D3 tadpole is given by
 $$
  Q_3 = - {\chi(Z) \over 24}.
 $$
The Euler characteristic $\chi(Z)$ of the fourfold say for the
asymmetric resolution can be computed for example as follows.
Removing the divisors $D_{i\alpha}$ from the base together with the
$D_4$ fibers on top gives a space which is a direct product with a
$T^2$ factor. This has Euler characteristic 0. The Euler
characteristic of the full space is therefore the sum of the Euler
characteristics of the $D_{i\alpha}$ times the Euler characteristic
of the $D_4$ fiber, which is 6 (it can be thought of as a collection
of 5 spheres connected along 4 double points according to the
extended $D_4$ Dynkin diagram). So $\chi(Z)=672$ and $Q_3=-28$. A
similar match can be made for the symmetric resolution after
properly taking into account the contribution to $\chi(Z)$ from
terminal $\IZ_2$ singularities \DasguptaCD.

The D3-tadpole thus produced can be canceled by adding 28
independent mobile D3-branes, or by turning on RR and NSNS 3-form
fluxes. This is further discussed in section 5.

In the symmetric resolution, we could also have chosen our O3-planes
to be exotic. This does not change the geometry; it merely
corresponds to turning on (torsion) twisted cohomology classes for
the field strengths $H_3$ and/or $F_3$ in $H^3(\IR\IP^5,\widetilde{\IZ})$,
where $\IR\IP^5 = S^5/\IZ_2$ surrounds the O3 in $Y/\IZ_2$
\refs{\WittenXY,\HananyFQ,\BergmanRP}. In the M-theory dual this
corresponds to turning on torsion $G$-flux around the terminal
$\IZ_2$-singularities \SethiZK.

The D3-charge of any exotic O3 has the opposite sign of a normal O3.
The total tadpole in this case is thus
 $$
  Q_3 = +16 - 12 = +4.
 $$
To cancel this, one needs 4 {\it anti}-D3 branes, which breaks
supersymmetry. Incidentally, the O3s are required to be exotic for a
consistent CFT description at the orbifold point, as we will discuss
further in section 4. But, as stressed at the beginning of this
section, it should not surprise us to find different consistent
models at large radius.

\newsec{Fourfold Geometry}

In this section we describe the geometry of the resolved Calabi-Yau
fourfold. We describe the symmetric and asymmetric resolutions of the
$T^8/\IZ_2^3$ orbifold, and the birational transformation relating
the two resolutions.

To simplify the presentation, we start with a lower dimensional
orbifold, $T^6/\IZ_2\times\IZ_2$, which is dual to IIB on an
orientifold of $T^4/\IZ_2$. We discuss the resolution of the
orbifold and the properties of the exceptional divisors introduced
in the blow-up process.

We then move to our main example, the $T^8/\IZ_2^3$ orbifold. We
present two distinct resolutions and discuss their elliptic
fibration structure. Starting from local models, we discuss the
birational factorizations of the transformations relating the
elliptically fibered Calabi-Yau fourfolds and their bases. This
somewhat technical analysis is necessary to prove that the
exceptional divisors ${\cal E}_{\bullet\bullet}$ in the (singular)
symmetric resolution have the right topological properties to
contribute to the nonperturbative superpotential: they have
holomorphic Euler characteristic $1$ and the higher cohomology
groups $H^{0,i}({\cal E}_{\bullet\bullet})$, $i=1,2,3$ vanish. We
also show that their third cohomology is trivial, which is important
for arguing that the instanton prefactor is nonvanishing.

\subsec{Lower dimensional orientifold}

It is instructive to consider first the lower dimensional analog,
namely F-theory on $T^6/{\IZ_2\times \IZ_2}$. The action of the
orbifold group $\Gamma=\IZ_2\times\IZ_2$ is presented in \toricdataA
. We can view $T^6/{\IZ_2\times \IZ_2}$ as an elliptic fibration
over $T^4/{\IZ_2\times \IZ_2}$: let $z_1$ and $z_2$ be the
coordinates on the base and $z_3$ be the coordinate on the elliptic
fiber. Then, the elliptic fiber degenerates to type $I_0^*$
fibers\foot{That is, along the fixed point set, the fiber
degenerates to $T^2/\IZ_2$, which is a rational curve with four
singular points.} \K\ along the fixed locus of $\alpha$ and
$\beta$ in the base. In F-theory, such a singularity
corresponds to an $SO(8)$ gauge group \refs{\CF,\BKKMSV}.

The base is $\IP^1\times\IP^1$ and there are $2\times 4$ lines of
$I_0^*$ fibers intersecting at $16$ points\foot{The corresponding Weierstrass
model describing the transverse collision of two $I_0^*$ fibers is not minimal.}.
In order to obtain a smooth Calabi-Yau threefold, we need first to blow-up the
base at these $16$ points and then resolve the singularities of the elliptic
fibration. Let us first discuss the blowing-up of the base $B=\IP^1\times\IP^1$.
\ifig\PonePonebase{Base of the elliptic threefold. There are $8$
lines of $I_0^*$ singularities intersecting at $16$ points, where the Kodaira
vanishing orders are $(4,6,12)$.}
{\epsfxsize2.65in\epsfbox{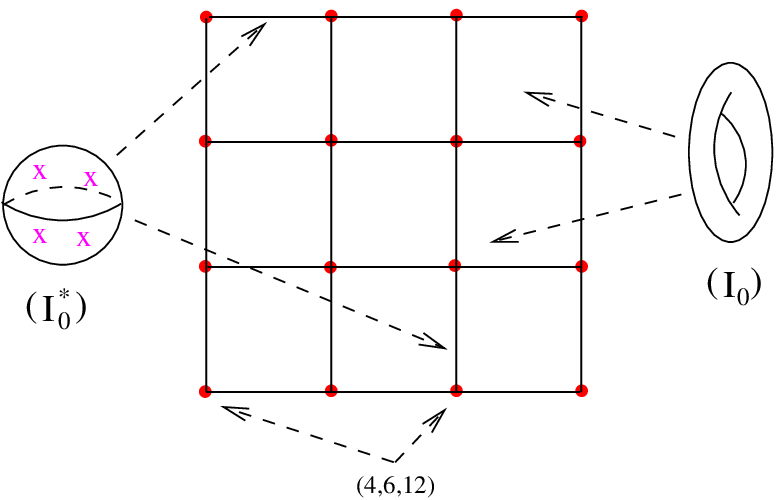}}

\noindent We can work in local coordinates in the fibration $(\IC^2\times T^2)/
\IZ_2\times\IZ_2$ around the point $P=(0,0)\subset\IC^2$, which lies at the
intersection of the fixed lines $z_1=0$ and $z_2=0$. To describe the blow-up,
introduce two coordinate patches $(t_1,z_2)$ and $(z_1,t_2)$ as follows:
$$
(z_1=t_1z_2,z_2=z_2)~{\rm and}~(z_1=z_1,z_2=t_2z_1)
$$
The coordinates $t_1$ and $t_2=1/t_1$ are homogeneous coordinates on the
exceptional $\IP^1$. The $\IZ_2$ actions lift to the blown-up threefold;
there are two fixed points on the exceptional $\IP^1$ given by $t_1=0$ and
$t_2=0$, where it intersects the unresolved singular divisors. The elliptic
fiber over the exceptional $\IP^1$ is smooth, except at the points $t_1=0$
and $t_2=0$, where there are $I_0^*$ singularities. We note that the elliptic
fibration over the blown-up base admits a section. The blow-up process is ilustrated
in the figure below.

\ifig\bup{Base blow-up at an $I_0^* - I_0^*$ collision point.}
{\epsfxsize4.35in\epsfbox{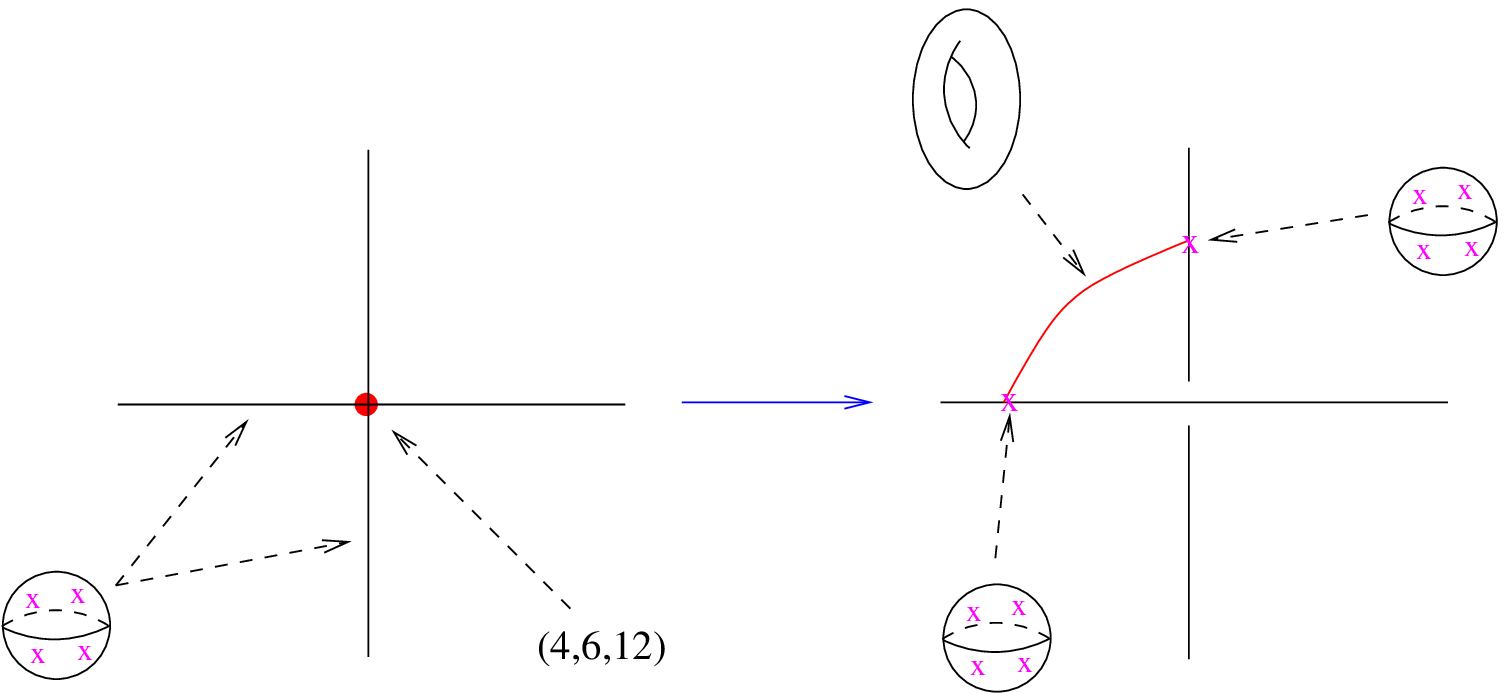}}

Blowing-up the base introduce $16$ new K\"ahler parameters. The next
step is to resolve the elliptic fibration and
this will introduce $8\times 4=32$ additional K\"ahler parameters
since after the blow-ups in the base there are $8$ isolated curves
on top of which the elliptic fiber is type $I_0^*$. Taking into
account the original 2 K\"ahler parameters of the base and the
section, we recover the $51$ K\"ahler parameters of the resolution.
We have obtained a smooth threefold that is elliptically fibered and
is one of the Borcea-Voisin models \refs{\B,\V}.

\ifig\bup{Fiber blow-ups.}
{\epsfxsize4.45in\epsfbox{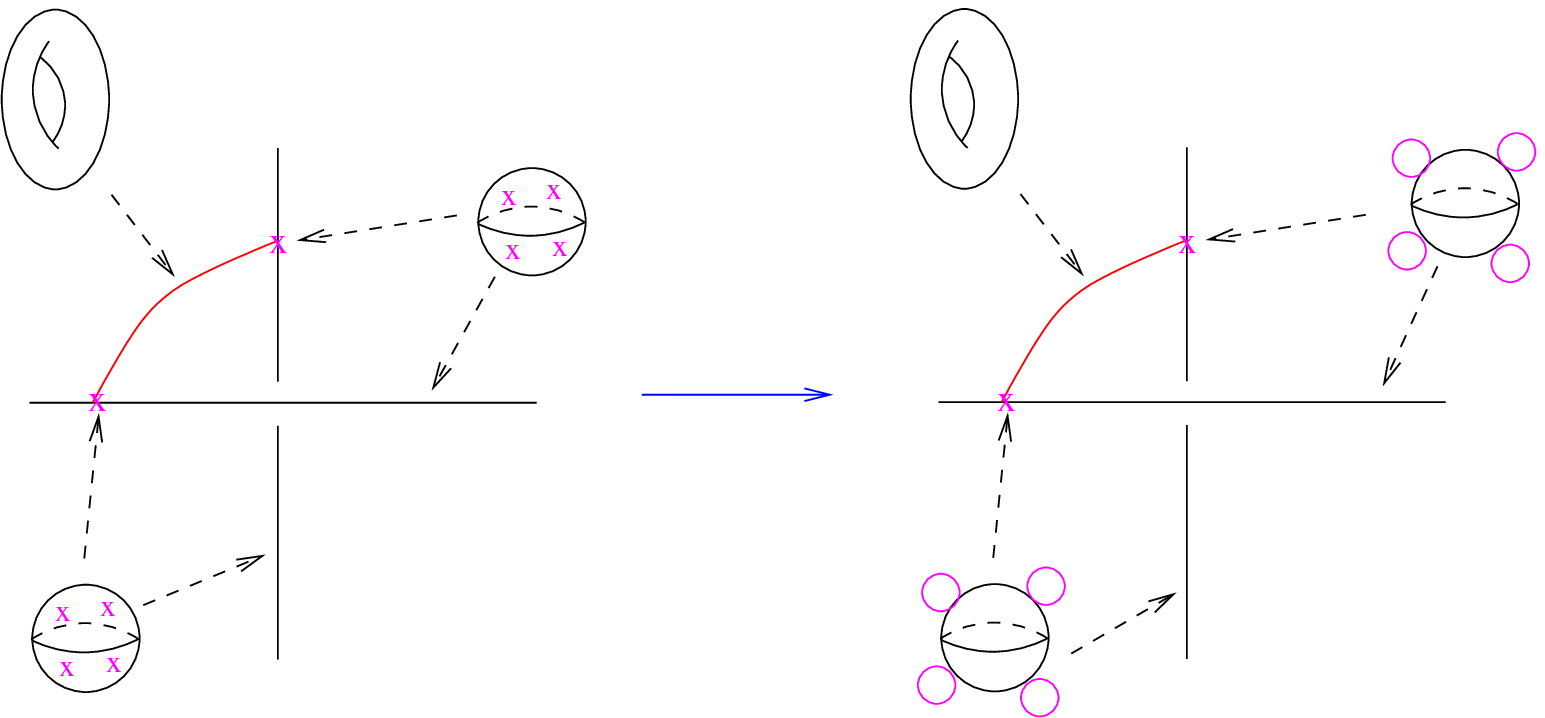}}

The process of resolving the elliptic fiber will turn the elliptic fibration over the
exceptional divisors in the base into rational elliptic surfaces $S_i$, $i=1,\ldots,16$,
that is del Pezzo surfaces $dP_9$. These have $h^{1,0}(S)=h^{2,0}(S)=0$.

In type ${\rm IIB}$ theory language, the description of the
orientifold of $T^4/\IZ_2$ is as follows \gopamukh. The fixed curves are
wrapped by D$7$-branes sitting on top of orientifold O$7$ planes,
and on the worldvolume of each $7$-brane there is an $SO(8)$ gauge
theory. After blowing-up the base, the gauge theory is $SO(8)^8$
with no matter.

\subsec{$T^6/\IZ_2\times\IZ_2$ orientifold}

Proceeding analogously to the previous section, consider now the
orbifold $Z=T^8/{(\IZ_2)^3}$. The orbifold group acts as
\eqn\toricdataA{\eqalign{ \matrix{ & z_1 & z_2 & z_3 & z_4\cr
        \alpha & + & - & - & +\cr
        \beta & - & + & - & + \cr
        \Omega & - & - & - & -. }\cr}}
\noindent This is another example of the Borcea-Voisin construction \refs{\B,\V}.
To get our Calabi-Yau fourfold start with the Calabi-Yau orbifold
$Y=T^6/{\IZ_2\times\IZ_2}$ with the orbifold action given by
\toricdataA\ and construct $Z=Y\times T^2/(\sigma,-\bbbone)$, where
$\sigma$ is an involution of $Y$ that changes the sign of the
holomorphic three-form.

The local singularities are of the form $\IC^4/(\IZ_2)^3$. The
figure below presents the toric resolutions of the singularities. We
see again that the local singularities do admit crepant resolutions
and it is possible to glue them together and get a smooth crepant
resolution ${\widetilde Z}\rightarrow Z$ with $h^{1,1}(\widetilde
Z)=100,h^{2,1}(\widetilde Z)=0, h^{3,1}(\widetilde
Z)=4,h^{2,2}(\widetilde Z)=460$ and $\chi(\widetilde Z)=672$.

\ifig\fancthztwztwztw{Symmetric and asymmetric resolutions of
$\IC^4/(\IZ_2)^3$.} {\epsfxsize4.5in\epsfbox{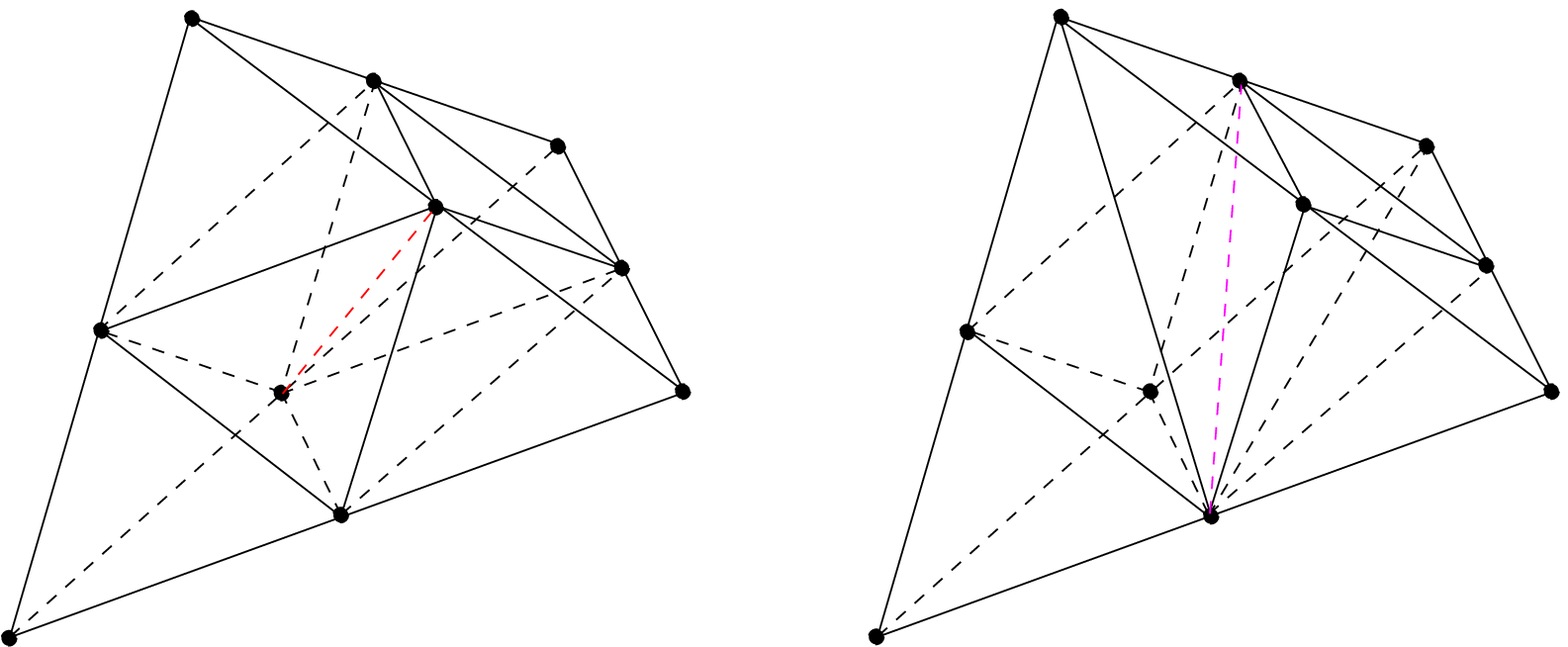}}

We can also think of $Z=T^8/(\IZ_2)^3$ as an elliptic fibration over
$T^6/(\IZ_2)^3$, with $I_0^*$ fibers along the fixed point set of
$\alpha\circ\Omega$, $\beta\circ\Omega$ and
$\alpha\circ\beta\circ\Omega$.
$$\xymatrix{
  T^2    \ar[r] &  T^8/(\IZ_2)^3 \ar[d]^{\pi}\\
  & T^6/(\IZ_2)^3.\\
}$$ \noindent The base is $\IP^1\times\IP^1\times\IP^1$ and the
structure of the fixed point set is presented in fig. $8$ below.
There are $12$ planes of $I_0^*$
singularities intersecting along $48$ lines where the Kodaira
vanishing orders are $(4,6,12)$ and the Weierstrass model is not
minimal. There are $64$ points where three such lines meet. This is
in fact an orientifold of $T^6/\IZ_2\times\IZ_2$.
\vfill\eject
\ifig\PonePonePonebase{Base of the elliptic fourfold. There are $12$
planes of $I_0^*$ singularities intersecting along $48$ lines where
the Kodaira vanishing orders are $(4,6,12)$. There are $64$ points where
three such lines meet.} {\epsfxsize2.3in\epsfbox{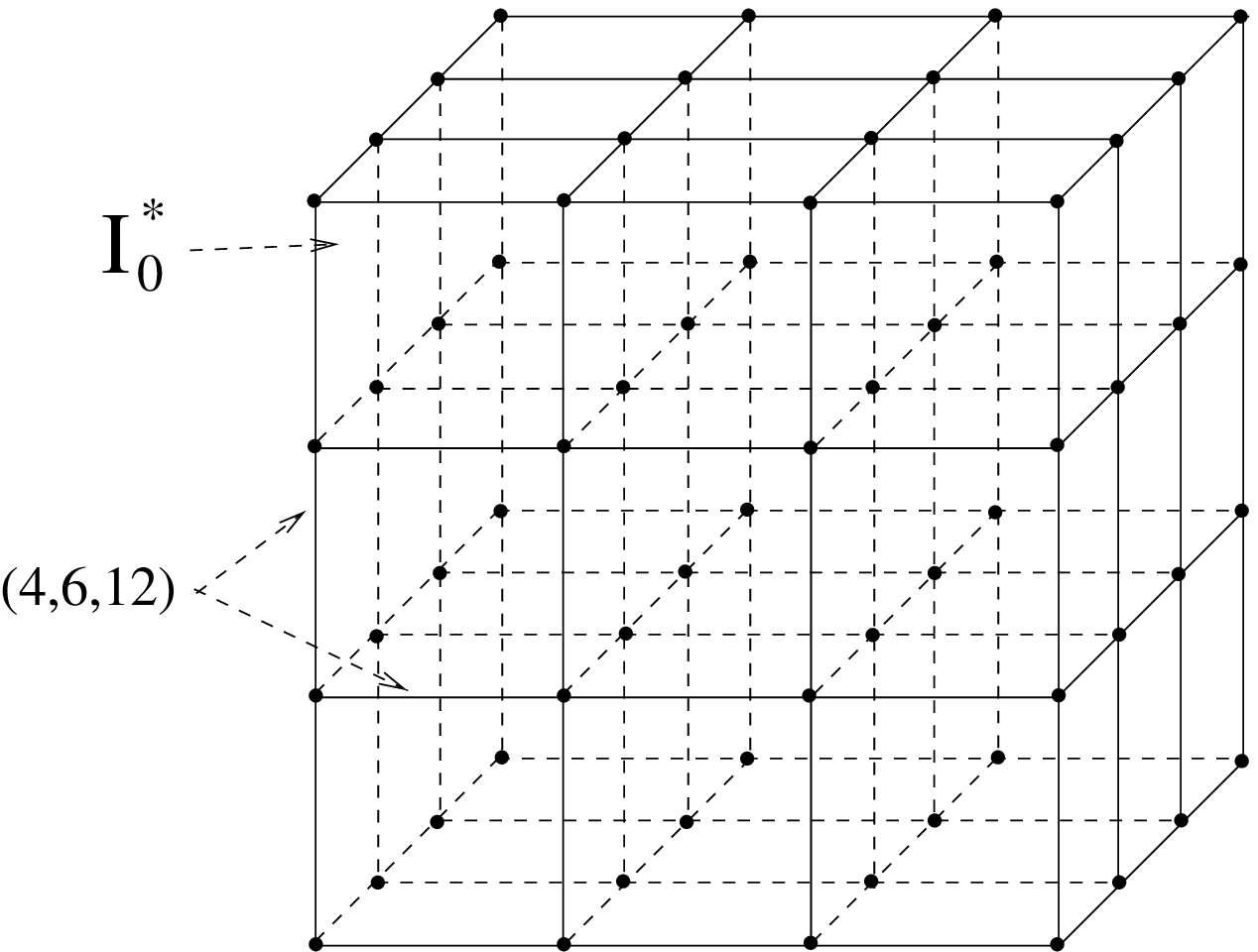}}

We would like to resolve the singularities of the fourfold while
preserving the elliptic fibration structure, such that we can
consider a compactification of F-theory on ${\widetilde Z}$. We
first analyze the local geometry of three intersecting planes of
$I_0^*$ singularities. In order to understand the behavior of the elliptic
fibration under base blow-ups, we consider the local geometry of three
intersecting planes of $I_0^*$ singularities modeled by the following Weierstrass
equation \eqn\wm{
y^2=x^3+xs_1^2s_2^2s_3^2+s_1^3s_2^3s_3^3, } where $s_1,s_2,s_3$ are
affine coordinates along the three coordinate lines. The situation
is presented in fig. $9$ below.

\ifig\PonePonePoneLocalbase{Local geometry of three intersecting
planes of $I_0^*$ singularities. Along the coordinate axes, the
Kodaira vanishing orders are $(4,6,12)$.}
{\epsfxsize2.3in\epsfbox{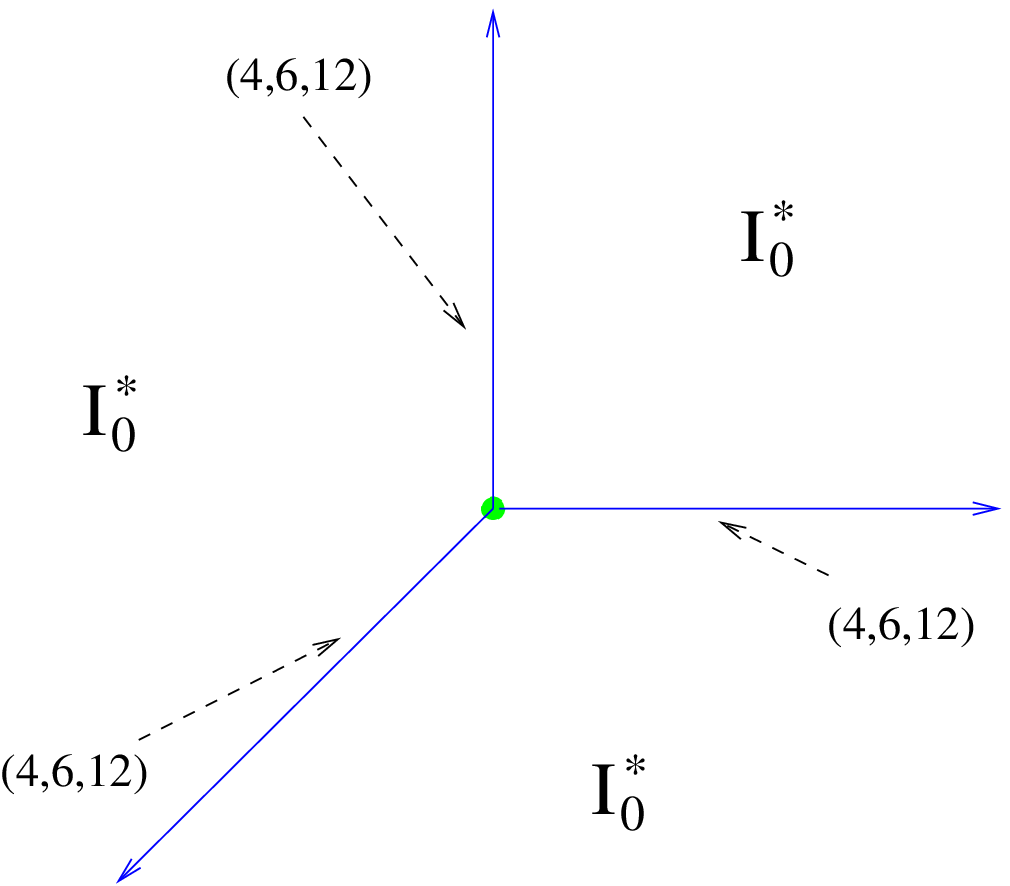}}

\subsec{The base threefolds}

In order to obtain a smooth Calabi-Yau fourfold, we need to blow-up
the $48$ lines in the base on top of which the Weierstrass model is
not minimal. In the local model, the base is $\IC^3$ blown-up along
the three coordinate lines. These blow-ups are toric and the
different triangulations correspond to different topologies of the
noncompact base and are obtained by performing different sequences
of birational transformations. The two different types of
triangulations are presented in the figure below. Note that the
asymmetric triangulation corresponds to a smooth geometry of the
base, which we denote $B_+$, while the symmetric one correspond to a
singular one, and we denote this base by $B$. To see this, note that
there is a $\IZ_2$ singularity in the coordinate patch associated
with the cone $E_{12}E_{23}E_{31}$.

\ifig\Localresbasetwo{Toric description of $\IC^3$ blown-up long the
$3$ coordinate axes. (a) Asymmetric and (b) symmetric
triangulations.} {\epsfxsize4.0in\epsfbox{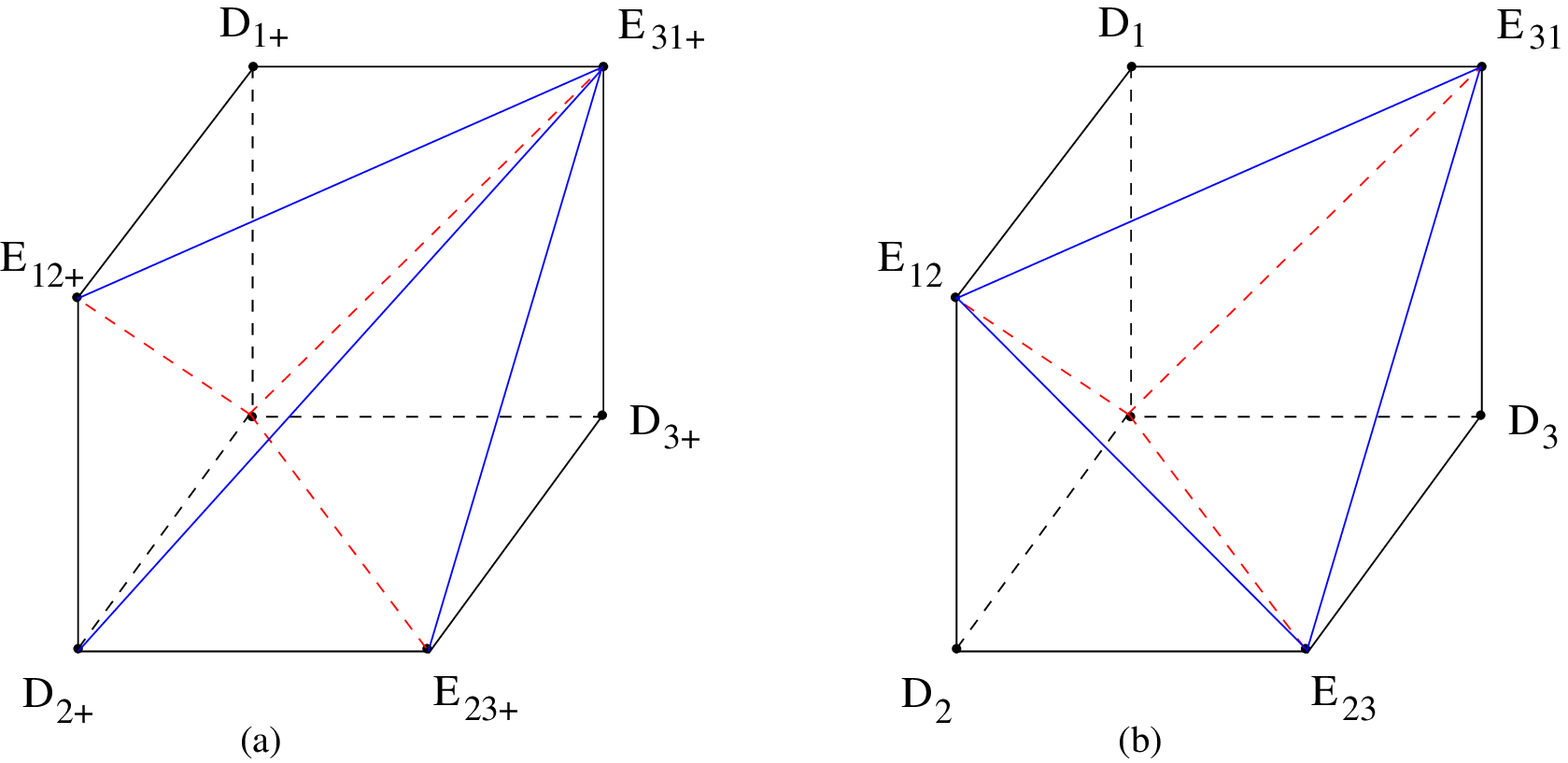}}

It is again useful to draw the dual diagrams illustrating the
compact and noncompact cycles of the local base geometries. These
are presented in the figure below, which is essentially a $\IZ_2$
quotient of \divscthztwztw. The $\IZ_2$ is the orientifold action
and $Y_+{\longrightarrow}B_+$ and $Y\longrightarrow B$ are $2:1$
branched coverings.

\ifig\divscthztwztwi{(a) Asymmetric and (b) symmetric blow-ups:
divisors and curves.} {\epsfxsize4.5in\epsfbox{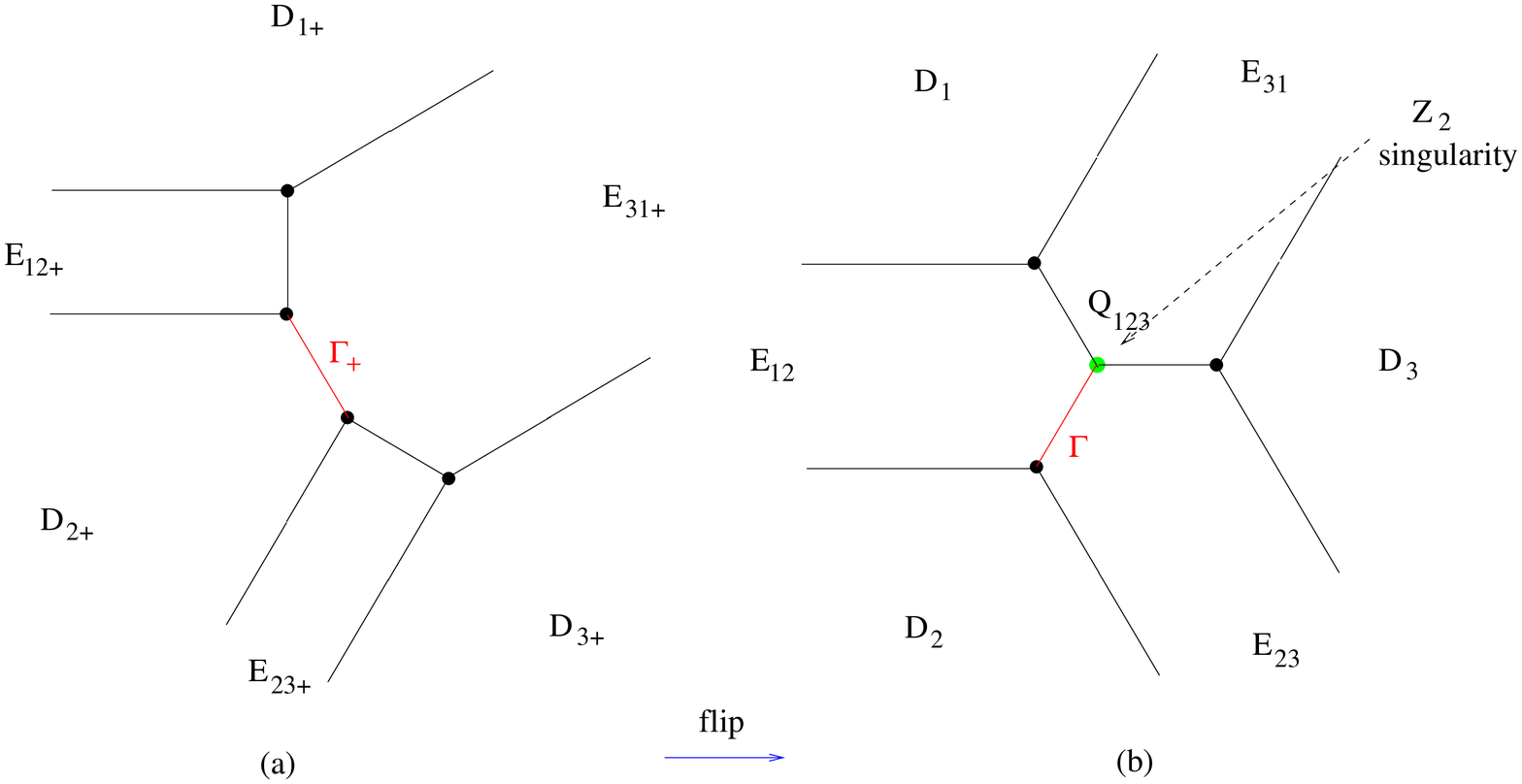}}

The birational transformation between $B_+$ and $B$ is no longer a
flop. To see this, note that the curve $\Gamma_+$ has normal bundle
$N_{\Gamma_+/B_+}= \CO_{\Gamma_+}(-1)\oplus\CO_{\Gamma_+}(-2)$ and
$$
K_{B_+}\cdot\Gamma_+=1.
$$
On the other hand,
$$
K_B\cdot\Gamma=-{1\over 2}.
$$
This is a {\it flip} transformation; there exist birational
morphisms $B \to \overline B$ and $B_+ \to \overline B$ contracting
$\Gamma$ and $\Gamma _+$ respectively .

We now recall some basic
facts about the flip as a $\IZ_2$ quotient of the flop. We follow
the construction in \KM\ and we first describe the flop. In the
following we will restrict to neighborhoods of the curves of
interest. We will abuse notation and denote these neighborhoods by
the same symbols used for the corresponding varieties discussed
until now, which contain these neighborhoods as subsets. Consider
$\overline Y$ defined by the equation $(xy-uv=0) \subset \IC ^4$.
The origin is an isolated singularity. We can resolve the
singularity by blowing up the origin in $\IC ^4$: the exceptional
locus in the resolved threefold $\widetilde Y$ is isomorphic to a
projective quadric in $\IP^3$, defined by the projective equation
$xy-uv=0$.

By blowing up the the plane $x=v=0$ we obtain a small resolution $Y$
of $\overline{Y}$; by blowing up the plane $x=u=0$ we have  another
small resolution $Y_+$ isomorphic to $Y$ outside the locus of the
exceptional curves.  The birational transformation
$$
 Y \leftarrow \cdots \rightarrow Y_+
$$
is a {\it flop}. The equation of $Y_+$ and $Y$ are respectively:
$$
\lambda v-\mu x=0,~~\lambda y-\mu u=0,
$$
$$
\lambda u-\mu x=0,~~\lambda y-\mu v=0.
$$

Consider now the $\IZ_2$ action on $\IC ^4$ defined by $(x,y,u,v)
\to (-x, y, -u, v)$. This involution induces an action on $\overline
Y$, $Y$, $Y_+$ and $\widetilde Y$. The fixed loci of the actions are
respectively:

\noindent$\bullet$  On $\overline Y$: The plane $x=u=0$.

\noindent$\bullet$ On  $\widetilde Y$: The strict transform of the
plane $x=u=0$ and  on the exceptional projective quadric the lines
$x=u=0$ (at the intersection with the fixed  plane) and $y=v=0$.

\noindent$\bullet$ On $Y$: The strict transform $D$ of the plane
$x=u=0$, and a point $Q$ on  the exceptional $\IP^1$ of the small
resolution $Y \to \overline Y$. This point is the  the image of the
fixed line  $y=v=0$ on the exceptional quadric.

\noindent$\bullet$ On $Y_+$: The strict transform $D_+$ of the plane
$x=u=0$; $D_+$ is a plane blown up at a point, which contains the
exceptional rational curve. The other fixed line on the exceptional
quadric has been mapped surjectively onto the line at the
intersection of the quadric and the fixed plane: the image curve is
the exceptional $\IP^1$ of the small resolution $Y_+ \to \overline
Y$.

Let ${\overline B}$, $B$, $B_+$ and ${\widetilde B}$ the quotient
threefolds by these actions:

\noindent$\bullet$  $B$ is singular at the fixed point $Q$ (which is
contained on the flipped curve); $Q$ is the image of the fixed line
$y=v=0$ on the exceptional quadric.

\noindent$\bullet$  $B_+$ is smooth.

This analysis can be trivially extended to the noncompact geometries
that are of interest to us and is in agreement with the toric
description in \divscthztwztwi. We now proceed with the study of the
elliptically fibered fourfolds over $B_+$ and $B$.

\subsec{The elliptic fourfolds}

We start with the elliptic fibration over $B_+$
$$\xymatrix{
  T^2    \ar[r] &  Z_+ \ar[d]^{\pi_+}\\
  & B_+\\
}
$$
and perform successively the
blow-ups along the three coordinate lines according to the sequence
presented below. The figure also indicates how the Kodaira type of
the elliptic fiber on top of the divisors in the base changes at
every step of the the blow-up sequence. We find that the blow-up
geometry contains three exceptional divisors on top of which the
elliptic fiber is smooth. The proper transforms of the surfaces of
$I_0^*$ singularities do not intersect any longer.

\ifig\Localresbase{Local sequence of blow-ups in the base,
asymmetric phase. The fiber is smooth on top of the exceptional
divisors and the surfaces of $I_0^*$ singularities do not intersect
any longer.} {\epsfxsize4.25in\epsfbox{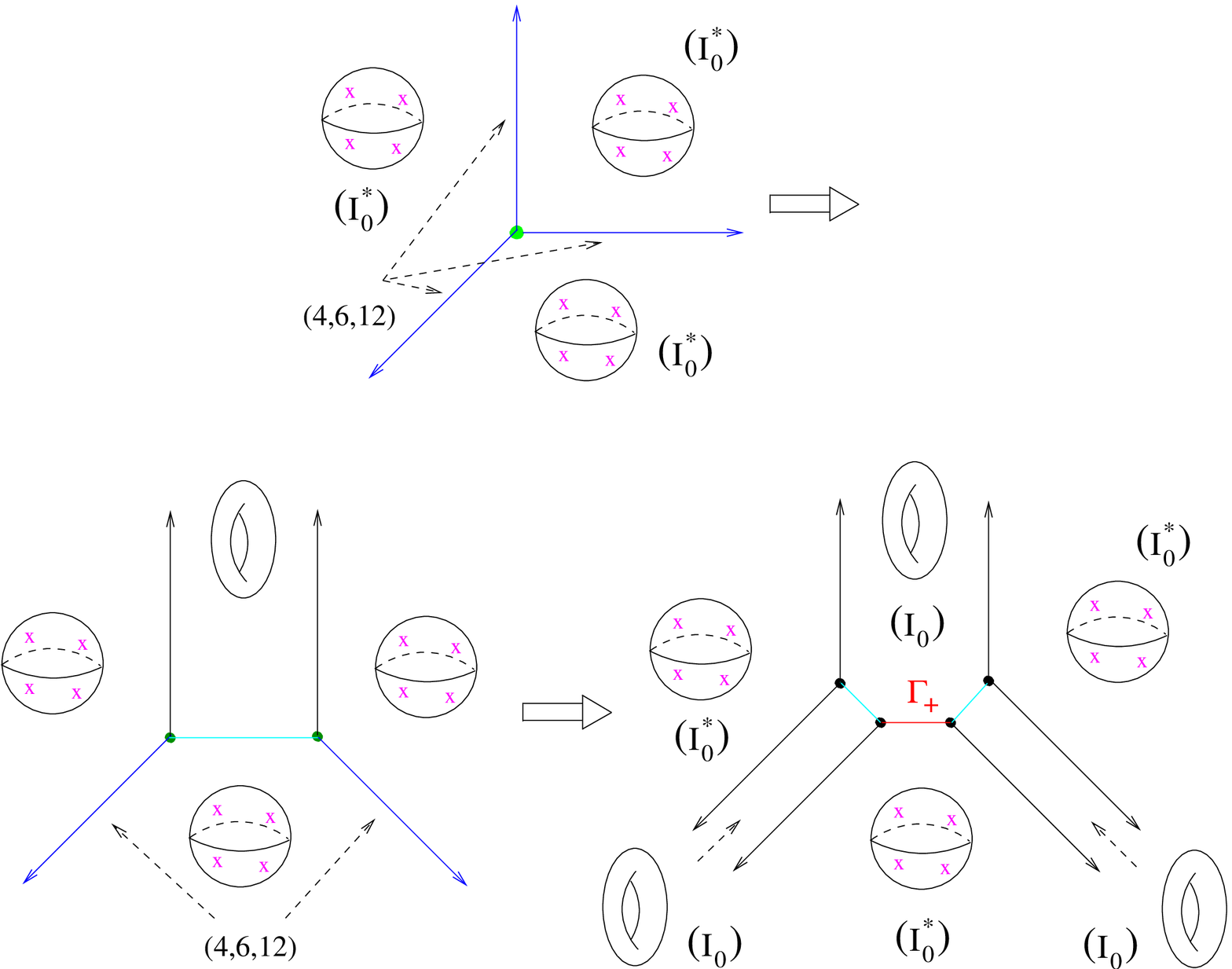}}

This local picture is suggestive of the strategy we need to follow
when blowing- up the base of the compact fourfold. That is, we first
perform the blow-ups along the $16$ curves in a given homology class
and then blow-up the remaining two sets of $16$ curves. Then, there
will be $16$ exceptional divisors that are del Pezzo surfaces $dP_9$
and $32$ exceptional divisors that are Hirzebruch surfaces $\IF_0$.
Moreover, the proper transforms of the $I_0^*$ surfaces are
Hirzebruch surfaces $\IF_0$ and $\IF_0$ blown-up at $16$ points.

The next step is to resolve the elliptic fibration. The analysis is
similar to the one performed in the lower dimensional example. We
find that the fourfold exceptional divisors ${\cal
E}_{\bullet\bullet +}$ are either threefolds isomorphic to
$\IP^1\times dP_9$ or blow-ups of $\IP^1\times dP_9$ along $8$
(reducible) rational curves. They are therefore rational and have
holomorphic Euler characteristic $1$. In fact, in the case of the
smooth elliptic fourfold we can arrive at this conclusion by
performing a Riemann-Roch computation as in \G.

We would like to understand the elliptic fibration over the singular
base $B$ along the same lines. It turns out that it is a bit subtle
to understand how the fibration behaves under the
sequence of birational transformations that lead to $B$. In order to
construct an elliptic fourfold, let us return to the local picture
used in the previous subsection. Consider the trivial elliptic
fourfolds $\overline Y \times T^2$, $Y \times T^2$, $Y_+ \times T^2$
and $\widetilde Y \times T^2$; let us consider a $\IZ_2$ action,
defined on the first factor as the action defined in the previous
subsection and on the second factor as the standard involution on
the torus.

Now, let $\overline W$, $W$, $W_+$ and $\widetilde W$ be the
fourfolds obtained by quotienting this action: they are naturally
elliptically fibered over ${\overline B}$, $B$, $B_+$ and
$\widetilde B$. The singular fibers map over the fixed locus of the
quotient action as follows.

In $Y_+  \times T^2 \to Y_+$ the fixed locus is $D_+ \subset Y_+$;
in the quotient $W_+  \to B_+$ the fibers are double rational curves
with $4 $ singular  points, over the points in $D_+$. The minimal
resolution of $W_+$ is a smooth fourfold $Z_+$, the singular fibers
over $D_2$ in $Z_+$ are of Kodaira type $I^*_0$.

In $Y \times T^2 \to Y$ the fixed locus is the strict transform $D$
of the  plane $x=u=0$ and the fixed point $P$: in the quotient $W
\to B$ the fibers are double  rational  curves with $4$ singular
points over each point in $D$ and the point $Q$. The (minimal)
smooth resolution of $W$ is a singular fourfold $Z$: $Z$ is smooth
over  $D$, the singular fibers are of  Kodaira type $I^*_0$, while
the singular isolated  fiber over $Q$  has $4$ singular points,
these $4$ singularities are terminal (already in $W$) \Reid. These
are the only singularities of $Z$.

Again, this local analysis extends trivially for the noncompact
geometries we are interested in. In the smooth case, it agrees with
the previous description of the elliptic fibration. In the singular
case, we learn that the elliptic fibration is as in the figure
below.

\ifig\Localresbases{Elliptic fibration over the singular base. The
fiber is generically smooth on top of the exceptional divisors, but
degenerates to a double rational curve with $4$ terminal
singularities over the $\IZ_2$ singularity in the base $B$. The
surfaces of $I_0^*$ singularities do not intersect any longer.}
{\epsfxsize3.0in\epsfbox{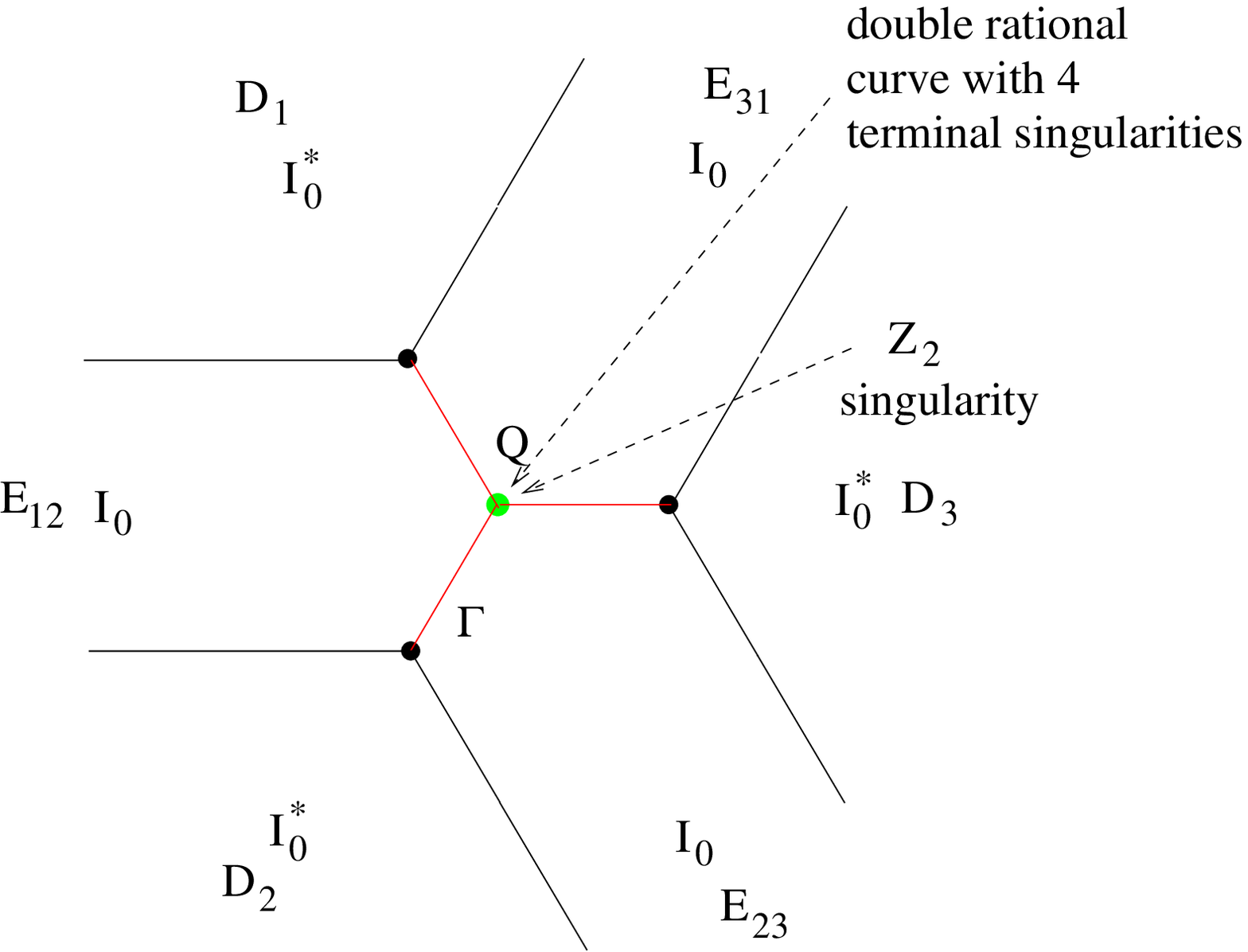}}

In the following sections we describe in detail a birational
factorization of the flop between the elliptic fourfolds $\pi: Z \to
B$ and $\pi _+: Z_+ \to B_+$ as a sequence of blow-ups and
contractions. As in the previous sections, it is sufficient to
consider the local situation; we start by describing the
factorization of the threefold flip.

The objects of particular interest are the threefold divisors in $Z$
and $Z_+$ of holomorphic Euler characteristic $1$: we saw that
${\cal E}_{31+}=\pi_{+} ^{-1}(E_{31+}) \subset Z_+$ is a smooth
threefold (in the smooth fourfold $Z_+$) with holomorphic Euler
characteristic $1$. From the explicit description of the fourfold flop
it will follow that  the strict trasform of
${\cal E}_{31+} \subset Z_+$  is ${\cal E}_{31}=\pi ^{-1}(E_{31})
\subset Z$  and that ${\cal E}_{31}$ is also a divisor (in the
singular fourfold $Z$) with holomorphic Euler characteristic $1$.

\subsec{A birational factorization of the threefold flip}
 We consider the asymmetric resolution $B_+$ and perform two
 blow-ups followed by two birational contractions. The composition
 of these morphisms gives a birational factorization of the
 flip. The geometry can be easily seen from the pictures below.
  The existence of the blow-ups is clear: to see the contractions we can either consider a
  toric picture, or invoke the {\it minimal model contraction algorithm}
  with its {\it log} variation. We describe the contraction algorithm because it  can  also be used to
  study the transformation between our fourfolds.

\ifig\flips {$B_+$ blown-up and contracted to $B$. (a) Threefolds
blow ups. (b) Threefold contractions.}
{\epsfxsize4.0in\epsfbox{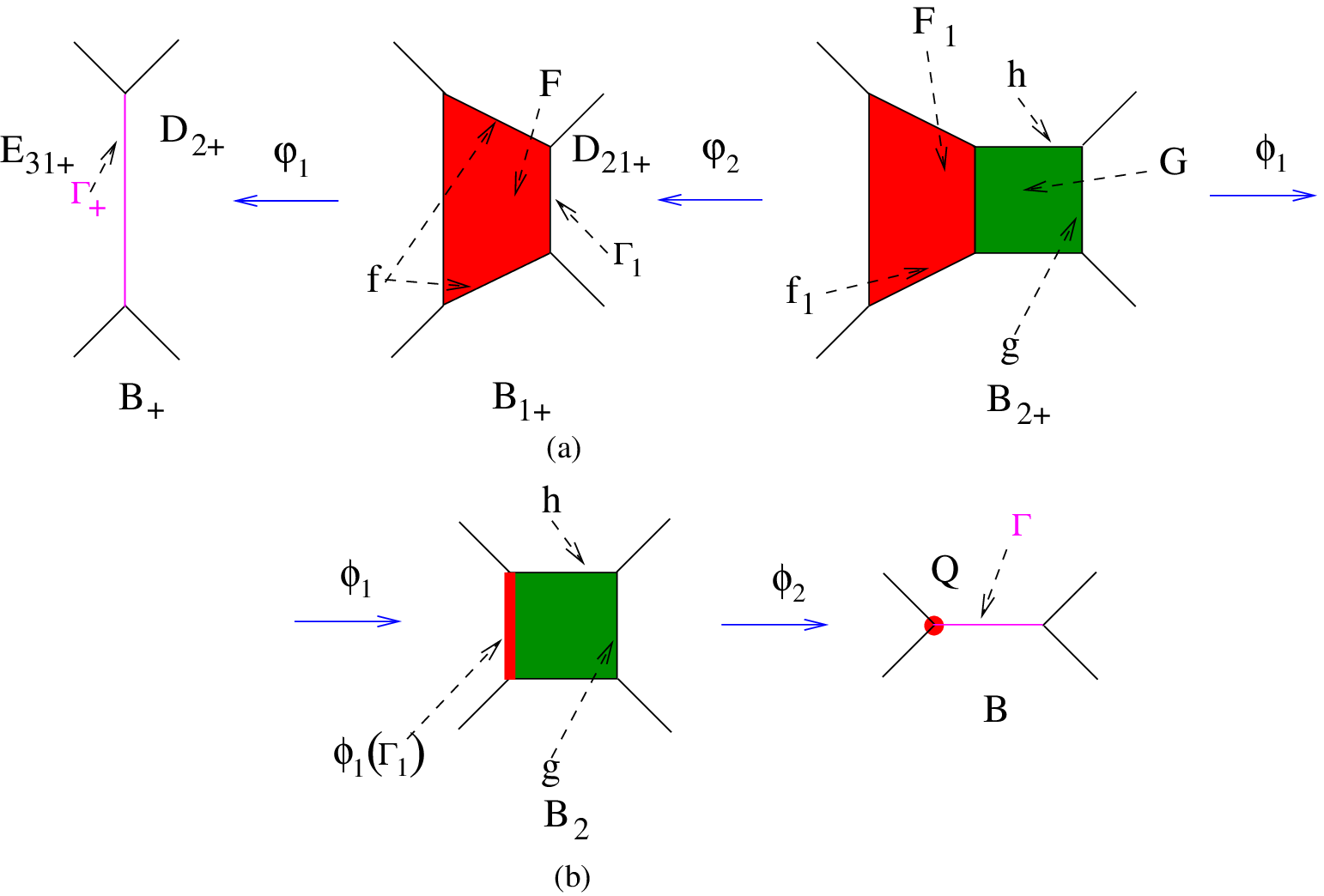}}

\noindent $\bullet$ (The first blow up.) In $B_+$ we first blow up $\Gamma _+$ and denote by
$\varphi_1: B_{1+} \to B_{+}$ the corresponding contraction
morphism; since $\Gamma _+$ is a smooth rational curve with normal
bundle $(-2,-1)$, the exceptional divisor $F$ is isomorphic to the
Hirzebruch surface $ \IF_1$; let $f$ denote the fiber of the ruling.
Let $D_{21+}$ be the strict transform of $D_{2+}$ in $B_{1+}$. Let
$\Gamma _1$ be the unique irreducible curve with negative self
intersection on $F$, then: ${(\Gamma _1) ^2 }_{| F}= -1$ and $\Gamma
_1= F \cap D_{21+}$. Note that $\Gamma _1$ is a smooth rational
curve with normal bundle $(-1,-1)$.

\noindent $\bullet$ (The second blow up.)
 In $B_{1+}$, let us blow
up $\Gamma _1$ and denote by $\varphi_2: B_{2+} \to B_{1+}$ the
corresponding contraction morphism; since $\Gamma _1$ is a smooth
rational curve with normal bundle $(-1,-1)$,  the exceptional
divisor $G$ of $\varphi_2$ is isomorphic to the Hirzebruch surface $
\IF_0$. Let $F_{1}$ denote the strict transform of $F$ in $B_{2+}$,
and $f_1 \simeq f$ the strict transform of the ruling $f$; $f_1$ is
a smooth rational curve with normal bundle $(0,-2)$. In fact,  if we
denote by $H$ the strict transform of $E_{12+}$ in $B_{2+}$,
$f_1=F_1\cap H$ and  ${(f_1)^2}_{|H}=-2$. By the adjunction formula
$K_{B_{2+}} \cdot f_1=0$; note also that $H \cdot f_1=-2$.

 \noindent $\bullet$  (The first contraction.)
  By the (log-)contraction theorem there exists a
birational morphism $\phi _1: B_{2+} \to B_{2}$ contracting all the
curves in the ruling $f_1$. The contraction $\phi _1$ is a
log-terminal contraction; $B_{2}$ has $\IQ$-factorial canonical
singularities along  $\phi _1 (\Gamma _1)$, which is a smooth
rational curve. ($f_1$ is a curve with normal bundle $(0,-2)$: which
means it is a $-2$ curve in a generic surface $S$ that meets $F_1$
transversely along a fixed curve in the ruling $f_1$. Contracting
the ruling $f_1$ will lead to an $A_1$ singularity in the image of
the surfaces $S$: in the threefold the singularity is  $A_1 \times
\IP ^1$, where  $ \IP ^1 \simeq \Gamma _1$.) This contraction is
allowed since $f_1$ is a negative extremal ray in the Mori cone
$\overline {NE}(B_{2+})$. To see this, note that
$f_1=\varphi^*_2f-h,$ where $h$
 is a ruling of $G$,
and ${(K_{B_{2+}}+ 1/2 F_1) \cdot f_1=-1}$. For further reference,
note also that
 $K_{B_{2+}}= \phi_1 ^*
(K_{B_{2}})$.

\noindent $\bullet$ (The second contraction.) Now we consider $\phi
_1(G) \simeq \IF_0$ and  its ruling $g$ which is homologous to $\phi
_1 (\Gamma _1)$ in $\phi_1(G)$ (in fact $[g]=[ \phi _1 (\Gamma
_1)]$).
 Again,
by the contraction theorem there exists a birational morphism $\phi
_2: B_{2} \to B$ contracting all the curves in the ruling $g$. In
fact, $g$ is an extremal ray in the Mori cone $\overline
{NE}(B_{2})$ and $K_{B_{2}} \cdot g =-1$ (to see this, recall that
$K_{B_{2+}}= \phi ^* (K_{B_{2}})$, the normal bundle of $g$ in
$B_{2+}$ is $(0,-1)$ and use adjunction on the smooth threefold
$B_{2+}$). In particular, $\phi _2(\phi _1 (\Gamma _1) )=Q$ is a
singular point: it can be verified that $B$ is smooth outside $Q$
(it follows from the Castelnuovo-Enriques contraction criterion, see
\GH) and that the singularity at $Q$ is terminal.

To see that the singularity is terminal is enough to compare the
pullback of the canonical divisor to the smooth resolution $\phi _2
\cdot \phi _1: B_{2+} \to B$ with that of $B_{2+}$: $K_{B_{2+}}=
(\phi _2 \cdot \phi _1)^*(K_B) + G$. For $h$, the other ruling of
$G$, we have that $\phi_2\cdot \phi_1( h)= \Gamma$ is a smooth
rational curve, and with the same methods it can also be verified
that $K_B \cdot \Gamma =-1/2$, as claimed in the previous
subsection.


\subsec{The fourfold flop}

We first study the induced elliptic fibration $\pi_{2+}: Z_{2+} \to
B_{2+}$;
 $Z_{2+}$ is smooth and isomorphic to $Z_+$ outside the exceptional locus.
 The  fiber over a point in $G$ is a smooth elliptic curve, while the  fiber over a point
 in $F_1$ is of type $I^*_0$.
 By construction, the fibers in $Z_{2+}$ over the points in $\Gamma _+, \Gamma _1$ and $f_1$
  are all isomorphic
  to each other, and so are all the fibers over the points in $g$.

 Note also that $K_{Z_{2+}}=\pi ^*_{2+}(K_{B_{2+}} + \Delta _{2+})$, where
 $\Delta _{2+} $ is supported on the ramification locus of the fibration, with  suitable
 coefficients, determined
 by the Kodaira type fibers; for example, the coefficient of $F_1$ in $\Delta_{2+}$ is $1/2$. The
  birational morphism $Z_{2+}
\to Z_+$ is induced by the resolution of the pullback of $Z_+$. We
want to show that there is a birational morphism  $Z _{2+} \to Z$
induced by the contraction morphisms $\phi _2 \cdot \phi _1 $.

\noindent $\bullet$ (The first fourfold contraction, following
$\phi _1$). The goal of these contractions if to construct a
birationally equivalent elliptic fibration over $B_2$; the threefold
over $F_1$ will be contracted
 to a singular surface over
$\phi_1 (\Gamma_1)$, in particular the surfaces in the fourfold over
the fiber $f_1$ are contracted to a double rational curve with 4
marked points.

To see this, note first that $\pi^{-1}_{2+}(f_1)$ is a reducible
surface, as illustrated in the picture below: the red vertical
surface is isomorphic to $\IF _0$ (this surface appears in the
fourfold with
 multiplicity two and it
is the strict transform of $T^2 / \IZ _2 \times \IP ^1$); the four
blue surfaces arise as resolution of the four fixed points in $T^2
$, and each of these surfaces is isomorphic to $\IF_1$. The negative
sections of the $\IF_1$ surfaces are exactly at the  intersections
of the blue surfaces with the (reduced) red surface.

\ifig\flips {The surface in $Z_{2+}$ over $f_1 \in F_1$ and the
first contraction, following $\phi_1$. The right hand side
represents the fiber over $\phi_1(\Gamma_1)$.}
{\epsfxsize3.2in\epsfbox{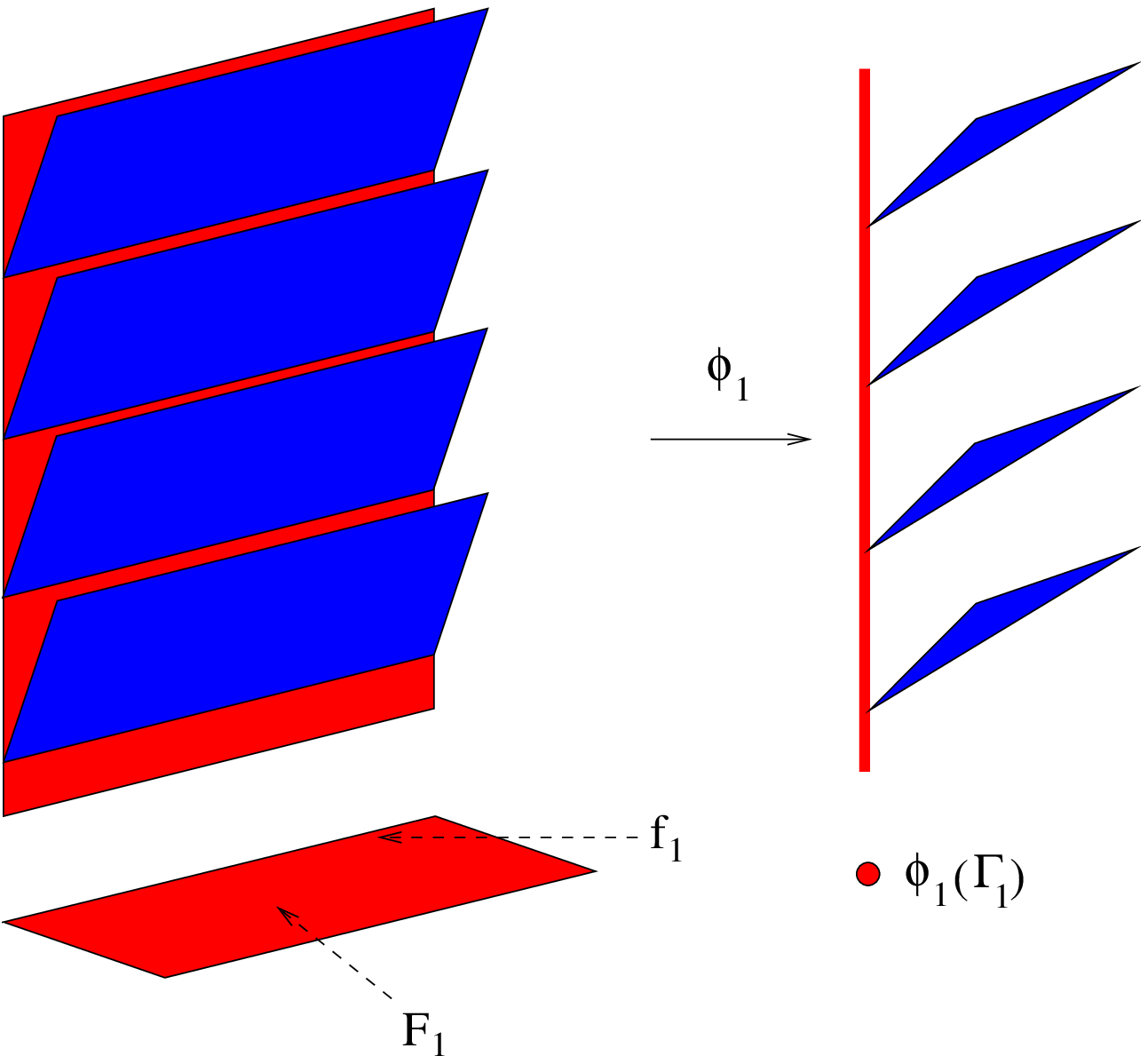}}

Consider now a surface $S_{2+}\hookrightarrow B_{2+}$ intersecting
$F_1$ and $G$ along their rulings and let ${\cal E}_{2s}$ be the
elliptic threefold defined by the following commutative diagram
$$\xymatrix{
  {\cal E}_{2s}    \ar[d]^{\pi_{2s}}\ar[r] &  Z_{2+} \ar[d]^{\pi_{2+}}\\
  S \ar [r] & B_{2+},\\
}$$ where the horizontal arrows are inclusions. We have $K_{{\cal
E}_{2s}}=\pi^*_{2s}(K_S+\Delta_S)$ where $\Delta_S=\Delta_{2+}\cap
S$. Let $\gamma$ be a curve in ruling of $\IF _0$ such that $\pi
_{2s}(\gamma)= f_1$; then by the projection formula we have
$K_{{\cal
E}_{2s}}\cdot\gamma=(K_S+\Delta_S)\cdot\pi_{2s*}(\gamma)=\Delta_S\cdot
f_1=-1$ since the fibers over $f_1$ are type $I_0^*$ and $f_1$ does
not intersect the proper transform of $D_{21+}$ under $\varphi_2$.
By adjunction on ${\cal E}_{2s}$ we then find that $N_{\gamma/{\cal
E}_{2s}}=\CO_{\gamma}(0)\oplus\CO_{\gamma}(-1)$, proving that the
blue surfaces are isomorphic to $\IF_1$. Now, in order to find the
birational contractions we use again the minimal model contraction
algorithm: by the projection formula: $K_{Z_{2+}} \cdot \gamma=
{\pi^*_{2+}} (K_{B_{2+}}+ \Delta _{2+}) \cdot \gamma= (K_{B_{2+}}+
\Delta _{2+}) \cdot f_1 = (K_{B_{2+}}+ 1/2 F_1) \cdot f_1 = -1$.
Therefore, $\gamma$ is a negative extremal ray in the Mori cone
${\overline {NE}}(Z_{{2+}})$, and by the contraction algorithm there
exists a birational morphism which contracts all the curves in the
homology class $\gamma$, to a  smooth fourfold. The image of the red
double threefold over $F_1$ is now a double surface isomorphic to
$\IF_0$. The blue surfaces become isomorphic to $\IP^2$, represented
by the blue triangles in fig. 15.

\ifig\flips {The surface in $Z_{2}$ over $\phi_1 (\Gamma_1) $
after the first contraction, following $\phi_1$.}
{\epsfxsize1.5in\epsfbox{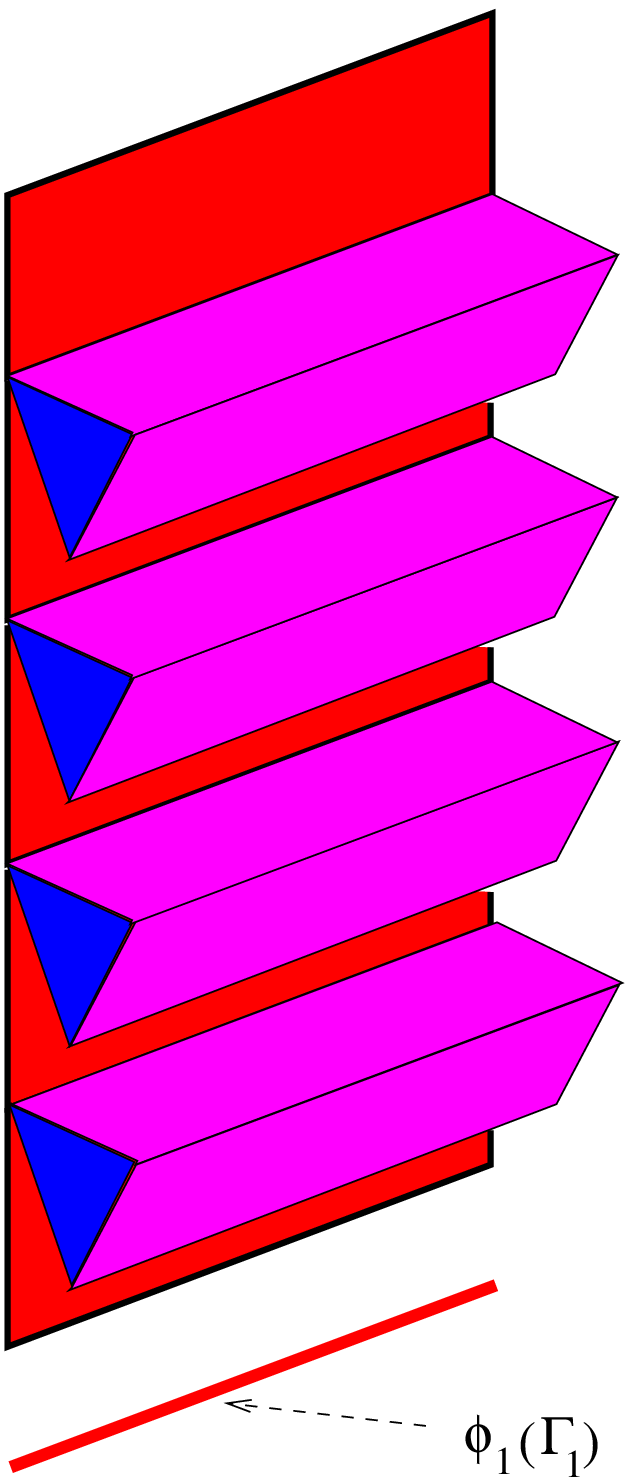}}

The contraction algorithm ensures then the existence of another
birational morphism contracting all these blue $\IP^2$'s, in the
fourfolds (it can be checked that $N_{\IP^2/{\cal
E}_{2s}}=\CO_{\IP^2}(-2)$), to four smooth rational curves in
$\IF_0$ (in the same ruling). The resulting fourfold  $Z_2$ is now
singular along these four curves and the singularities are terminal;
we also have an elliptic fibration $\pi _2: Z_2 \to B_2$. The fiber
over $\phi _1 (\Gamma _1)$ is the product of $\IP^1$ and a double
rational curve with 4 marked points.

\noindent $\bullet$ (The second fourfold contraction, following $\phi
_2$.) Using again the birational contraction algorithm we can
contract all the (product) surfaces over the fibers $g \in G$: the
resulting fourfold $Z$ is now elliptically fibered over $B$. The
elliptic fibration admits a section away from the singular point in
the base, where it becomes a double section. $Z$ is smooth  outside
the fiber over  point $Q=\phi _2 \cdot \phi _1 (\Gamma _1)$: over
$Q$ there is a double rational curve with four terminal quotient
singularities, as described in the previous section with the local
quotient analysis.

\subsec{ Two divisors with holomorphic Euler characteristic $1$:
${\cal E} _{31+} \subset Z_+$ and  ${\cal E}_{31} \subset Z$}

The fourfold birational transformation described in the previous
section exchanges the two divisors ${\cal E} _{31+} \subset Z_+$ and
${\cal E}_{31}  \subset Z$; the first threefold is the smooth
resolution of the second, which has terminal singularities. We can
compute directly $\chi_h ({\cal E} _{31+})=1$ and then deduce
$\chi_h ({\cal E} _{31})=1$.

In particular the flop among the fourfolds restricts to  a
birational morphism $\rho: {\cal E} _{31+} \to {\cal E}_{31}$, which
consists of the birational contraction of
  the reducible divisor ${{\pi^* _+}(\Gamma _{+})}$ to a double rational curve with four marked points, which
  are terminal singularities. This completes the proof that the (vertical) exceptional divisors in the singular fourfold also
have holomorphic Euler characteristic $1$. Moreover, since the
exceptional divisors in the asymmetric resolution have $h^{0,i}
({\cal E}_{\bullet\bullet+})=0$, $i=1,2,3$, it follows immediately
that the same holds true for the exceptional divisors in the
symmetric resolution and therefore they do contribute to the
superpotential.

\subsec{Third cohomology of the exceptional divisors}

In this subsection we show that the third cohomology of the
exceptional divisors is trivial, both for the smooth and singular
Calabi-Yau fourfolds. Therefore, there is no theta function entering
into the instanton prefactors. We recall that in the smooth
resolution the exceptional divisors are either $\IP^1\times dP_9$ or
blow-ups of $\IP^1\times dP_9$ along $8$ (reducible) rational
curves. It is clear that these are rational and that the third
cohomology is trivial, $H^3({\cal E}_{\bullet\bullet +},\IZ)=0$.

We would now like to prove that the same is also true for the
exceptional divisors in the singular fourfold. The sequence of
birational transformations taking one of the exceptional divisors in
the smooth fourfold to the corresponding exceptional divisor in the
singular fourfold is presented in the figure below.

\ifig\Divseq{Sequence of birational transformations of the
exceptional divisors.} {\epsfxsize4.8in\epsfbox{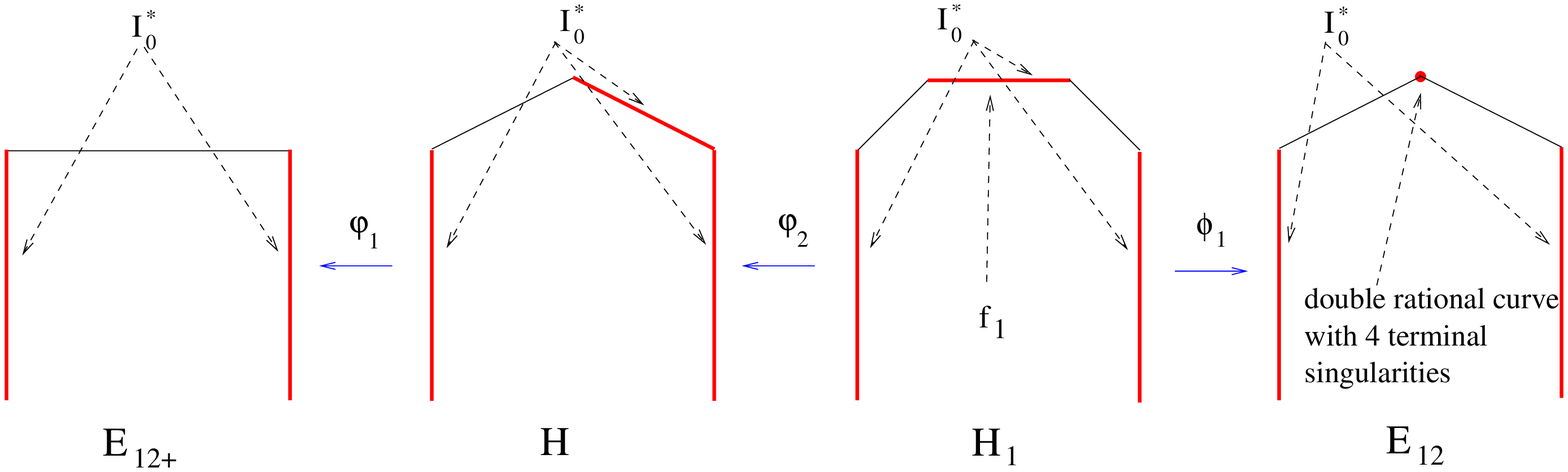}}

A topological Euler characteristic computation shows that $H^3({\cal
H}_1,\IZ)=0$, where ${\cal H}_1=\pi^*_{2+}(H_1)$. The group
$H^3({\cal E}_{12},\IZ)$ is nontrivial only if the relative homology
group $H^3({\cal H}_1,\pi^*_{2+}(f_1);\IZ)$ is nontrivial. We have
the following long exact sequence in homology
$$
\cdots\rightarrow
H^2({\cal H}_1,\IZ)\mathop{\rightarrow}^{i^*}
H^2(\pi^*_{2+}(f_1))\rightarrow H^3({\cal H}_1,\pi^*_{2+}(f_1);\IZ)
\rightarrow H^3({\cal H}_1,\IZ) \mathop{\rightarrow}^{i^*}
H^3(\pi^*_{2+}(f_1))\rightarrow
\cdots,
$$
where $i:\pi^*_{2+}(f_1)\hookrightarrow {\cal H}_1$ is the
inclusion. But $i^*:H^2({\cal H}_1,\IZ)\rightarrow H^2(\pi^*_{2+}(f_1),\IZ)$
is a surjection, and therefore $H^3({\cal H}_1,\pi^*_{2+}(f_1);\IZ)$ is trivial.
This proves that $H^3({\cal E}_{12},\IZ)=0$ and this will be true for all
the exceptional divisors in the Calabi-Yau fourfold $Z$ by symmetry.

\newsec{Relation to Toroidal Orientifold Models}

At this point the world-sheet definition of our model should be
clear -- we take the supersymmetric sigma model with Calabi-Yau
target space $Y$ and orientifold as discussed in \S2, and then add
background Neveu-Schwarz and Ramond-Ramond flux chosen to stabilize
the complex structure and dilaton, as we will discuss in \S5.  While
backgrounds with RR flux are not easy to handle in the NSR
formalism, there is abundant evidence that the difficulties are just
technical, and there is steady progress in developing formalisms
which solve them \berkovits.

While the model is well defined, it would be even better to obtain it
starting from a solvable world-sheet definition, such as an
orientifold of $T^6$.  Unfortunately, we cannot expect to obtain our
model in this way, for the simple reason that it contains wrapped
branes which become massless in the orientifold limit.  These are
D$3$-branes wrapped on the exceptional two-cycles obtained by
resolving the $\IC^3/\IZ_2\times\IZ_2$ singularities; they give
rise to BPS strings charged under RR two-forms, which become tensionless
in this limit.  Such states lead to divergences in the limit and
the breakdown of world-sheet conformal field theory.  Since we will be
stabilizing the K\"ahler moduli away from this point, this is not a
problem for our purposes, but it does preclude using the toroidal
orientifold as the definition.

On the other hand, one can construct a very similar toroidal
orientifold model which we will call $T'$, suggested by Gopakumar and
Mukhi \gopamukh\ and first studied by Aldazabal {\it et al} \alda.  In
this section, we outline its construction, filling in some gaps in
previous work, but leaving most of the details to future work.
Besides explaining how the spectrum we found in F-theory arises, the
main points we wish to see here are how the problematic states are
given mass, and whether any variation of the construction might
provide a solvable model.

The $T^6/\IZ_2^3$ orientifolds were the first with ${\cal N}=1$ supersymmetry
in $D=4$ to be constructed \refs{\bianchi,\berkoozleigh}, and have
been much studied since (see \angelreview\ for a review).  Other
important and relevant references on orientifold and orbifold
compactification include
\refs{\gimonpol,\dougmoore,\gopamukh,%
\blumzaf,\dabpark,\poltensor,\BerkoozIZ,%
\alda,\DasguptaSS,\angelantonj,\KleinTF,\KleinHF,\KleinQW,\blumentwo}.

We start with IIB string theory on $T^6\cong (T^2)^3$, with
coordinates $z_i$ (with $i=1,2,3$) and complex structure moduli
$\tau_i$ (so $z_i\cong z_i+1\cong z_i+\tau_i$).  Following
\refs{\gimonpol,\dougmoore}, we then define an orientifold by choosing
a set of Dirichlet branes, and then choosing an action of the
orientifold group $\Gamma$ on both the compactification space and on
the Chan-Paton factors of the branes.  We will denote the orientation
preserving subgroup of $\Gamma$ as $\Gamma_0$; the group $\Gamma$ is
then an extension of $\Gamma_0$ by $\IZ_2$.  Let $\Gamma_1$ be the set
of orientation reversing elements.

Let $\Omega$
be the world-sheet orientation reversal, and $R_i$ be
the reflection $z_i\rightarrow -z_i$, {\it i.e.}
$$
R_1:(z_1,z_2,z_3) \rightarrow (-z_1,z_2,z_3)
$$
resp. $R_2$, $R_3$.

The model $T'$ is an orientifold of $(T^2)^3$ by
$\Gamma\cong(\IZ_2)^3$ with the generators
$$ g_1=\Omega R_1;\ g_2=\Omega R_2;\ g_3=\Omega R_3 .
$$
The fixed sets of these group elements are $12$ O$7$ planes.
The four planes fixed by $g_i$ are $z_i=0$,
$z_i=1/2$, $z_i=\tau_i/2$ and $z_i=(\tau_i+1)/2$;
let us denote the corresponding O$7$ planes
as O$7_{ia}$, O$7_{ib}$, O$7_{ic}$, and O$7_{id}$ respectively.
Continuing, the group elements
$g_1g_2$, $g_2g_3$ and $g_3g_1$ fix $48$ lines of
$\IC^2/\IZ_2$ orbifold points, while the group element $g_1g_2g_3$ fixes
$64$ O$3$ planes.
There are twelve groups of D$7$ branes, which will be taken coincident
with the O$7$-planes, and denoted analogously as D$7_{iA}$ with
$i=1,2,3$ and $A=a,b,c,d$.  Finally, we will need D$3$ branes.

In further specifying the $\Gamma$ action, one must make
several discrete choices.  In the closed string sector,
there are two such choices.  First, one must choose an element $\epsilon$
in $H^2(\Gamma_0,U(1))$, here $\IZ_2$, usually called (Neveu-Schwarz)
discrete torsion.  The model $T'$ and our model $T$ have $\epsilon$
trivial, {\it i.e.} no discrete torsion (in the conventions
of \vafawitten; some early references on this orientifold used the opposite
convention).  Second, one can modify the usual geometric actions by
using symmetries which appear after twisting, say a symmetry which
acts as $-1$ on all sectors twisted by a given $\IZ_2$ and $+1$ otherwise,
as discussed (for example) in \poltensor\ and used in
the $T^4/\IZ_2\times \IZ_2$ orientifold of \refs{\blumzaf,\dabpark}.
However, this is not required to define the models $T'$ and $T$.

The closed string spectrum is obtained following standard techniques.
In the untwisted sector, there is the ${\cal N}=1$ graviton multiplet;
$7$ chiral multiplets corresponding to the dilaton,
$3$ complex structure moduli and the $3$ K\"ahler moduli
$dz_i\wedge d\bar z_i$.  The K\"ahler moduli are paired with
RR scalars which transform as four-forms in $T^6$.

Twist sectors in a $D=6$
IIB orbifold without discrete torsion lead to $N=2$ hypermultiplets (and
correspond geometrically to K\"ahler moduli).  The $\Omega$
projection then keeps a single NS scalar and a single RR scalar in each of
the $48$ twist sectors.  The hypermultiplet contained two RR scalars,
one transforms as a scalar under the discrete symmetries of
$T^6$; the other as a two-form; for example as
$dz_3\wedge d\bar z_3$ in the $g_1g_2$ twist sector.  The
projections $g_i=\Omega R_i$ keep the two-form (which corresponds
geometrically to a four-form integrated over an exceptional two-cycle).

The final spectrum
is equivalent to that obtained by compactification on the smooth
Calabi-Yau threefold with $h^{1,1}=51$ and $h^{2,1}=3$ discussed in
\S 2.  This would change if we modify the projections in the
twisted sector, so we conclude that our models use the standard projection.

We next need to specify the $\Gamma$ action on the Chan-Paton factors for
the Dirichlet branes.
This is done by postulating matrices $\gamma(g)$ which
form a projective representation of the orientifold group, in the
following sense:
$$
\gamma(g) \gamma(h) = \epsilon(g,h) \gamma(g~h) \qquad g,h \in \Gamma_0
$$
for the orbifold elements, where $\epsilon(g,h)$ is
a two-cocycle in $H^2(\IZ_2\times\IZ_2,U(1))\cong \IZ_2$
of discrete torsion.
The orientifold elements satisfy
$$
\gamma(g) \gamma(u) = \epsilon'(g,u) \gamma(g~u) \qquad
 g\in\Gamma_0;\ u\in\Gamma_1.
$$
and
$$
\gamma(u) \gamma(v) = \epsilon''(g,u) \gamma(v~u) \qquad
 u,v\in\Gamma_1 ,
$$
where the factors $\epsilon'$ and $\epsilon''$ are like cocycles, but
must carry extra signs to cancel phases produced by the action $R(u)$
of orientation reversal on the world-sheet, as discussed in \gimonpol\ and
subsequent work.

The case which will concern us applies to open strings
between pairs of branes with relative codimension 4 (in our models, all
pairs D$7_i$--D$7_j$ with $i\ne j$ and all D$7_i$-D$3$ combinations).
In this case, one can show that generally $R(\Omega R_i)^2=-1$,
so that one must have
$\gamma(\Omega R_i)\gamma(\Omega R_i)^T=-1$ for these open string sectors to
be non-empty, which is required for consistency.  For example, this
is responsible for the $USp$ gauge groups of D5 branes in the
type I string in flat space; the combination $\gamma(\Omega)=1$ acting
on D9 branes (leading to $SO(32)$ gauge symmetry) with
$\gamma(\Omega)=i\sigma_2\otimes 1$ acting on D5 branes satisfies
$\gamma\gamma^T=-1$.
However, there is a loophole in this argument, as we discuss below.

One then projects the open strings of the underlying toroidal
compactification as
\eqn\openpro{\eqalign{
\gamma(g) \phi \gamma(g)^{-1} = R(g) \phi ; \qquad g \in \Gamma_0 ;\cr
\gamma(u) \phi^{tr} \gamma(u)^{-1} = -(R(g) \phi) ; \qquad u \in \Gamma_1 ;\cr
}}
where $R(g)$ is the world-sheet action of the group element $g$.

Finally, one must check the resulting theory for consistency; in
particular tadpoles for unphysical gauge potentials sourced by
space-time filling branes must cancel.  In principle all such tadpoles
can be seen as one-point functions on the disk and $\IR\IP^2$
diagrams, but this check is usually done at one loop, as these
diagrams (the torus, Klein bottle, M\"obius strip and cylinder) have
canonical normalizations (being traces over one string Hilbert spaces).
Perhaps the clearest discussions of this related to the models at hand
(actually, their double T-duals) appear in \angelreview\ and
\KleinQW, which we refer to for details.

Given that our models do not have discrete torsion, we
can start by considering linear (non-projective) representations
of $\Gamma$.  Since $\Gamma$ is abelian, the irreps are one-dimensional,
each characterized by the action of the generators on the Chan-Paton
factors,
\eqn\Cirreps{
\gamma(g_i) = r_i; \qquad r_i = \pm 1.
}
Let us denote such an irrep as $R(r_1,r_2,r_3)$; we can
think of each D-brane as transforming in one of these irreps.

We will then take $8$ D$7$-branes in each group, transforming as
\eqn\CPrep{\eqalign{
D7_{1A} \qquad R(-1,+1,-1) \cr
D7_{2A} \qquad R(-1,-1,+1) \cr
D7_{3A} \qquad R(+1,-1,-1) .
}}
This leads to the gauge group $SO(8)^{12}$, and would seem to be
in direct contradiction to the condition
$\gamma(\Omega R_i)\gamma(\Omega R_i)^T=-1$ we discussed above.

The loophole which allows this was pointed out in
\refs{\dabpark,\blumzaf}\ -- while the rule $\Omega^2=-1$ applies
to massless open strings, in some theories the massless sector is
projected out.  This is the case here; consider open strings between D$7_{1A}$
and D$7_{2B}$.  The choice \CPrep\ implies that
$$
\gamma(g_1g_2) \phi_{12} \gamma(g_1g_2)^{-1} = - \phi_{12}
$$
and thus the massless strings in this sector are projected out.
On the other hand, strings at half integral mass levels,
for example $\psi_{-1/2}\ket{0}$, survive this projection.
These get an extra sign from $g_3$ and thus we have
$$
R(g_3)^2 = +1
$$
on this sector.
Thus, one instead needs
$\gamma(\Omega R_i)\gamma(\Omega R_i)^T=1$ as is satisfied by the
representations \Cirreps.

The same considerations hold for all of the $7_i-7_j$ sectors
with $i\ne j$, so there is no massless bifundamental matter in this
theory.  The massless $7_i-7_i$ adjoint matter is removed as
well.  For example, consider $X^2$ on $D7_{1A}$; the $g_2$ projection
has the opposite sign from the $g_1$ projection, so together they
project out the zero modes.

Another physical choice which is tied to this is the choice between
``normal'' or ``exotic'' O planes.  The RR charge of an O plane
can be inferred from the M\"obius diagram with boundary on a D-brane $D_i$.
This is proportional to
$$\Tr_i \gamma(u) \gamma(u)^T$$ for $u=\Omega R$, where $R$ is a group
element fixing the plane $D_i$.  If $\gamma(u)^2=1$, this correlates
the sign of the O charge to the symmetry of the matrix $\gamma(u)$.  A
symmetric $\gamma(u)$ has negative charge and is a ``normal'' O plane,
while antisymmetric $\gamma(u)$ has positive charge and is an
``exotic'' O plane.  The corresponding projection leads to $SO$ and
$Sp$ gauge groups respectively for the corresponding D-branes.  Of
course, the gauge groups could be broken further by the other
projections.

The choice \CPrep\ then corresponds to the choice of ``normal''
O$7$-planes in the model $T'$.  This also follows from tadpole
cancellation; the O$7_{iA}$ plane tadpoles are each cancelled by the
corresponding stack of $8$ D$7_{iA}$ branes.

The D$7$ branes also contribute to RR twisted tadpoles, one associated
with each of the 48 fixed lines.  Consider the $g_1g_2$ twisted tadpoles;
these receive contributions from branes extended along $z_3$,
D$7_{1A}$ and D$7_{2A}$, proportional to $\gamma(g_1g_2)$.
One can check that in \CPrep,
$\Tr\gamma(g_1g_2)=\Tr\gamma(g_1)\gamma(g_2)=0$, and that the
other tadpoles cancel as well.  It is this condition which forces
the D$7$'s to be taken in groups of $8$ and thus the $SO(8)^{12}$
gauge group.

This leaves the D3 tadpole.  The $64$ O$3$ planes contribute
through the $\IR\IP^2$ world-sheet diagram constructed using the
$g_1g_2g_3$ identification.  One can also check that the
D$7$ branes do not contribute to this tadpole.
This suggests that
the total result for the tadpole is the same as what would
be obtained in the $T^6/\IZ_2$ orientifold dual to type I theory,
and will cancel if $\Tr_{D3} \gamma(1)=32$, {\it i.e.} if we add
32 D$3$ branes in some $\Gamma$ representation.

However, to get a model with an F-theory
dual, we must take the D$3$ branes in a regular representation.  Then,
massless open $3-7$ strings will survive the projection, and
$\Omega^2=-1$ in the $3-7$ sector.  This will require us to take
$\gamma(\Omega R_1 R_2 R_3)$ antisymmetric, leading to $USp$ gauge
groups on the D$3$ branes, and again corresponding to exotic O$3$ planes.
These contribute to the tadpole with the opposite sign, and thus the
tadpole cannot be cancelled with D$3$-branes.  One can obtain a
consistent model by using anti-D$3$-branes,
now with $\Tr_{anti\ D3} \gamma(1)=32$.

Such a result is known from the world-sheet analysis of these
orientifolds \refs{\angelantonj,\KleinQW} and actually can be
seen just by considering factorization of the Klein bottle amplitude.
In fact, the consistency condition found there is weaker.
Let $\epsilon_i$
be the types of the O7$_i$ planes ($\epsilon_i=-1$ is normal and
$\epsilon_i=+1$ is exotic) and $\epsilon_0$ be the type of the O3 plane.
Then, taking $\epsilon_3=-1$ to define the conventions of charge, one
must have
\eqn\clocon{
\epsilon = \epsilon_0 \epsilon_1 \epsilon_2
}
Thus, with our present choices $\epsilon=+1$ and $\epsilon_1=\epsilon_2=-1$,
we are forced to take exotic O$3$ planes with $\epsilon_0=+1$.
Their tadpole can be cancelled with 32 anti D$3$-branes,
to obtain model $T'$.

Another way to satisfy the constraint \clocon\ is to take
$\epsilon_0=-1$ (normal O$3$ planes) with $\epsilon=-1$ (non-trivial
discrete torsion).  However, one cannot make the choices \CPrep\ in
this case, as taking $SO$ groups for both the D3's and D7's
conflict with the $\Omega^2=-1$ constraint.
It is argued in \refs{\berkoozleigh,\KleinQW} that the only
model solving these constraints is the one found in
\refs{\berkoozleigh,\bianchi} with $USp(16)^4$ gauge group.

Finally, another possibility which may be consistent (we have not
checked this in all detail) would be to choose the
D$3$ branes in the fourth representation $R(+1,+1,+1)$, which would
kill the $3-7$ matter and lead to gauge group $SO(8)^{16}$.  This
model, $T''$, is in many ways the most symmetric, though
it does not have an F-theory dual.

Let us conclude by describing the world-volume theory of the
anti-D3 branes in theory $T'$, as it has some interesting features,
especially in how its moduli space reproduces the resolved Calabi-Yau
geometry discussed in \S 2.
The $\Gamma$ representation for these branes is $4$ copies of
a projective version of the regular
representation, which acts by group multiplication
on an $8$ dimensional Hilbert space $\CH\cong\IC\Gamma$ with basis $\Gamma$,
\eqn\CDthree{
\gamma(g) \ket{h} = \rho(g,h) \ket{g~h}
}
with a cocycle $\rho(g,h)$ defined by
$\rho(\Omega^p g_0,\Omega^q h_0) = (-1)^{pq}$.  Equivalently, it
is $4$ copies of the regular representation of $\Gamma_0$ tensored
with $\gamma(\Omega)=i\sigma_2\equiv\sigma$,
so that the resulting gauge group
is $USp(8)^4$ (where $8$ is the dimension of the fundamental representation).
We will subsequently denote the different gauge factors as $USp(N_i)$
(all $N_i=8$).

As anti-D3-branes in Calabi-Yau compactification break supersymmetry,
we must consider the fermions separately.  From the world-sheet point
of view, an anti-D3 can be defined by reversing the GSO projection;
this changes the sign of the $\Omega$ projection \openpro\ on the fermions
and replaces the adjoint representation with the representation with
the opposite symmetry, for $USp$ the antisymmetric representation.

The rest of the anti-D3 and anti-D3--D7  open string spectrum can be derived
along the lines of \refs{\alda\KleinQW}.  We can start with the usual
probe theory of the $\IC^3/\IZ_2\times\IZ_2$ singularity \Greene,
an $\CN=1$ $U(N)^4$ theory with 12 bifundamental chiral multiplets,
one $(\bar N_i,N_j)$ for each $1\le i\ne j\le 4$, denote these
$Z_{ij}$.  To obtain theory $T'$, we apply the $\Gamma_1$ projection,
which sets
\eqn\spconst{
Z_{ij}=\sigma~Z_{ji}^t~\sigma
}
(with $\sigma\equiv i\sigma_2$) for both fermions and bosons,
to obtain six complex bosons and Weyl fermions
in the $(N_i,N_j)$ for $1\le i<j\le 4$.

Substituting these projected
fields into the potential and Yukawa couplings of the underlying
${\cal N}=1$ theory, one obtains a Lagrangian with a non-trivial potential.
Perhaps the most interesting case of this is the fate of the D-terms
of the ${\cal N}=1$ theory, which were
$$
V_D = \sum_{i=1}^4
  \tr \left(\zeta_i\cdot\bbbone - \sum_{j\ne i} Z_{ij} Z_{ij}^\dagger\right)^2
$$
where $\zeta_i$ are four Fayet-Iliopoulos parameters controlled by the
K\"ahler moduli as in \dougmoore, and satisfying $\sum_{i=1}^4 \zeta_i
= 0$.  These terms also survive the projection, leading to a potential
quadratic in the combinations $H_{ij}=Z_{ij} Z_{ij}^\dagger$.  Most importantly,
it depends on the $\zeta_i$, as must be the case for the resulting moduli
space to depend on the K\"ahler moduli.  This would have been impossible
for an ${\cal N}=1$ supersymmetric probe theory with $USp$ gauge groups, as
Fayet-Iliopoulos terms are not possible for semisimple groups.

To find the moduli space of the resulting theory, we can again
start from the analysis of the underlying $\IC^3/\IZ_2\times\IZ_2$
probe theory, and then study the effect of the orientifold quotient.
In \Greene, it is shown that the $U(1)^4$ theory of a single probe
indeed reproduces the two resolutions of $\IC^3/\IZ_2\times\IZ_2$
discussed in \S 2, depending on the signs of the Fayet-Iliopoulos
parameters.  Since the potential was preserved by the constraints,
a set of $8$ such configurations satisfying \spconst\ will be a solution
of the $T'$ anti-D3 theory.  Furthermore, the orientifold action
$z_i\rightarrow -z_i$ on each anti-D3 is realized as the
gauge transformation
$Z_{ij}\rightarrow \sigma_2 Z_{ij} \sigma_2^{-1}$
in its $USp(2)^4$ subgroup.

To complete the theory, the anti-D3--D7 bosons transform as scalars
under rotations in the dimensions transverse to the D7, while the
fermions transform non-trivially.  Taking this into account, one
should find complex bosons $\phi_i$ in the $(8_i,N_i)$, one for each
of the {\it three} of the $USp(N_i)$ factors $i=1,2,3$ which carry the
same representations appearing in \CPrep, one fermions in the
$(8_i,N_i)$ for $i=1,2,3$, and three fermions in the $(8_i,N_4)$.
This spectrum allows for mass terms for all 3--7 strings proportional
to the 3--7 distance parameters $|Z_{ij}|$, the bosons as
$Z_{ij}Z^\dagger_{ij} |\phi_{iA}|^2$
and the fermions as $\psi_{Ai} Z_{i4} \psi_{A4}$.

From this discussion, another way to search for variations on the
orientifold which would have negative D$3$ tadpole and thus admit
supersymmetric flux vacua would be to look for a candidate
supersymmetric probe theory whose moduli space is the orientifold.
While this can be done in the strict orientifold limit ($\zeta_i=0$)
and even with one non-zero FI parameter \parkrabur,
after some attempts, we suspect there is no weakly coupled
supersymmetric probe theory which realizes this moduli space with
all three of the resolution parameters.

Indeed, a similar result was already found in six dimensions by \ahar,
who found difficulties in constructing a supersymmetric probe theory
for the orientifold of \refs{\blumzaf,\dabpark}.  That orientifold is
closely related to what one obtains by moving away from one D7 plane
here, say $z_3\sim 0$, so perhaps an anti D3-brane probe is also the
appropriate probe there.  More generally, it would be interesting to
have a simple argument for why certain orientifold geometries can be
obtained from antibrane probes and not brane probes.

\medskip

The upshot of this discussion is that, while we can define
consistent $T^6/\IZ_2\times\IZ_2$ orientifolds containing the
$SO(8)^{12}$ gauge group predicted by the F-theory construction,
we have not found a construction of the model $T$ with D3 tadpole $-28$
discussed in \S 2; instead we have a model $T'$ with D3 tadpole $+4$
cancelled by anti D3-branes with gauge group $USp(8)^4$, and another
model $T''$ with D3 tadpole $-4$ cancelled by ``fractional'' D3-branes
with gauge group $SO(8)^4$ and no matter  (we have not fully checked
its consistency however).

One might attempt to start with the second of these models, and
replace the D3-branes with flux, to obtain a fully stabilized model
from a perturbative world-sheet starting point.  This may be
possible, but one needs to stabilize the $51$ complex structure
moduli of the orbifold with discrete torsion (and the dilaton) and
obtain small $|W_0|$ using only $4$ units of flux.  This seems very
optimistic, though not obviously impossible.  Other tricks might
help, for example increasing the flux tadpole by adding $D9$--$\bar
D9$ pairs along the lines of \refs{\blumentwo,\marshiu}.

As we said in the beginning, the failure to find the model $T$ as a
toroidal orientifold is not entirely surprising (in retrospect) as it
has massless nonperturbative states at the orientifold point.
Conversely, the orientifold CFT's incorporate discrete choices which
give mass to all of these states, either exotic O planes or lack
of vector structure \sensethi\ forcing $B\ne 0$, or discrete torsion.
While we see no argument which in principle forbids the existence
of a toroidal or CFT orientifold with flux stabilizing all moduli,
for the models at hand these choices decrease the D3 tadpole to where
such vacua are few or non-existent.

\newsec{Complex Structure Stabilization}

The fourfold $Z$ we are studying has $h^{3,1}(Z)=4$ -- in IIB
language, the Calabi-Yau manifold $Y$ contributes three complex
structure moduli, and the fourth modulus is identified with the
axio-dilaton $\phi$. Achieving moduli stabilization in a controlled
manner in the spirit of \KKLT\ requires a small number $e^K|W|^2$,
where $W \sim \int_Y (F-\phi H) \wedge \Omega$ is the
Gukov-Vafa-Witten superpotential \GVW. Therefore, we proceed in this
section to outline the construction of explicit flux vacua in this
model, that provide the small number. It should be quite clear that
a similar strategy could be applied in more general models, the only
difficulty being that as the number of complex moduli increases,
explicit computations become difficult. In fact, other examples of
explicit flux vacua with small $e^K |W|^2$ ($\sim 10^{-3}$ and less)
were explicitly constructed in \refs{\GKTT,\DDF}, and the
statistical results suggest that much smaller values are attainable
\ddone, with the fraction of vacua having $e^K |W|^2 \leq \epsilon$
scaling like $\epsilon$.

Before blowing up the singularities of the base, its periods are (up
to simple issues of normalization) just a subset of those of the
$T^6/\IZ_2$ orientifold. Resolving the space does not change the
periods of the holomorphic three-form. Therefore, we can use any
flux vacua found in the $T^6/\IZ_2$ orientifold (which has been
studied in some detail in e.g. \refs{\KST,\DGKT}), subject to the
constraints that:

\noindent
$(1)$ The total D3 charge in the fluxes, on the covering space
of the orientifold, should
satisfy:
 \eqn\chargecond{N_{\rm flux} = {1\over (2\pi)^4 (\alpha^\prime)^2}
 \int_{Y} H \wedge F \leq 56~.}
Any difference can be compensated by adding D3 branes. We shall
actually choose to saturate the tadpole condition entirely with flux
in our most explicit example.

One can understand the appearance of the 56 as follows.  We have
taken the integral over all of $Y$; the integral over the
orientifold would reduce this by a factor of 2, yielding the correct
tadpole charge of 28, as derived in \S 2.4. However, for
further ease, we can relate the condition \chargecond\ to one
formulated on the $T^6$ covering space of the orbifold $Y=T^6/\IZ_2
\times \IZ_2$ as well. If we choose a standard basis $C_i$ for
$H^3(T^6,\IZ)$ with e.g. $C_1 = dx^1 \wedge dx^2 \wedge dx^3$ and so
forth, then a symplectic integral basis for $H^3(Y,\IZ)$ is given by
considering $B_i = 2C_i$ (for those forms which are projected in
only, of course). Then $\int_{Y} B_i \wedge B_j = {1\over 4}
\int_{T^6} 4 C_i \wedge C_j = \int_{T^6} C_i \wedge C_j$. So in
fact, we can easily translate any flux vacua satisfying
\eqn\chargeagain{{1\over (2\pi)^4 (\alpha^\prime)^2} \int_{T^6} H
\wedge F \leq 56} on $T^6$, with fluxes in the $C_i$ basis, to flux
vacua on the orbifold which also satisfy the necessary tadpole
condition.

An important caveat is that if we wish to avoid the need to
introduce exotic O3 planes, which contribute differently to the
tadpole than standard O3 planes, we should use only ${\it even}$
fluxes in the construction \Frey.  We will adhere to the use of even
fluxes in our examples; it is quite possible that more general
configurations with odd fluxes and exotic planes would also yield
interesting models.

\noindent
$(2)$ We can only turn on the fluxes in the covering space, which are
consistent with the orbifold action.  This is an 8-parameter family
of choices for both $H$ and $F$, where one is allowed to turn on
fluxes in cohomology classes with precisely one leg along each $T^2$
in the $T^6$.  In fact, for our explicit example, we will restrict
to the even smaller subclass where
\eqn\his{\eqalign{F ~=&~a^0 dx^1 \wedge dx^2 \wedge dx^3 + a (dx^2 \wedge
dx^3 \wedge dy^1 + dx^3 \wedge dx^1 \wedge dy^2 + dx^1 \wedge dx^2 \wedge
dy^3) \cr
&- b (dy^2 \wedge dy^3 \wedge dx^1 + dy^3 \wedge dy^1 \wedge dx^2 +
dy^1 \wedge dy^2 \wedge dx^3) + b_0 dy^1 \wedge dy^2 \wedge dy^3}}
and with similar integer choices $c^0, c, d, d_0$ defining $H$.
As explained above, we will choose only even integers here to
avoid the subtleties explained in \Frey.

The flux vacua of class \his\ were studied in \KST\ and in fact
completely classified in \DGKT\ (while powerful analytical results which are
applicable both to this model and to a wider class were also derived
in \Moore, where the connection between flux
vacua and the attractor mechanism was exploited).
It is a simple matter\foot{At least
for A. Giryavets, whom we thank for significant help!}
to do a computer
search at fixed $N_{\rm flux}$ to find vacua of this form with small $g_s$
and moderately small
$e^K|W|^2$.  Here we present two examples:

\noindent Example A):
At the relevant value $N_{\rm flux} = 56$, the explicit
flux choice
\eqn\fluxis{(a^0,a,b,b_0) = (0,10,-10,28),~~(c^0,c,d,d_0) = (2,2,-2,4)}
yields an interesting vacuum for our purposes.  It saturates the
entire tadpole constraint with flux, so no mobile space-filling
D3s need be
introduced.  The resulting
complex structure is of the form $(T^2)^3$ with all two-tori sharing
the same modular parameter; explicitly
\eqn\values{\tau = .46 + .84i,~~\phi = 7 + 3.64i~}
Using the $SL(2,\IZ)$ S-duality symmetry one can easily shift $\phi$
into the fundamental domain; the resulting string coupling is
$g_s \sim .27$.  The value of $e^K |W|^2$ in this vacuum is $0.100$,
in units where $(2\pi)^2 \alpha^\prime = 1$.
\medskip

\noindent Example B):
Also at $N_{\rm flux} = 56$, the choice
\eqn\fluxtwo{(a^0,a,b,b_0) = (0,8,8,28),~~(c^0,c,d,d_0) = (2,0,0,2)}
is interesting.  The resulting vacuum has $g_s = .21$ and
$e^K |W|^2 \sim .350$.

Both examples have only moderately small tuning parameter
$e^K |W|^2$.  We shall
see in \S6.5\ that
with these values of $g_s$ and the gravitino mass, one
can nevertheless
argue that the leading potential (after including Euclidean D3
instantons and gaugino condensates), which exhibits stabilization of
all moduli, receives only rather small corrections from the known
higher order effects.

We note that following the estimates of \refs{\AD,\ddone},
this model actually has
${\cal O}(10^{13})$ flux
vacua, and most likely yields a minimal $e^{K/2}|W| \sim 10^{-5}$.
By generalizing our ansatz somewhat, we could therefore improve the
values of $g_s$ and $e^K |W|^2$
even in this example with few complex moduli.  Of course
in models with larger numbers of complex structure moduli,
the number of flux vacua is vastly larger, and the tuning
parameter should correspondingly attain smaller values.

\newsec{K\"ahler Moduli Stabilization}

K\"ahler moduli can be stabilized by nonperturbative contributions
to the superpotential, as outlined in \KKLT. There are two sources
of such contributions: gaugino condensation and IIB Euclidean
D3-brane instantons wrapping suitable divisors. In the M-theory
picture, both effects are associated to the existence of divisors of
holomorphic Euler characteristic 1 in the fourfold
\refs{\W,\Vafagcond}.\foot{In certain circumstances, divisors of
$\chi_h > 1$ can contribute \refs{\GorlichQM,\Renata,\Sandip}. The
precise criteria for this to occur are still being determined, and
we will see in \S6\ that any such contributions can be
self-consistently neglected in our example.}

In the following we will show in detail that these effects do indeed
fix the K\"ahler moduli in a controlled regime in our model.
Essentially the idea is that the 48 exceptional divisors are rigid,
so they give 48 D3 instanton contributions to $W$, and the
$SO(8)^{12}$ gauge theory has no matter, so this produces 12 gaugino
condensation contributions. Together this gives enough independent
terms in the superpotential to fix all K\"ahler moduli. If $W_{\rm
flux}$ is sufficiently small (in fact as we will see just below
order ${\cal O}(1)$ is already sufficient in our model), the radii
will be stabilized at values such that $\alpha'$ corrections are
small. Moreover, in this regime the effect of K\"ahler moduli
variations on the complex structure moduli is exponentially
suppressed, justifying integrating out the former first and treating
K\"ahler moduli stabilization separately.

\subsec{Divisors}

As discussed in sections 2 and 3, there are 51 K\"ahler moduli, 3 of
which descend from the radii of $B=Y/\IZ_2=(\IP^1)^3$ we have before
blowing up, and 48 from blowing up the fixed lines. The moduli space
is holomorphically parametrized by complexified divisor volumes
 \eqn\kahlerpar{
  \tau_n = {1 \over \ell_s^4} \int_{D_n} dV + i \, C_4.
 }
Here $\ell_s \equiv 2 \pi \sqrt{\alpha'}$, and we work in ten
dimensional Einstein frame. Our normalization conventions are
described in detail in appendix A. Unless otherwise specified, we
will work directly in the quotient $B=Y/\IZ_2$. A suitable basis of
divisors is given by $\{ R_i, E_{i\alpha,j\beta} \}$, where
$i,j=1,2,3$ and $\alpha,\beta=1,\ldots,4$. The $R_i$ are ``sliding''
divisors, descending from the $(\IP^1)^2 \subset (\IP^1)^3$, and
essentially given by $z_i = {\rm const.}$. The $E_{i\alpha,j\beta}$
($i < j$) are the exceptional divisors obtained by blowing up the
fixed line at the intersection of $z^i=z^{\rm fix}_{\alpha}$ and
$z^j=z^{\rm fix}_{\beta}$, where the $z^{\rm fix}_{\alpha}$ label
the four fixed points of $T^2/\IZ_2$.

The D7-brane stacks wrap the divisors obtained from the fixed planes
$z^i=z^{\rm fix}_{\alpha}$ after blowing up. In line with the
notations of section 2 and 3, we denote these by $D_{i\alpha}$. The
difference the sliding divisors $R$ and these 7-brane divisors $D$
is given by the sum of the exceptional divisors $E$ obtained from
the fixed lines inside the fixed plane under consideration (see
section 2 and below for more discussion). For example
 \eqn\DdivER{
  D_{1\alpha} = R_1 - \sum_{\beta} E_{1\alpha,2\beta} - \sum_{\gamma}
  E_{3\gamma,1\alpha}.
 }
This relation is independent of the chosen resolution.

The contribution to the superpotential from a D3 instanton wrapping
an exceptional divisor $E_{i\alpha,j\beta}$ is
 \eqn\instcontr{
  \Delta W \sim e^{- 2 \pi \tau_{i\alpha,j\beta}}
 }
while gaugino condensation in the $SO(8)$ gauge theory associated to
the divisor $D_{i\alpha}$ contributes
 \eqn\gauginocontr{
  \Delta W \sim e^{-2 \pi \tau_{i\alpha}/6},
 }
which can be expressed in terms of the basis coordinates using
\DdivER. Here we used that $c_2(SO(8))=6$. We will discuss these
contributions and various subtleties in much more detail further on.

\subsec{K\"ahler potential}

After integrating out the complex structure moduli, the classical
K\"ahler potential is given by
 \eqn\Kpot{
  K = K_0 - 2 \log V
 }
where $K_0$ is a constant (the contribution to $K$ coming from the
complex structure part) and $V$ is the volume of the threefold $B$
in Einstein frame and in units of $\ell_s = 2 \pi \sqrt{\alpha'}$.

Denoting our basis of divisors collectively by $\{S_a\}$,
$a=1,\ldots,51$, and expanding the K\"ahler form as $J=y^a S_a$, the
volume is
 \eqn\Vexpr{
  V = {1 \over 6} S_{abc} y^a y^b y^c
 }
where $S_{abc} \equiv S_a \cdot S_b \cdot S_c$ is the triple
intersection product of the divisors.

To compute these triple intersection numbers, we proceed as in the
local model of section 2. First we deduce the triple intersection
numbers between {\it distinct} divisors from the (overcomplete) set
$\{ S_A \}_{A=1}^{63} = \{ R_i, E_{i\alpha,j\beta}, D_{i\alpha} \}$.
This can be done using a local toric model. Then we compute the
triple intersections with two and three equal divisors, i.e.\
$S_{AAB}$ and $S_{AAA}$, using the 12 linear relations (such as
\DdivER) between the $S_A$. These are of the form $r^A S_A = 0$, and
thus give $63 \times 63 \times 12$ equations of the form $S_{ABC}
r^C = 0$. The number of independent equations (far) exceeds the
number of unknowns, so solving them determines all $S_{AAB}$ and
$S_{AAA}$, and gives a nontrivial consistency check of the input of
step 1. The system is sparse, so it can be solved very fast using a
computer.

The reason we can use a local model for step 1 is that a triple
intersection involving indices $i\alpha$ and $i\alpha'$ with
$\alpha'\neq\alpha$ will automatically be zero, because the
corresponding two divisors are manifestly disjoint (this is clear
from \PonePonePonebase). Thus to compute intersections we can fix
$(1\alpha,2\beta,3\gamma)$ and model the local geometry near
$z^i=(z^{\rm fix}_{\alpha},z^{\rm fix}_{\beta},z^{\rm
fix}_{\gamma})$ by $\IC^3$ blown up along three lines meeting at the
origin, as described in section 2 and 3. To model also the sliding
divisors, we compactify this geometry to $(\IP^1)^3$ blown up along
three lines, where the divisors at infinity model the sliding
divisors $R_i$. Toric methods can now be used to derive the triple
intersections between distinct divisors. This depends on the chosen
resolution, asymmetric or symmetric. Both cases are presented in the
figure below.

\ifig\Localresbase{$(\IP^1)^3$ blown-up along three lines that meet
in a point. (a) Asymmetric and (b) symmetric triangulation of the
fan.} {\epsfxsize5.2in\epsfbox{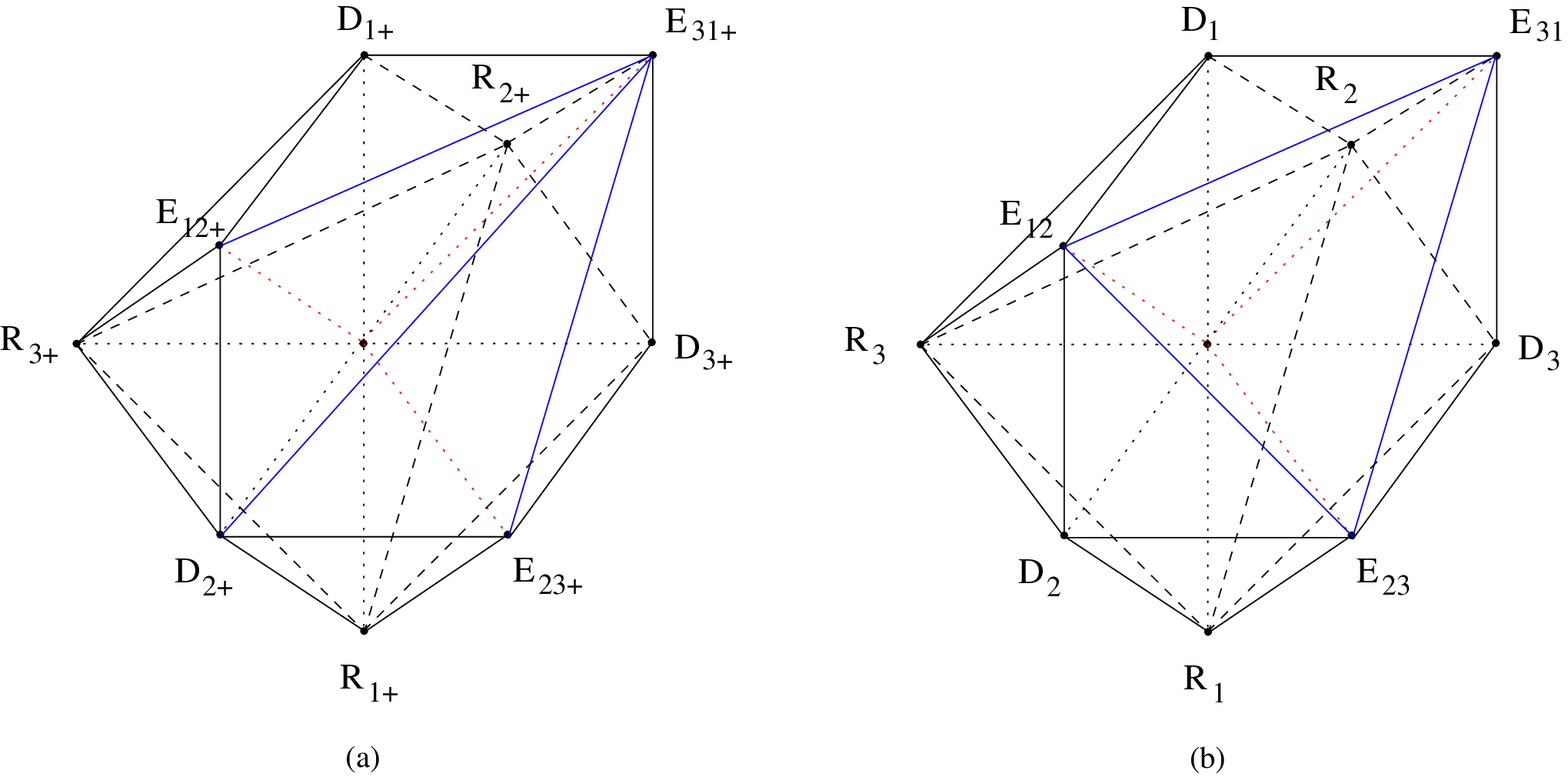}}

Recall that divisors are represented as points (or rather rays from
the origin to these points), with lines between pairs of points
representing intersections, and triangles representing triple
intersections. All such triple intersection products equal 1, with
the exception of $E_{12} \cdot E_{23} \cdot E_{31} = 1/2$ in the
symmetric resolution because of the $\IZ_2$ singularity at the
intersection point. This is in agreement with
\moriconesymmresquotient.

This data suffices to compute all triple intersection products.
Parametrizing the K\"ahler form as
 \eqn\kahlerpar{
  J = r_i R_i - t_{1\alpha,2\beta} \, E_{1\alpha,2\beta}
  - t_{2\beta,3\gamma} \, E_{2\beta,3\gamma} - t_{3\gamma,1\alpha} \, E_{3\gamma,1\alpha}
 }
gives the following expression for the volume $V=J^3/6$:
 \eqn\volsymm{\eqalign{
  V_{\rm symm} = &r_1 \, r_2 \, r_3 - {1 \over 2} \bigl( r_1 \sum_{\beta\gamma} t_{2\beta,3\gamma}^2 + \cdots \bigr)
  - {1 \over 3} \bigl( \sum_{\alpha \beta} t_{1\alpha,2\beta}^3 + \cdots
  \bigr) \cr
  &+ {1 \over 4} \bigl( \sum_{\alpha\beta\gamma} t_{1\alpha,2\beta} \, t_{2\beta,3\gamma}^2
  + t_{1\alpha,2\beta} \, t_{3\gamma,1\alpha}^2 + \cdots \bigr)
  - {1 \over 2} \sum_{\alpha\beta\gamma} t_{1\alpha,2\beta} \,
  t_{2\beta,3\gamma} \, t_{3\gamma,1\alpha}
 }}
if everything is resolved symmetrically, and
 \eqn\volasymm{\eqalign{
  V_{\rm asymm} = &r_1 \, r_2 \, r_3 - {1 \over 2} \bigl( r_1 \sum_{\beta\gamma} t_{2\beta,3\gamma}^2 + \cdots \bigr)
  - {2 \over 3} \bigl( \sum_{\alpha \beta} t_{1\alpha,2\beta}^3 + \sum_{\beta \gamma} t_{2\beta,3\gamma}^3
  \bigr) \cr
  &+ {1 \over 2} \bigl( \sum_{\alpha\beta\gamma} t_{3\gamma,1\alpha} \, t_{1\alpha,2\beta}^2
  + t_{3\gamma,1\alpha} \, t_{2\beta,3\gamma}^2 \bigr)
 }}
if everything is resolved asymetrically with distinguished direction
$i=2$. Here ``$+\cdots$'' means adding $(1,2,3)$ cyclically permuted
terms. Many other mixed symmetric/asymmetric resolutions are
possible of course, but we will restrict to these two.

As explained in the previous subsection, the proper holomorphic
coordinates on moduli space are the complexified divisor volumes,
rather than complexified $y^a$. The divisor volumes are given in
terms of the latter by
 \eqn\divvol{
  V_a = \partial_{y^a} V = {1 \over 2} S_{abc} y^b y^c.
 }
It is not possible in practice to invert these relations explicitly,
but fortunately this is not necessary to find critical points.

We will discuss corrections to the K\"ahler potential below.

\subsec{Curve areas and K\"ahler cone}

Let $\{ C_r \}$ be a basis of the Mori cone, i.e.\ the cone in
$H_2(B,\IZ)$ of effective holomorphic curves. The geometric moduli
space for a fixed resolution is given by the dual cone in
$H^2(B,\IR)$, i.e.\ the space of K\"ahler forms $J$ for which the
curve areas $A_r \equiv C_r \cdot J > 0$. This is the K\"ahler cone.

When finding critical points of $W$, it is important to verify that
they actually lie in the K\"ahler cone. To do this, we need a set of
generators $C_r$ for the Mori cone, or equivalently their areas
$A_r$ as a function of the coordinates $r_i$, $t_{i\alpha,j\beta}$.
We used the algorithm outlined in appendix B of \DDF, which is
essentially the procedure followed for the local model in section 2,
with the following results. For the symmetric resolution:
 \eqn\morisymm{\eqalign{
  A_{i,j\beta} &= r_i-\sum_{\alpha} t_{i\alpha,j\beta} \cr
  A^{++-}_{\alpha\beta\gamma} &=
  {1 \over 2}(t_{1\alpha,2\beta}+t_{2\beta,3\gamma} - t_{3\gamma,1\alpha}),
 }}
plus cyclic permutations $+-+$ and $-++$ of the latter. The curves
of the second line are exceptional curves produced by the blowup and
will be small near the orbifold point, while those of the upper line
are related to the $\IP^1 \subset (\IP^1)^3$ we have before blowing
up and stay finite in the orbifold limit. Note that the
$A_{\alpha\beta\gamma}$ curves are nothing but the exceptional
curves $\widetilde{C}_i$ of the local model $X/\IZ_2$ discussed in
section 2. For the asymmetric resolution we have
 \eqn\moriasymm{\eqalign{
  A_{i,j\beta} &= r_i-\sum_{\alpha} t_{i\alpha,j\beta} \qquad {\rm except} \quad (i,j)=(1,2) {\rm ~and~ } (3,2) \cr
  A_{\alpha\beta\gamma} &=
  -t_{1\alpha,2\beta}-t_{2\beta,3\gamma} + t_{3\gamma,1\alpha} \cr
  A_{1\alpha,2\beta} &= t_{1\alpha,2\beta} \quad {\rm ~and~} \quad A_{2\beta,3\gamma} =
  t_{2\beta,3\gamma}.
 }}

\subsec{Superpotential}

The contributions to the superpotential from D3 instantons wrapping
the exceptional divisors are of the form \instcontr, and those from
gaugino condensation of the form \gauginocontr. We now discuss these
in more detail.

We start with the D3 instantons. They wrap the exceptional divisors,
which as discussed in section 2 and 3 have $h^{i,0}=0$ for $i>0$.
This is true both for the divisor in the threefold as well as for
its lift to the M-theory fourfold $\widetilde{Z}$ (where the D3
becomes an M5). Therefore, these instantons do give contributions to
the superpotential \W. Moreover, each such divisor is unique in its
homology class, so there can't be cancellations between different
divisors in the same homology class.

The full M5-brane instanton partition function also involves a theta
function, essentially from summing over worldvolume 3-form fluxes
$h$ on the M5 \refs{\WittenHC,\MooreJV}. This could lead to a
problematic suppression of the instanton amplitude by a factor
$e^{-c/g_s}$, or even to complete cancellation between contributions
at different values of $h$. A similar phenomenon is observed for D3
instantons.\foot{We thank G.~Moore for several discussions on these
issues and on the topological constraints discussed below.}
Fortunately, in our case, as argued in section 2.3 and proven in
section 3.8, the third cohomology of the M5 divisors is trivial, so
there is no sum over $h$, and this problem does not occur. From the
IIB point of view (related by straightforward KK reduction on the
elliptic fiber) the sum is absent because the $\IZ_2$-twisted second
cohomology group of the D3 divisor is trivial.

Additional zeros of the partition function can also arise when
mobile D3-branes are present or when $h^{2,1}$ of the fourfold is
nonzero \GaZ, but since we saturate the tadpole by fluxes, there are
no D3-branes, and as we saw in section 2 and 3,
$h^{2,1}(\widetilde{Z})=0$.

The M5 worldvolume theory can have global anomalies \refs{\Wi,\DFM},
related to the Freed-Witten D-brane anomalies \FW. The anomaly
analysis for D-branes in orientifolds is quite subtle and does not
follow directly from the results of \FW. A systematic analysis has
not yet been carried out in the literature.\foot{Note added: we were
informed by D.~Freed and G.~Moore that they obtained the general
anomaly cancellation conditions for orientifolds \fmorient, and that
for our model, it appears that indeed the anomaly can be cancelled
by turning on a suitable $B$-field $\sum_{i\alpha} D_{i\alpha}/2$
mod 1. This value of $B$ also cancels the unwanted ``anomalous''
tadpole contributions, closely related to the Freed-Witten anomaly,
which in general appear on non-spin D-branes \MinasianMM.} From the
M5-brane point of view it is clear though that there are no
anomalies in the case at hand. The M5 analog \Wi\ of the half
integer shift in the quantization of the D-brane gauge field
strength \FW\ is absent because $H^3(M5,\IZ)=0$. The torsion anomaly
\refs{\Wi,\DFM}, which is the analog of the $W_3=H$ anomaly for
D-branes, vanishes as well. One way to see this is that when
$H^3(M5,\IZ)=0$, the function $\Omega:H^3 \to \IZ_2$ of \Wi\ is
trivial, and therefore the anomaly constraint (5.10) of \Wi\ becomes
$G|_{M5} = 0$ (in integral cohomology), whose de Rham part is just
the usual tadpole cancellation condition for the M5 2-form field.
This is clearly satisfied in our case as the only 4-fluxes we turn
on are Poincar\'e dual to flat sliding Lagrangian cycles in the
$(T^2)^4/\IZ_2^3$ limit, and these manifestly have zero intersection
with any of the divisors we are considering. An alternative and in
some cases more accurate way of seeing the absence of this anomaly
is by using the criterion in terms of the 8-cohomology class
$\Theta_X$ of \DFM.

Next, consider the contributions from gaugino condensation. For
brevity we will restrict to the symmetric resolution from now on.
(It turns out that if one proceeds with the asymmetric resolution as
we will do for the symmetric one in the following, no critical
points of $W$ are found within the K\"ahler cone, or at least no
simple ones, so we do not lose much by making this restriction.) As
explained in section 2, the 7-brane stacks wrap 12 disjoint rigid
divisors of topology $\IP^1 \times \IP^1$, and each stack consists
of an O7 with four D7's on top, giving rise to an $SO(8)^{12}$ gauge
theory without any light charged matter. The latter is true at least
if no mobile D3-branes are close to the 7-branes and if no gauge
flux is turned on in the D7-branes. But since we saturate the entire
D3 tadpole with 3-form flux, and because nothing forces us to turn
on gauge flux in the branes,\foot{In general, the Freed-Witten
anomaly \FW\ may force turning on gauge flux on branes, but since
$\IP^1 \times \IP^1$ is spin, there is no anomaly here.} these
provisos are indeed met.

Thus, for the symmetric resolution, in our normalization
conventions, the dimensionless superpotential $\widetilde{W}$ (see
appendix A) we consider is:
 \eqn\superpotential{
  \widetilde{W} = \widetilde{W}_0 + \sum_{i<j,\alpha,\beta}
  b_{i\alpha,j\beta} \, e^{-2 \pi \tau_{i \alpha,j \beta}} + \sum_{i,\alpha} c_{i\alpha} \, e^{-2 \pi \tau_{i
  \alpha}/6},
 }
where $\widetilde{W}_0$ is the flux contribution discussed in
section 5. There can be higher order terms, for example from
multi-wrapped instantons. But we will see below that at the critical
point, these can be neglected.

After integrating out the dilaton and complex structure moduli,
$\widetilde{W}_0$, $b_{i\alpha,j\beta}$ and $c_{i,a\alpha}$ are
constants. Because of the rich dependence of $W$ on the K\"ahler
moduli, one expects critical points for generic values of these
constants.

To verify this explicitly, let us first try to simplify the general
form of $W$. First, of course, we can pick a K\"ahler gauge in which
$\widetilde{W}_0$ is real and negative. Because of the symmetry
between the divisors,\foot{The complex structures could break this
symmetry, but in the explicit examples given in the previous
section, this is not the case.} we can also assume that the absolute
values of all D3 instanton and all gaugino condensation coefficients
are equal, i.e.\ $|b_{i\alpha,j\beta}|=b$ and $|c_{i\alpha}|=c$.
There could be phase differences, as long as the phases form a
representation of the symmetry group. However, because the K\"ahler
potential does not depend on the imaginary parts of the 51
coordinates $\tau_a$ (i.e.\ the axions), we can redefine these
imaginary parts by shifts $\Im \, \tau_a \to \Im \, \tau_a + c_a$
together with compensating shifts of the phases of the coefficients
without altering the critical point equations,\foot{In the presence
of nonperturbative corrections to the K\"ahler potential, it is no
longer true that the critical point equations remain invariant under
phase shifts. We will show though that these corrections are small
at the critical point.} and so we can effectively put the phases of
any subset of 51 coefficients equal to 1 (keeping in mind the
modified relations between $\Im \, \tau_a$ and the physical axions
defined by the periods of $C_4$). This together with the symmetry is
sufficient to make all 60 phases of the coefficients in
\superpotential\ effectively equal.

A more detailed argument for the last assertion goes as follows. We
can certainly set the 48 phases of the D3-instanton contributions to
zero by shifting 48 axions, since all 48 exceptional divisors are
linearly independent. Furthermore we can use the remaining 3 axion
degrees of freedom to put the phase of one gaugino condensate in
each of the 3 distinct groups of four to 1. Given the $SL(2,\IZ)$
symmetry within a given group of four, of which the phases must
furnish a representation, the four resulting phases in a group can
be either all $+1$ or two $-1$ and two $+1$. Now note that the
phases of the $SO(8)$ gaugino condensates can still be shifted by
sixth roots of unity. This corresponds to shifting the 51 axions by
various multiples of $2 \pi$, which will not change the phases of
the D3-instantons, but gives more than enough freedom to change any
gaugino condensate phase by an arbitrary sixth root of unity. Since
$-1$ is a sixth root of unity, we can use this to flip the remaining
$-1$ phases to $+1$. Thus we conclude that, after suitable shifts of
the axion coordinates, we can take all $b_{i\alpha,j\beta}=b>0$ and
$c_{i\alpha}=c>0$.

We don't know exact expressions for $b$ or $c$, but fortunately the
stabilization problem is not very sensitive to the precise values,
as long as they are larger than $W_0$. One could assume the
coefficients to be essentially of order 1 in the K\"ahler gauge
where $K_0 \equiv 0$ (with $K_0$ as in \Kpot\ being the complex
structure and dilaton contribution to the K\"ahler potential). A
potential problem in general is that there could be suppression
factors of the form $e^{-c/g_s}$ for the D3 instantons, but as
discussed earlier, these are absent in our model. In fact, if one
takes into account curvature corrections in the Dirac-Born-Infeld
action for the D3 instantons, which induce a negative $(-1)$-brane
charge $-\chi(D)/24$ \refs{\GreenDD,\CheungAZ} and therefore by
supersymmetry a negative contribution to the effective volume, there
will be an {\it enhancement} of the coefficient by a factor
 $$b_{g_s} \sim e^{2 \pi \, \chi(D) \over 24 \, g_s}.$$
At weak string coupling this quickly becomes very large. In the
following we will not take this enhancement into account and just
put $b \equiv c \equiv 1$, keeping in mind that this is likely to be
a significant underestimate. Since larger coefficients give larger
volumes (as $DW=0$ roughly sets $b e^{-2 \pi V_D} \sim
\widetilde{W}_0$), this implies that the volumes at the critical point
we will compute are likely to be significantly below their exact
values.

To summarize, in the K\"ahler gauge $K_0 \equiv 0$, we will take our
superpotential to be
 \eqn\superpotentialansatz{
  \widetilde{W} = \widetilde{W}_0 + \sum_{i<j,\alpha,\beta}
  e^{-2 \pi \tau_{i \alpha,j \beta}} + \sum_{i,\alpha} \, e^{-2 \pi \tau_{i
  \alpha}/6},
 }
with $\widetilde{W}_0<0$ given by the values of $-e^{K/2} |W|$ in
the notation of section 5, and appropriate redefinitions of the
coordinates $\tau_a$ understood to incorporate the required axion
shifts.


\subsec{Critical points}

To simplify things, we will look for candidate critical points on
the locus of moduli space maximally respecting the symmetries
between the divisors. For the symmetric resolution this amounts to
setting
 \eqn\transatz{
 t_{i\alpha,j\beta} \equiv t, \quad r_i \equiv r.
 }
Then the compactification volume, relevant divisor volumes, and
areas of the Mori generators are given by
 \eqn\ansatz{\eqalign{
  V &= r^3 - 24 \, r t^2 + 48 \, t^3 \cr
  V_{i\alpha,j\beta} = V_E &= r t - 3 \, t^2 \cr
  V_{i\alpha} = V_D &= r^2 - 8 \, r t + 16 \, t^2 \cr
  A_{i,j\beta} = A_r &= r - 4 \, t \cr
  A_{\alpha\beta\gamma} = A_t &= {t \over 2}.
  }
 }
To further simplify the problem, we will also restrict our search of
vacua to the (shifted) axion-free case ${\rm Im}\, \tau_a =
0$.\foot{Relaxing this would presumably provide a large number of
additional vacua.} Under these assumptions, the superpotential
\superpotentialansatz\ becomes
 \eqn\Wansatz{
  \widetilde{W} = \widetilde{W}_0 + 48 \, e^{-2 \pi (r t - 3 \, t^2)}
  + 12 \, e^{-2 \pi (r^2 - 8 \, r t + 16 \, t^2)/6}.
 }
Supersymmetric vacua are given by K\"ahler covariant critical points
of $W$, i.e.\ solutions to $D_i W \equiv \partial_{\tau_i} W +
(\partial_{\tau_i} K) W = 0$, or equivalently ordinary critical
points of $e^{K/2}|W|$. Clearly, $e^{K/2}|W|$ expanded around ${\rm
Im}\, \tau = 0$ is quadratic in $\delta \, {\rm Im}\, \tau$, so the
axion-free case is automatically critical in the axion directions.

\ifig\numerics{Numerical results for critical point moduli as a
function of the flux superpotential $|\widetilde{W}_0|$. The
$x$-axis shows everywhere $- \log_{10} |\widetilde{W}_0|$ and the
$y$-axis in (a): $t$, (b): $r$, (c): $V_D$, (d): $V_E$ and (e): $V$,
as defined in \ansatz. Finally (f) shows the estimated relative size
of perturbative and instanton corrections to $V$ and its derivatives
$\partial_t^k V$, $k=1,2,3$, for $g_s \sim
1/4$.}{\epsfxsize14cm\epsfbox{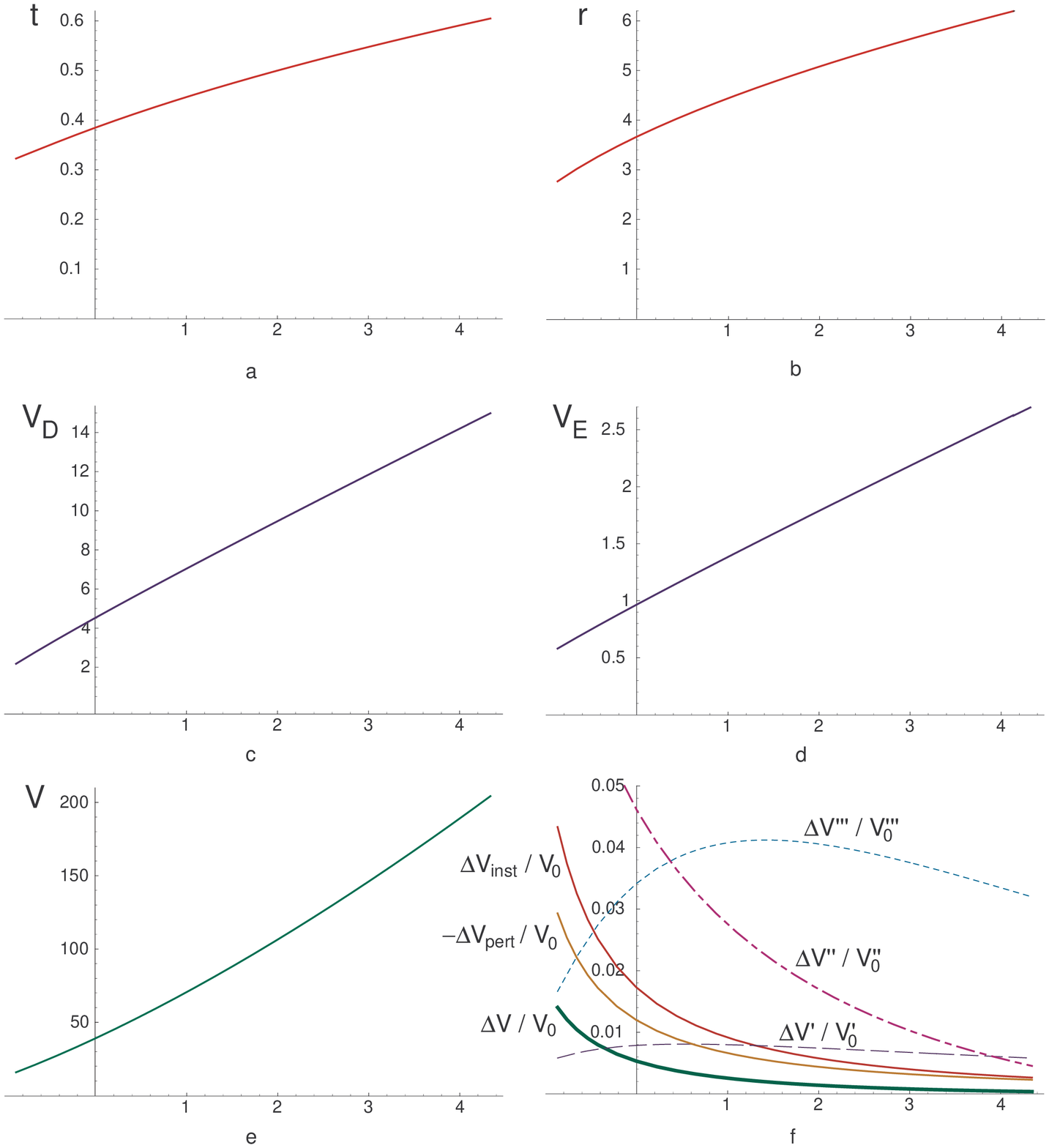}}

The critical points of $e^{K/2}|W|$ as a function of $r$ and $t$ are
easily found numerically. The results for various values of
$\widetilde{W}_0$ are shown in \numerics\ (a)-(e). Because the
critical point equation roughly fixes $e^{-2 \pi V_D/6} \sim e^{-2
\pi V_E} \sim \widetilde{W}_0$, the volume of a $2n$-dimensional
cycle grows as $(-\log |\widetilde{W_0}|)^{n/2}$, which is clearly
reflected by the figure. Note in particular the rather weak
dependence on $\widetilde{W}_0$, allowing solutions with moderately
large volumes up to $\widetilde{W}_0 \sim 10$, which is the order of
magnitude of the effective coefficients in \Wansatz. Similarly, as
we verified numerically, there is only weak logarithmic dependence
on the relative size of the coefficients of the exponentials.

For the explicit example A in section 5, we have $\widetilde{W}_0
\sim 0.3$, so
 \eqn\solution{
  r \approx 4, \quad t \approx 0.4
 }
and the various volumes in \ansatz\ are
 \eqn\solutionvols{
    V \approx 55, \quad V_E \approx 1, \quad V_D \approx 6, \quad A_r \approx 2.5, \quad A_t
    \approx 0.2.
 }
Recall that these are measured in units of $\ell_s = 2 \pi
\sqrt{\alpha'}$. As discussed in the previous subsection, we
actually expect the coefficients of the exponentials in $\widetilde{W}$
to be significantly larger than we assumed at weak string coupling.
Since increasing these coefficients is equivalent to decreasing
$\widetilde{W}_0$, this could easily give values of the volumes
corresponding to an effective $\widetilde{W}_0$ a few orders of
magnitude smaller than what we assumed.

Finally, we should still verify that the critical points we found
are also critical for variations away from the ansatz \transatz, but
because of the symmetry between the divisors, this turns out to be
automatic. We checked this explicitly numerically by computing the
full $DW$ at the candidate critical point. We also checked that
there are no flat directions by computing the full $51 \times 51$
mass matrix. In fact, we found that the supergravity potential is at
a local minimum at this critical point in all directions.

\subsec{Corrections}

There may be higher order nonperturbative corrections to the
superpotential for example from multi-wrapped instantons, from more
complicated divisors of holomorphic Euler characteristic 1, or even
from divisors of holomorphic Euler characteristic $> 1$ \GorlichQM.
Because the divisors we took into account form a basis (in the sense
that all other effective divisors are linear combinations of those
with positive coefficients, and in particular have larger volumes),
these will be suppressed by positive integral powers of the original
exponentials. For the values of \solutionvols, we get $e^{-2 \pi
V_E} \sim 5 \times 10^{-4}$, and $e^{-2 \pi V_D/6} \approx 2 \times
10^{-3}$. Therefore these higher order contributions can reasonably
be expected to be negligible.

Because of the smallness of these exponentials, it also follows that
the dependence of the coefficients $b,c$ on the dilaton and complex
structure moduli will not significantly affect the critical point of
the flux superpotential, as long as the relative derivatives such
as $b'/b$ and $b''/b$ are not huge \KKLT. This justifies integrating
out these moduli first.

The K\"ahler potential receives perturbative and nonperturbative
corrections. The leading order perturbative correction appears at
order $\alpha'^3$ and was derived in the orientifold context in
\BeckerNN. In our conventions, i.e.\ defining $V={\rm vol}(B) = {\rm
vol}(Y)/2$ in units of $\ell_s \equiv 2 \pi \sqrt{\alpha'}$ and in
Einstein frame:
 \eqn\Kpotcorr{
  K = - 2 \ln\left(V + {\xi \over g_s^{3/2}}\right), \qquad \xi \equiv - {\chi(Y) \zeta(3) \over 8 (2
  \pi)^3} \approx - 0.06.
 }
Here we used $\chi(Y)=2 (h^{1,1} - h^{2,1}) = 96$. Since in our
example $V \sim 55$ and $g_s \sim 1/4$, this correction is at the
1\% level and therefore negligible (see \numerics\ (f) for more
general values of $\widetilde{W}_0$). This remains true as long as $g_s
\gg 1/100$.

Finally, there will also be instanton corrections to the K\"ahler
potential. The most important corrections come from fundamental
string worldsheet instantons, whose amplitude (in Einstein frame) is
proportional to
 \eqn\fundws{
  \Delta K \sim e^{-2 \pi \sqrt{g_s} A},
 }
with $A$ a holomorphic curve area on the covering Calabi-Yau space
$Y$, expressed in terms of the areas of the Mori generators of the
quotient $B$ as $A=m A_r + 2 n A_t$, $m,n>0$. The factor of 2 is due
to the fact that the area of the exceptional curves is twice as
large in the covering space than in the quotient. That this is the
appropriate area rather than the quotiented one was explained in
section 2. (At any rate, this does not qualitatively alter the
conclusions.)

Since $e^{-2 \pi A_r} \sim 10^{-7}$ and $e^{-4 \pi A_t} \sim 0.1$
for the values of \solutionvols, it is reasonable (at least for
$g_s$ not too small) to assume that the main contribution will come
from instantons with $m=0$, i.e.\ worldsheets wrapping the
exceptional curves. To estimate their contribution, we use the
multicover formula for the corresponding corrections to the $\CN=2$
prepotential ${\cal F}$ (of which the leading part is essentially
the string frame volume of $Y$, i.e.\ ${\cal F} \sim 2 g_s^{3/2} V$)
\HosonoAV:
 \eqn\instcorr{
  \Delta {\cal F} = {1 \over (2 \pi)^3} \sum_{n=1}^{\infty} {1 \over
  n ^3} e^{-2 \pi \sqrt{g_s}(2A_t) n} \approx 10^{-3}.
 }
The factor $1/(2 \pi)^3$ can be understood from the fact that the
third derivative of the prepotential should give a $q$-expansion
with coefficients equal to the Gromov-Witten invariants. The
prefactor of the exponentials is 1 because the exceptional curves
are isolated and unique. There are $64 \times 3 = 192$ such minimal
exceptional curves, so for the example this gives a total
contribution at the one percent level compared to ${\cal F} = 2
g_s^{3/2} V \sim 14$. Similarly the relative corrections to the
first three $t$-derivatives of ${\cal F}$ are at most at the few
percent level in the entire range of interest.

Actually the above estimate is a bit imprecise, in that the $\CN=2$
type II relation between $e^{-K}$ and ${\cal F}$ is somewhat more
complicated in the presence of corrections, and that there are also
contributions from curves wrapping the sums of two and three
exceptional curves associated to the same triple intersection point
(with Gromov-Witten invariants $-1$ resp.\ $+1$). It is not hard
though to make a more precise estimate taking this into account. The
resulting relative correction is very close to the less refined
estimate given above at the values of example (but unlike the less
refined estimate is actually analytic in $t$ at $t = 0$). We used
this more precise estimate in our numerical computations for
\numerics(f).

These estimates indicate that even for our explicit examples with
modest values of $\widetilde{W}_0$, $\alpha'$ corrections are small.

One could also worry about tensionless strings appearing, for
example from D3-branes wrapping the exceptional curves. However,
since in the range of parameters of interest these curves all have
string scale area, the effective tension of such strings will be
comparable to that of a fundamental string.

Less is known about string loop corrections, but there is no reason
to believe that these will destabilize the solution for the values
of $g_s$ we are considering, and since anyway the number of flux
vacua with $g_s$ less than $g_*$ scales only linearly with $g_*$
\ddone, there should be many more flux vacua with significantly
smaller $g_s$ once we generalize the ansatz of section 5 to allow
the $\tau_i$ to vary independently.

Thus we conclude that all evidence points towards the existence of a
supersymmetric minimum near the value we obtained.

One could also argue more generally that a critical point of the
supergravity potential should exist. It is straightforward to verify
that this potential approaches zero from below in the limit of large
radii and small $g_s$, where we trust the leading approximation.
Therefore, assuming the supergravity potential stays bounded below
in the interior of moduli space, a critical point should exist
(although in general it may be nonsupersymmetric).

\newsec{Models with a single K\"ahler parameter}

There has been some confusion in the literature about the status of
KKLT models with a single K\"ahler parameter \refs{\DDF,\RS}. Here
we show that there exist models for which it is possible in
principle to stabilize the single K\"ahler parameter.

We start with $\pi:Z\rightarrow B$ the elliptic fibration over the base $B=\IP^3$. The corresponding Weierstrass model $W$ is given
by the following hypersurface in $\IP(\CO_B\oplus\CO_B(-2K_B)\oplus\CO_B(-3K_B))$
$$
y^2=x^3+xf+g,
$$
where $f\in H^0(B,-4K_B)$ and $g\in H^0(B,-6K_B)$. The pullback of
the hyperplane divisor in the base has holomorphic Euler
characteristic $\chi_h(\pi^*D)=-2$, and there is no nonperturbative
superpotential generated by threebrane instantons. We can now
enforce ${\tt ADE}$ type singularities of the Weierstrass model
along a hyperplane divisor in the base, say $D_1$ given by $z_1=0$,
where $(z_1:z_2:z_3:z_4)$ are homogeneous coordinates on $B$. In
general, there exists a crepant resolution $\pi:{\widetilde
Z}\rightarrow W$ of the Weierstrass model and the exceptional
divisors contracted by $\pi$ have holomorphic Euler characteristic
$1$.

As an example, we consider a Weierstrass model where we enforce $I_3$ fibers (corresponding to an $A_2$ singularity) along the
divisor $D_1$. This model has a toric crepant resolution; the vertices of the $\nabla$ polyhedron are given by
$$
\eqalign{
&{\tt (1,0,0,2,3),~(1,0,0,1,2),~(1,0,0,1,1),~(0,1,0,2,3),~(0,0,1,2,3),(-1,-1,-1,2,3)},\cr
&{\tt (0,0,0,-1,0),~(0,0,0,0,-1).}
}
$$
The first three divisors are vertical and of holomorphic Euler characteristic $1$. They are elliptically fibered over
$D_1\simeq \IP^2$. However, $D_1$ is not spin, but rather ${\rm spin^c}$. Therefore, in order to cancel the Freed-Witten anomaly
\FW,\ we need to turn on a half-integrally quantized flux on the worldvolume of the D3 brane that wraps the divisor\foot{We
thank S. Sethi for a discussion on this point.}. The requirement
that this flux satisfies the DUY equations can then be satisfied only if the classical volume of the divisor vanishes.

However, there are other $1$-parameter threefolds whose anticanonical
divisor is very ample. Over these threefolds there exists a smooth
Weierstrass model and we can therefore repeat the above analysis. The
only requirement is that the hyperplane divisor in the base is
spin. It is easy to see that the smooth quadric hypersurface
$Q\subset\IP^4$ has this property. Therefore, for this model it is
possible in principle to generate a nonperturbative superpotential for
the single K\"ahler modulus and stabilize it at moderately large
values, provided that turning on fluxes fixes the complex structure
moduli of the Weierstrass model at the singular locus.
The F-theory compactification on this manifold was
described in \gkp.

\newsec{Discussion}

We showed that with existing techniques, one can go a very long way
towards demonstrating the existence of string vacua with all moduli
fixed by a combination of fluxes and nonperturbative effects, by
considering a particular model, explicitly computing the effective
potential and finding its minima.

A next step in developing a proof would be to compute the D3-instanton
prefactors in section 6.  As discussed there these are expected to be
$O(1)$, they are clearly $O(1)$ for the corrections with a gauge
theory interpretation, and the construction will work for a wide range
of $O(1)$ values.  While it would be surprising if there were a
problem at this stage, one would need explicit results to eventually
prove that subleading corrections can be neglected.

Indeed, while one might think that proving the existence of vacua
would require computing the exact scalar potential, all known
corrections to the effects we used can be estimated by analogy to
those found in simpler models, and we were able to argue that they would
spoil the stability of our vacua only if they were far smaller or
larger than expected.  Thus, a proof might only require establishing
rather weak bounds on the corrections.

Here we carried out the construction for a supersymmetric AdS ground state.
It is also quite easy to construct dS vacua in our model at the
four dimensional effective field theory level, by adding a number
of anti-D3 branes and minimizing the total classical potential thus
obtained.  For suitable values of $W_0$, this will give
a local minimum with positive energy.  Unfortunately, in our model,
with no large warp factors and in which $W_0$ is not so small,
this would break supersymmetry near the string scale, so there would be no
small parameter suppressing further corrections to the potential, and these
vacua would not be trustworthy.  However, in similar examples,
one could in principle establish the existence of
better controlled dS vacua at this level with existing techniques.
Besides studying models with a conifold singularity and
adding anti-D3 branes in the warped region \refs{\KS,\gkp,\KPV},
one could also gain extra positive energy at the minimum
by using critical points of the flux potential with
a non-vanishing IASD component for the three-form flux \Saltman.
Since such vacua are generic
and their number scales like $F^6$ \refs{\ddtwo,\dine}, the
number of such dS vacua with small enough F-terms to admit
parametric control is expected to be very large, although such
models may be a small fraction of all candidate vacua.

The model discussed here has several advantages which made a concrete
treatment possible, in particular it starts with $55$ massless fields,
many fewer than the $O(1000)$ typical of ${\cal N}=1$ models with many gauge
groups, and these are further related by known discrete symmetries.
Furthermore, since $b_3$ was small, and the special geometry for this
CY is inherited from that of $T^6$, we could easily find explicit flux
vacua.  However none of this seems essential, nor did any other
features clearly particular to this model enter explicitly into our
discussion.

What must be clearly distinguished are the technical difficulties of
the problem of exhibiting explicit vacua, from the actual constraints
which determine their number.  Even the best posed part of the
problem, finding explicit supersymmetric flux vacua with small $W_0$, is very
challenging.  First, setting up the problem by solving the
Picard-Fuchs equations and finding the explicit change of basis to a
symplectic integral basis for $H^3$ in problems with hundreds of
moduli is highly nontrivial.  Then, since the number of flux vacua
grows exponentially with $b_3$, listing them explicitly or even
finding those with small $W_0$ is computationally difficult.
While the study of K\"ahler stabilization is still in its early stages,
it is very optimistic to think that once that is understood it will be
easier; one expects this will be of comparable difficulty.

What could save the situation to some extent is the general idea that
the problem of stabilizing moduli is largely a problem of satisfying
many nearly independent conditions, which can be studied separately,
as illustrated by the explicit constructions here and in \DDF.  For
example, we have seen that if one turns on only three-form fluxes in
the IIB compactification manifold (and e.g. no field strengths on
various branes), the problem of stabilizing K\"ahler moduli can be
dealt with almost independently of the flux stabilization of complex
structure moduli; the two problems talk to each other, at leading
order, only through the existence of the constant $W_0$ generated by
the flux potential.  Similarly, the problem of stabilizing K\"ahler
moduli depends on the existence of a suitably large set of instanton
corrections, but the existence of one instanton correction does not
directly influence the existence of another.

This rough independence between the conditions is what makes a
statistical approach reasonable, in which one estimates the fraction
of vacua satisfying each condition, and multiplies these estimates to
estimate the total fraction of stable vacua \Douglas.  Indeed, this
approach appears far more widely applicable; for example in finding
flux vacua, while the many fluxes and complex structure moduli do
interact, to a good approximation the different components of the
defining equations ($DW=0$ for supersymmetric vacua) can be treated
independently at each point in moduli space, to estimate the total
number of flux vacua \AD.  The general principle is simply that most
of the many choices and structures in the problem are not directly
relevant to the final outcome (stabilizing moduli, or finding other
observables) and can with due care be approximated or left out to
get these estimates.

In this spirit, the present results provide a start for evaluating the
general claim that stabilized vacua are common, by making the recipe
of \KKLT\ sufficiently concrete that it could be tried in a large
list of models.  The work so far \refs{\DDF,\Sandip}
suggests that stabilizing K\"ahler moduli is generic, but this
remains to be verified.

The model we studied actually has many stringy avatars, and was
identified early in the duality revolution as one of the simplest
fourfolds with many dual descriptions \gopamukh.  It could be
worthwhile to flesh out the physics of our model from suitable
heterotic or type I dual descriptions; perhaps by combining the
insights from various dual pictures, one can obtain a better
understanding of the physics of such stabilized models. These models
also provide simple laboratories where the growing knowledge about
flux-induced soft-breaking terms in D3-D7 systems
\refs{\ibanez,\GL,\JL,\lrstwo,\lmrs} can be combined with full
moduli stabilization to study how the spectrum of soft masses is
modified. In addition, close relatives of this model (which exist at
different discrete choices of the $B$-field in the orbifold
singularities) have played an important role in constructions of
semi-realistic particle physics models
\refs{\blumenone,\blumentwo,\casu,\cvet,\fontib,\lrs,\marshiu}, and
such constructions may also be possible in simple extensions of our
setup.

Another recent suggestion for stabilizing F-theory models with several
K\"ahler moduli in a controllable regime, using the same ingredients as
in \KKLT\ but in a different scaling regime, appeared
in \refs{\BB,\BBtwo};
it would be interesting to see if similarly explicit constructions
are possible in that context.

For the AdS vacua presented here, it would be very interesting to proceed
along the lines of \EvaCFT\ and try to determine a dual CFT.  Since we
only use a few fluxes, the brane constructions
invoked in \EvaCFT\ may be amenable to explicit study.  However,
it is important to keep in mind that for theories of this sort where
the geometrical or gravitational picture is under reasonable control,
one expects the CFT to be strongly coupled.  We have no powerful
techniques for studying isolated, strongly coupled conformal fixed
points, and so far such dualities have mostly been useful
in using the gravity picture to learn about the CFT, as in \EvaCFT.

The existence of the landscape would seem to require
there to be a huge set of such $3D$ CFTs.  Unlike the
classes of vacua which are constructed by a Freund-Rubin ansatz using
a single $p$-form flux threading a $p$-sphere, which are prominent in
the simplest AdS/CFT dualities, the IIB flux vacua involve two different
three-forms threading many cycles of the compact geometry.  For
this simple reason, the degeneracy of expected vacua is much larger
than in the most familiar examples of the Freund-Rubin ansatz.

We also note that in contrast to the simple examples of Freund-Rubin
vacua, in our constructions the compact dimensions can be parametrically
small compared to the curvature scale of the noncompact dimensions (which
is of course required for any model of phenomenological interest).
In particular, there is no KK tower of modes with masses given by
the inverse AdS radius -- while in un-tuned constructions the lightest
moduli will have masses in this range, they do not arise as the
lowest modes in infinite towers.   This means that the $3D$ CFTs which
will arise are rather different than e.g. the theory on $N$ M2 branes
at large $N$.
This is not a surprise;
the properties of weakly coupled theories are always astonishing when
viewed from the strongly coupled side of a dual pair.  In particular,
the properties of weakly coupled or even free field theory seem truly
amazing as predictions about strongly coupled, highly curved gravitational
backgrounds!  Similarly, here the more controlled side of the duality involves
the compactification geometry, and the properties of the strongly
coupled dual field theories are likely to be interesting.
\bigskip
\smallskip
\centerline{\bf {Acknowledgements}}
\medskip

We are grateful to P.~Aspinwall, D.-E. Diaconescu, A.~Giryavets and
especially G.~Moore for many helpful discussions. We also benefited
from discussions with N.~Berkovits, O.~DeWolfe, C.~Florea,
R.~Kallosh, S. Katz, S.~Sethi, E.~Silverstein, W.~Taylor, S. Trivedi
and J.~Walcher. The work of F.D., M.R.D. and B.F. is supported in
part by DOE grant DE-FG02-96ER40949. S.K. was supported in part by a
David and Lucile Packard Foundation Fellowship for Science and
Engineering, the NSF under grant PHY-0244728, and the DOE under
contract DE-AC02-76SF00515.

\appendix{A}{Normalizations}

In this appendix we give our normalization conventions, and give
expressions for the superpotential and K\"ahler potential adapted to
the standard $\CN=1$ four dimensional supergravity framework, where
the four dimensional Planck mass $m_p$ is the only dimensionful
constant appearing explicitly.

We define the string length $\ell_s$ as
 $$
  \ell_s \equiv 2 \pi \sqrt{\alpha'}.
 $$
Then we have in the conventions of \polchinski:
 $$
  {1 \over 2 \kappa_{10}^2} = {2\pi \over \ell_s^8}, \quad
  \mu_p = {2 \pi \over \ell_s^{p+1}}, \quad
  T_p = {\mu_p \over g_s}
 $$
and
 $$
  \widetilde{F}_k \equiv {1 \over \ell_s^{k-1}} F_k \in H^k(Y,{\IZ}).
 $$
We want to write the four dimensional effective action in the
standard $\CN=1$ supergravity form appropriate for dimensionless
scalars, with $W$ of mass dimension 3 and the only dimensionful
parameter appearing the four dimensional Planck mass $m_p$:
 $$
  S = \int {m_p^2 \over 2} R - {1 \over m_p^2} e^K (|DW|^2 -
  3|W|^2) + \cdots
 $$
The relation between $m_p$ and $\ell_s$ is given by
 \eqn\mpdef{
  m_p^2 = {4 \pi \over \ell_s^2} {V_s \over g_s^2} = {4 \pi \over \ell_s^2} {V_e
  \over g_s^{1/2}},
 }
where $V_s$ is the string frame compactification volume in units of
$\ell_s$ and $V_e$ the ten dimensional Einstein frame volume in
units of $\ell_s$. In the bulk of the paper we work in Einstein
frame (more precisely we rescale the metric of the compactification
manifold with $g_s^{-1/2}$). Note that this absorbs for example a
factor $1/g_s$ in the action of a D3-brane instanton wrapping a
divisor $D$:
 $$
  S_{\rm inst} = {2 \pi \over g_s} V_s(D) = 2\pi V_e(D),
 $$
where $V(D)$ denotes the volume of $D$ in units of $\ell_s$. The
following K\"ahler potential and flux superpotential give the
correctly normalized four dimensional action:
 $$
  K = - \ln(-i(\tau - \bar{\tau})) - \ln i \int \Omega \wedge \bar{\Omega} -
  2 \ln V_e
 $$
and
 $$
  W_{\rm flux} = {m_p^3 \over \sqrt{4 \pi}} \widetilde{W}_{\rm flux};
  \qquad
  \widetilde{W}_{\rm flux} \equiv \int (\widetilde{F}_3 - \tau
  \widetilde{H}_3) \wedge \Omega.
 $$
This is written in such way that $W$ has mass dimension 3, and such
that only $m_p$ appears as a dimensionful parameter, as it should in
an effective four dimensional theory. To avoid dragging along the
prefactor in $W_{\rm flux}$, we use the dimensionless superpotential
$\widetilde{W}_{\rm flux}$ in the bulk of this paper.

To verify the correctness of the normalization, it is sufficient to
check the tension of a BPS domain wall between two flux vacua, say
with different RR flux $\widetilde{F}_3$. In the four dimensional
effective theory this is
 \eqn\tension{
  T = 2 e^{K/2} |\Delta W| = {2 m_p^3 \over \sqrt{4 \pi} \sqrt{2/g_s}
  V_e} {|\int \Delta \widetilde{F}_3 \wedge \Omega| \over (\int i \Omega
  \wedge \bar{\Omega})^{1/2}}
 }
In string theory the BPS domain wall can be thought of as a D5-brane
wrapped around a special Lagrangian submanifold $L$ Poincar\'e dual
to $\Delta \widetilde{F}_3$ (so $\int \Delta \widetilde{F}_3 \wedge \Omega =
\int_L \Omega$). It can be shown \BilloIP\ that for a special
Lagrangian $L$,
 $$
  {|\int_{L} \Omega| \over (\int i \Omega
  \wedge \bar{\Omega})^{1/2}} = {V_s(L) \over \sqrt{8 V_s}}
 $$
where $V_s(L)$ is the string frame volume of $L$ in units of
$\ell_s$. Plugging this in \tension\ and using \mpdef\ gives
 $$
  T = {2\pi V_s(L) \over g_s \ell_s^3} = T_5
  {\rm Vol}_s(L),
 $$
which agrees with the BPS domain wall tension obtained in the
D-brane picture, confirming the correctness of our normalizations
including all prefactors.

\listrefs

\end